\documentclass[compsoc,conference,a4paper,10pt,times]{IEEEtran}
\usepackage{enumitem}
\usepackage{setspace}
\usepackage{epsfig}  % for figures
\usepackage{amsmath}  % for math spacing
\usepackage{lscape}  % Useful for wide tables or figures.
\usepackage{mathtools}
\usepackage{bbm}
\usepackage{pifont}
\usepackage[table]{xcolor}
\usepackage{hyperref}
\usepackage{listings}
\usepackage{xcolor}
\usepackage{algorithm}
\usepackage{algpseudocode}
\usepackage{amsfonts}
\usepackage{amsmath}
\usepackage{algpseudocode}
\usepackage{boxedminipage}
\usepackage{stmaryrd}
\usepackage{amsmath}
\usepackage{multirow}
\usepackage{tikz}
\usepackage{bm}
\usepackage{mathpazo}
\usepackage{lineno}
\usepackage{dirtytalk}

\begin{document}
\title{Publicly Auditable MPC-as-a-Service with succinct verification and universal setup}

\author{\IEEEauthorblockN{Sanket Kanjalkar}
\IEEEauthorblockA{Blockstream Research\\sanket1729@blockstream.com}
\and
\IEEEauthorblockN{Ye Zhang}
\IEEEauthorblockA{New York University\\yezhang@nyu.edu}
\and
\IEEEauthorblockN{Shreyas Gandlur}
\IEEEauthorblockA{Princeton University\\sgandlur@princeton.edu}
\and
\IEEEauthorblockN{Andrew Miller}
\IEEEauthorblockA{University of Illinois,\\ Urbana Champaign\\soc1024@illinois.edu}
}
\maketitle
\thispagestyle{plain}
\pagestyle{plain}

%--- Uncomment these to remove notes
% \newcommand{\anote}[1]{{\color{magenta}}}% [AM: #1]}}
\newcommand{\snote}[1]{{\color{blue}}}% [SK: #1]}}

\newcommand{\sanket}[1]{{\color{red} [Sanket: #1]}}

\newcommand{\anote}[1]{{\color{magenta} [AM: #1]}}
% \newcommand{\snote}[1]{{\color{blue} [SK: #1]}}

% --- Macros from babynark
\newcommand{\Gen}{\mathsf{Gen}}
\newcommand{\Enc}{\mathsf{Enc}}
\newcommand{\Dec}{\mathsf{Dec}}
\newcommand{\pk}{\mathit{pk}}
\newcommand{\sk}{\mathit{sk}}
\newcommand{\bbZ}{\mathbb{Z}}
\newcommand{\bit}{\{0,1\}}
\newcommand{\la}{\leftarrow}
\newcommand{\ninN}{{n \in \mathbf{N}}}
\newcommand{\cF}{\mathcal{F}}
\newcommand{\cG}{\mathcal{G}}
\newcommand{\RF}{\mathsf{RF}}
\newcommand{\Half}{\frac{1}{2}}
\newcommand{\F}{\mathbb{F}}
\newcommand{\Adv}{\mathcal{A}}
\newcommand{\Ext}{\mathcal{E}}

\newcommand{\ignore}[1]{}

\newcommand{\samples}{\overset{\$}{\leftarrow}}
\newcommand{\hash}{\ensuremath{\mathcal{H}}}
\newcommand{\doubleplus}{+\kern-1.3ex+\kern0.8ex}
\newcommand{\mdoubleplus}{\ensuremath{\mathbin{+\mkern-10mu+}}}

\newcommand{\share}[1]{ \ensuremath{{ \llbracket {#1} \rrbracket }} }
\newcommand{\bigO}{\mathcal{O}}
\newcommand{\overbar}[1]{\mkern 1.5mu\overline{\mkern-1.5mu#1\mkern-1.5mu}\mkern 1.5mu}

\newtheorem{thm}{Theorem}[section]

\newcommand{\vect}[1]{\boldsymbol{#1}}
\newcommand{\todo}[1]{{\color{red} TODO: #1}}
%----Abstract-------------------------------------------------------------------
\begin{abstract}
In recent years, multiparty computation as a service (MPCaaS) has gained popularity as a way to build  distributed privacy-preserving systems like blockchain trusted parameter setup ceremonies, and digital asset auctions.
We argue that for many such applications, we should also require that the MPC protocol is \textit{publicly auditable}, meaning that anyone can check the given computation is carried out correctly --- even if the server nodes carrying out the computation are all corrupt.
In a nutshell, the way to make an MPC protocol auditable is to combine an underlying MPC protocol with a verifiable computing proof (in particular, a SNARK).
Building a general purpose MPCaaS from existing constructions would require us to perform a costly ``trusted setup" every time we wish to run a new or modified application.
To address this, we provide the first efficient construction for auditable MPC 
that has a one time universal setup.
Despite improving the trusted setup, we match the state-of-the-art in asymptotic performance: the server nodes incur a linear computation overhead and constant round communication overhead compared to the underlying MPC, and the audit size and verification  are logarithmic in the application circuit size. 
We also provide an implementation and benchmarks that support our asymptotic analysis in example applications.
Furthermore, compared with existing auditable MPC protocols, besides offering a universal setup our construction also has a 3x smaller proof, 3x faster verification time and comparable prover time. 
\end{abstract}
%-------------------------------------------------------------------------------

% \begin{IEEEkeywords}
% \sanket{Do we do this?}
% babySnark do do do
% \end{IEEEkeywords}

%-------------------------------------------------------------------------------
% Introduction
%-------------------------------------------------------------------------------
\section{Introduction:}
\label{sec:intro} 
The past few years have seen increasing interest in the Secure-Multiparty-Computing-as-a-Service (MPCaaS) model. 
MPCaaS is a distributed system where a quorum of servers provide confidential computing service to clients. All its security guarantees (including confidentiality, integrity, and optionally availability), rely on an assumption that at least some of the servers are honest (either a majority of the servers or even just one, depending on the protocol).
This model is flexible and well-suited to a range of applications including auctions and digital asset trading~\cite{massacci2018futuresmex,cartlidge2019mpc}, anonymous messaging systems~\cite{alexopoulos2017mcmix,lu2019honeybadgermpc}, computing statistics on confidential demographics data~\cite{lapets2016secure,rajan2018callisto}, and for trusted parameter generation in other cryptography applications~\cite{williamson2018aztec}.

An important research focus in making MPCaaS practical has been to reduce the necessary trust assumptions to a minimum. Malicious-case security for confidentiality and integrity guarantees has now become a standard feature of most implementations~\cite{keller2020mp,chida2018fast,wang2017global,barak2018end}, and protocols like HoneybadgerMPC~\cite{lu2019honeybadgermpc} and Blinder~\cite{abraham2020blinder} furthermore guarantee availability in this setting as well. 

% The MPCaaS model has also benefited from the significant successful research efforts to improve the underlying MPC protocols and implementations.
% There are now several generic MPC compilers and frameworks like MP-SPDZ~\cite{mpspdz} that are easily customized for different applications and support different protocol back-ends. 
% There are tools like MATRIX for orchestrating and deploying them in cloud environments.

%  \todo{move to related work: why not just use blockchains?}
% The use of MPC here is also complementary to other techniques like blockchains and zero knowledge proofs. Although MPCaaS may make use of a bulletin board, which could well be instantiated using a blockchain, MPC essentially provides a missing confidentiality layer. For applications like those mentioned above, where confidential inputs to the computation are provided by different parties, zero-knowledge proofs alone cannot be used.

%This model become more practical since Protocols for this model are based on secure multi-party computation (MPC) protocols, except the clients delegate the computation to dedicated server nodes rather than computing it directly. 
%providing availability, integrity and confidentiality guarantees,

\textbf{The need for public auditability.}
The present work aims to reduce the trust assumptions for practical MPCaaS even further. Our focus is \emph{publicly auditable MPC}, which can best be understood as a form of graceful degradation for MPC security properties, as summarized in Table~\ref{table:mpc_compare}. 
In the ordinary MPC setting, both confidentiality (\textsf{Conf}) and integrity (\textsf{Int}) only hold when the the number of corrupted parties is less than a threshold $t$; in Robust MPC~\cite{lu2019honeybadgermpc}\cite{abraham2020blinder}, availability (\textsf{Avail}) holds under these conditions too. Let $f$ refer to the number of parties \emph{actually} corrupted, such that $f > t$ means the ordinary assumptions fail to hold.
Auditable MPC enables anyone to verify the correctness of the output, ensuring that the integrity guarantees hold even when $t < f$.
Note that this notation describes equally well both the honest majority setting $t<n/2$ or $t<n/3$ (like Viff~\cite{damgaard2009asynchronous}, HoneyBadgerMPC\cite{lu2019honeybadgermpc}, HyperMPC~\cite{barak2018end}, or any Shamir sharing based MPC), as well as the dishonest majority setting $t<n$ (like SPDZ~\cite{damgaard2013practical} and related protocols).
The complementary relation between auditability and other MPC qualities is summarized in Table~\ref{table:mpc_compare}.
To give more context, the integrity guarantee is that the computation, if it completes, is performed correctly, i.e. the correct function is applied to the specified inputs. For blockchain applications which use SNARK, it is important ensure the setup ceremony\cite{williamson2018aztec} for parameter sampling is carried out correctly even if all participants are compromised. As another example, in an digital asset auction, we would want to know that the quantity of digital assets is conserved. Note that in these applications, integrity may matter even to users who did not themselves provide input (randomness or bids) to the service. Assuming a \textit{robust offline phase}\cite{abraham2020blinder}( Section~\ref{sec:mpc_prelim}), we show a construction of robust auditable MPC.

\begin{table}[!htb]
    \caption{Graceful Degradation of MPC protocols depending on number of actual faults $(f)$, versus \hspace{\textwidth} fault tolerance parameter $(t)$}
    % \begin{minipage}{.45\linewidth}
      \centering
        \begin{tabular}{c | c | c}
            & $f \leq t$ & $f > t$\\
            \hline
            non-robust MPC & \textsf{Conf, Int} &  \\
            \hline
            \rowcolor{lightgray} 
            non-robust auditable MPC & \textsf{Conf, Int} & \textsf{Int}\\
            \hline
            robust MPC & \textsf{Conf, Int, Avail} &  \\
            \hline
            \rowcolor{lightgray}
            robust auditable MPC & \textsf{Conf, Int, Avail} & \textsf{Int}\\
        \end{tabular}
    \label{table:mpc_compare}
\end{table}

%A loss of Confidentiality and Availability can lead to loss of privacy and denial of service attacks, but a compromise in Integrity can lead to the wrong election outcome, which might have far worse consequences. To summarize, we want to maintain Confidentiality and Availability when possible, but in the case that it is not possible, we would still like to preserve Integrity.

\begin{figure} 
% \begin{minipage}[t]{1\textwidth}

\centering
\label{fig:overview}
\includegraphics[scale=0.49]{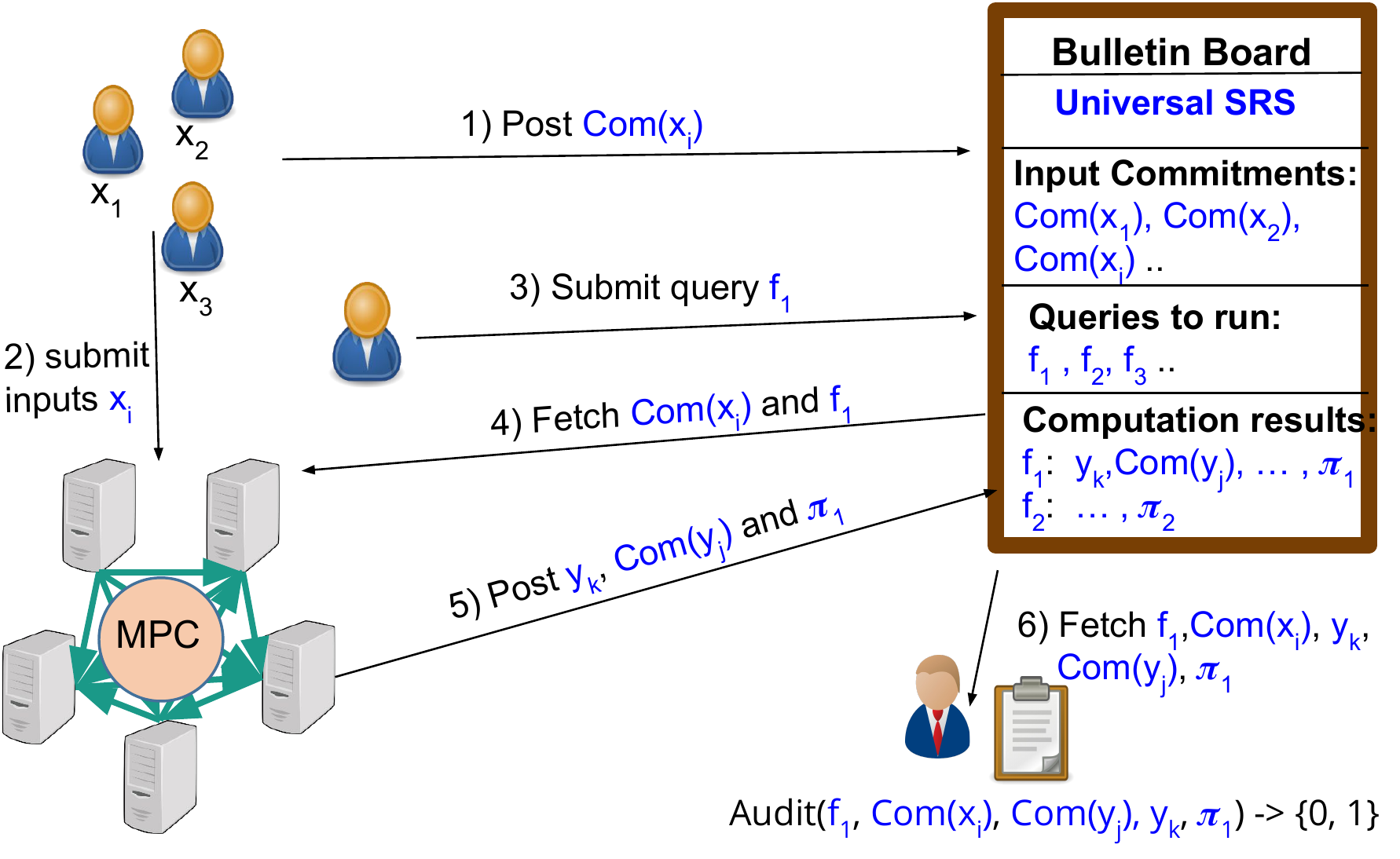}
\caption{Auditable MPC-as-a-Service with one-time trusted setup. Clients with inputs $x_i$ post commitments $Com(x_i)$ and query $f$ to the bulletin board. The servers produce output commitments $Com(y_j)$ and SNARK proof $\pi_1$ onto the bulletin board. The auditor collects the data from bulletin board and verifies the correctness execution of $f$}
\vspace{-5mm}
% \end{minipage}
\end{figure}

\textbf{Auditable MPC with one-time trusted setup.}
In a nutshell, auditable MPC is built from an underlying non-auditable MPC, composed with commitments and zero-knowledge proofs~\cite{stadler1996publicly,baum2014publicly}.  
The resulting arrangement is illustrated in Figure~\ref{fig:overview}. In addition to providing input to the servers, clients also publish commitments to their inputs to a public bulletin board that can be realized by a blockchain. 
The servers, in addition to computing MPC on the secret shared data, also produce a proof that resulting output is computed correctly. Any auditor can verify the proof against the input commitments to check the output is correct. 

The initial version of auditable-MPC by Baum\cite{baum2014publicly} examines the entire protocol transcript to audit the computation. Later Veeningen showed how to construct an efficient auditable MPC from an adaptive Commit-and-Prove zk-SNARK~\cite{veeningen2017pinocchio} (CP-SNARK) based on Pinocchio~\cite{parno2013pinocchio} which is more efficient than \cite{baum2014publicly}.
Adaptive roughly means that it is secure even when the relations are chosen after the inputs (statements) are committed to. 
However, like many SNARKs, Pinocchio relies on a difficult to carry out trusted setup to generate parameters.~\cite{bowe2017scalable,bowe2018multi,ben2015secure}.
Since Pinocchio's trusted setup depends on the particular circuit, there is no way to update the program once the setup is complete. Any bug fix or feature enhancement to the MPC program would require performing the trusted setup ceremony again.

\emph{The goal of our work is to remove this barrier to auditable MPC, by enabling a single trusted ceremony to last for the lifetime of a system, even if the programs are dynamically updated.}
Our approach makes use of recent advances in zk-SNARKs, especially the Marlin zk-SNARK~\cite{chiesa2019marlin}, and adapts it to the auditable MPC setting.
% Here succinct means that both the size of the proofs, as well as the cost to verify them, is sublinear in the size of computation.

\textbf{Technical challenges and how we overcome them.}
First, to summarize our approach, we follow Veeningen and build auditable MPC from an adaptive CP-SNARK. 
To achieve this, we follow the generic framework of {LegoSNARK}~\cite{campanelli2019legosnark}, and compose several existing CP-SNARK gadgets, namely ones for sum-check~\cite{lund1992algebraic}, linear relations, and opening of polynomial commitments, resulting in a new construction we call \emph{Adaptive Marlin}. 

%While a direct transformation of these techniques results in a CP-SNARK that is not zero-knowledge, we can fix this by borrowing query bound zero-knowledge techniques from Marlin.
% The auditor in the MPC runs the verification directly, and the integrity guarantees follow from the CP-SNARK knowledge soundness definition.\sanket{This sentence does not seem to fit here.}

From Adaptive Marlin, to build an auditable MPC requires two more steps. First, the prover algorithm must be replaced with a distributed MPC alternative, leaving the verifier routine essentially the same. Fortunately this turns out to be straightforward; Adaptive Marlin supports distributed computation in a natural way, and the soundness proof remains intact for the verifier.
Second, we must provide a way to combine the input commitments contributed by different clients. 
This poses a greater challenge; in particular, Veeningen's approach to combining input commitments does not work in Marlin since it makes use of the circuit-dependent structure of the Pinocchio CRS.

Our solution is based on a new primitive, \emph{polynomial evaluation commitments}(PEC)\ref{def:pec}, polynomial commitments that can be assembled in a distributed fashion. Each party contributes one evaluation point and they jointly compute a commitment to the resulting interpolated polynomial. This primitive serves as a bridge between LegoSNARK and Veeningen's auditable MPC.
%\todo{Now that we have polynomial commitments, we could make adaptive MPC from any of the snarks from legosnark?}

% In a sense, we combine the ideas of Veeningen and LegoSNARK.
% To bridge between these two approaches, we need to define a new primitive, a Polynomial Evaluation Commitment scheme, which is a an MPC-friendly generalization of Polynomial Commitments.
% These allow the servers to jointly produce a commitment to a polynomial where each party only holds one share of the polynomial. 
% %This commitment scheme borrows core ideas of ``interpolate in the exponent" from VSS literature.
% We instantiate this by adapting the polynomial commitment scheme from Lipmaa~\cite{lipmaa2017prover}.

To summarize our contributions:

- \textbf{Adaptive zk-SNARK with universal reference string and constant verification time.}
     Adaptive Marlin is the first adaptive zk-SNARK for general arithmetic circuits that has $O(\log N)$ verification time, $O(1)$ proofs and relies on a universal reference string.
     This is an asymptotic improvement over LegoUAC, the only known adaptive zk-SNARK with universal reference string~\cite{campanelli2019legosnark}, which has $O(\log^2 N)$ sized proofs and verification time where $N$ is the size of the circuit.
    
- \textbf{Auditable Reactive MPC with one-time trusted setup.}
    Informally, reactive MPC is a type of MPC where the computations to perform may be determined dynamically, even after inputs are provided. By constructing Auditable MPC based on Adaptive Marlin, we avoid the need to run a new trusted setup each time a new program is defined, removing an important obstacle to deployment. We provide our formal security analysis using the same ideal functionality setting as Veeningen, 
    except that we go further in considering the full universal composability environment.

   % \item \textbf{Generation of Pre-processing arguments by MPC:} Marlin gave a new compiler design for constructing zk-SNARKs from Algebraic Holographic proofs and polynomial commitments. However, in Marlin~\cite{chiesa2019marlin}, the prover knows the entire witness and thus easily construct such a marlin proof for R1CS(Rank-1 Constraint System). We show how to create Marlin proof for a relation where the witness and statements are secret shared across a set of parties. In order to facilitate the creation of Marlin proof via MPC,

- \textbf{Implementation.} 
    We implement and evaluate our auditable MPC construction. In our experiments with 32 MPC servers, over 1 million constraints, and 8 statement size, our prover time is about 678 seconds, auditing time less than 40ms, proof size is $\approx1.5$\textsf{Kb}, total MPC communication overhead is a constant $700$\textsf{Kb} with five additional rounds of communication. As an additional contributions, we also implement and evaluate sample application workloads, including an auction and a statistical test (logrank). As a representative figure, in the auction application with 125 bidders (an R1CS with 9216 constraints), with 32 MPC servers, the auditor time is about 50ms, while the time to compute the proof is about 20 seconds, or a total of 4 minutes when including the underlying MPC computations --- overall the auditable MPC is an overhead of 10\% compared to plain (non-auditable) MPC.
    
    In terms of performance, despite not relying on a circuit specific setup, our prover time is comparable to Veeningen's. In some settings, our auditor time and proof size show asymptotic and concrete improvements. For applications where each client contributes only a small input, our auditor has a constant pairing cost and constant proof size, which is asymptotically better than Veeningen's construction that has linear pairing cost and linear proof size in terms of number of clients (input commitments).
    %\sanket{Should we also highlight that our experiments confirm  the asymptotics?
    %}

\ignore{
\todo{The comparison with more protocols below can be moved to related work / appendix}
MPC toolkits like Viff~\cite{damgaard2009asynchronous}, SPDZ~\cite{damgaard2013practical}, Mascot~\cite{keller2016mascot}, Overdrive~\cite{keller2018overdrive}, EMP~\cite{wang2017global}, SCALE-MAMBA~\cite{aly2019scale}, HYPERMPC~\cite{barak2018end} provide Confidentiality and Integrity if number of faults $f$ is less than or equal to the threshold $t$. Toolkits like HoneyBadgerMPC~\cite{lu2019honeybadgermpc} provide Confidentiality, Integrity and Availability if $f\leq t$. However, none of the above mentioned toolkits provide any security guarantees if $f > t$. In this work. we show to how to add auditability to already existing protocols in order to provide Integrity if $f>t$. 
}

%-------------------------------------------------------------------------------
% Background
%-------------------------------------------------------------------------------
\section{Preliminaries}
\label{sec:prelim}
\subsection{Notation}

Let $\langle group \rangle = 
(\mathbb{G}_1, \mathbb{G}_2, \mathbb{G}_T, q, g, h, e)$ where $\mathbb{G}_1, \mathbb{G}_2, \mathbb{G}_T$
are groups of a prime order $q$, $g$ generates $\mathbb{G}_1$,
$h$ generates $\mathbb{G}_2$, and $e: \mathbb{G}_1 \times \mathbb{G}_2 \rightarrow \mathbb{G}_T$ is a (non-degenerate) bilinear map with a security parameter $\lambda$.
Bold letters like $\mathbf{v}$ denotes a vector of elements $[v_i]_{i=1}^n$. $|\mathbf{x}|$ denotes the cardinality of $\mathbf{x}$, while $|A|$ denotes the number of non-zeros elements when $A$ is a matrix.
$\mathbf{x} \circ \mathbf{y}$ denotes the element wise product of $\mathbf{x}$ and $\mathbf{y}$. $\mathbbm{F}_p$ denotes the finite field of prime order $p$ (usually we leave $p$ implicit and write $\mathbbm{F}$), $g(X) \in \mathbbm{F}^{d}[X]$ denotes a polynomial of degree at most $d$.
If $g$ is a function from $H\rightarrow\mathbbm{F}$, where $H \subseteq \mathbbm{F}$ then $\hat{g}$ denotes a low degree extension of $g$ (the smallest polynomial that matches  $g$ over all of $H$).
%$\mathsf{b}$ denotes the bound on number of queries to the polynomial.

%\anote{Preliminaries shouldn't rely on an auction example, lets figure out where else to put this}
% Consider an auction example where parties want to compute the winning bid such that only the winning bid and nothing else is revealed.
\ignore{
We first informally define the security properties of interest for this application:
\begin{itemize}
    \item \textsf{Correctness/Integrity:} The winning bid is correctly computed. A person with a lower bid cannot win
    \item \textsf{Privacy:} Participants only learn the winning bid and nothing else
    \item \textsf{Independence of Inputs:} Participants cannot bid 0.01\$ more than the highest bid to win the auction
    \item \textsf{Fairness:} Participants cannot abort the auction if their bid is not the highest. If one party learns the outcome, all parties must learn the outcome.
    \item \textsf{Guaranteed Output delivery:} Parties cannot abort the protocol.
\end{itemize}
}

\subsection{Secret Sharing and MPC}
\label{sec:sss_prelim}
\label{sec:mpc_prelim}
\label{sec:prelim_mpc_operations}

Secure multi-party computation (MPC) enables parties to jointly compute a function over secret shared inputs, while keeping those inputs confidential --- only disclosing the result of the function.
%The MPC protocol must preserve  computation must preserve security properties, even if some(less than a threshold) of the parties collude and maliciously attack the protocol.

%We discuss the system model and formal security definitions of MPC and auditable MPC in Section~\ref{sec:auditable_overview}.
%Secret sharing is a distributed encoding technique for storing confidential data. 
%The sharing of $s$, denoted $\share{s}$, means that one share is stored by each of $n$ servers, such that the secret can be reconstructed by combining a sufficient number (more than $t$) of shares.
We present our construction for Shamir Secret Sharing (for honest majority MPC), although it is also compatible with other linear secret sharing such as SPDZ (for dishonest majority MPC).
For prime $p$ and a secret $s \in \mathbb{F}_p,$  $\share{s}$ 
denotes Shamir
Secret Sharing~\cite{shamir1979share}(SSS) in a  ($n, t$) setting. 
%In more detail, a degree-$t$ polynomial $\phi : \mathbb{F}_p \rightarrow \mathbb{F}_p$ is sampled such that $\phi(0) = s$. The share $\share{s}^i_t$ is same as evaluation $\phi$ at evaluation point $i \in \mathbb{F}_p$.
% In the honest majority $t<n/2$ setting, we can verify the decoding by checking whether at least $2t + 1$ points.
We omit the superscript and/or subscript when it is clear from context.
For a concrete instantiation in our benchmarks (Section~\ref{sec:evaluation}), we assume a robust preprocessing MPC using Beaver multiplication~\cite{beaver1991efficient} and batch reconstruction~\cite{beerliova2008perfectly,damgaard2007scalable}, similar to HoneyBadgerMPC~\cite{lu2019honeybadgermpc}

\ignore{
\anote{don't need illustration of this}
\begin{figure}
\centering
\label{first}
\includegraphics[scale=0.3]{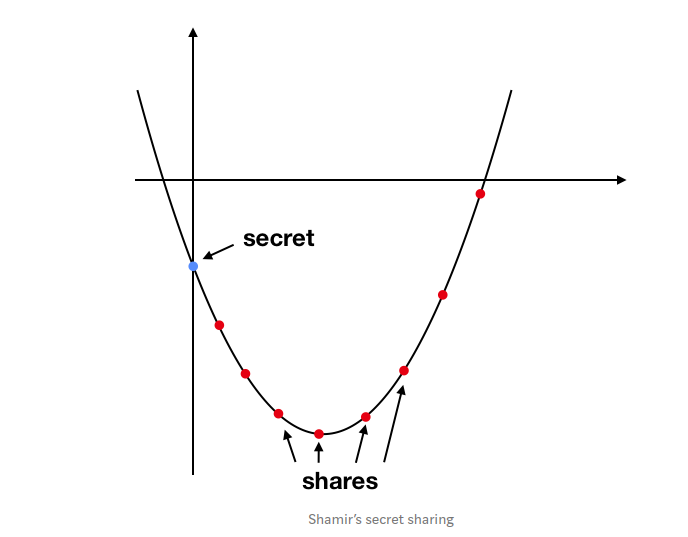}
\caption{Shamir Secret Sharing illustration}
\end{figure} 
}

% \begin{figure*}[!htbp] 
% \begin{boxedminipage}[t]{1\textwidth}
%   {\centering \textbf{Client input processing for ensuring input independence}\\}
% % \quad\emph{// Compute the function}
% Data-clients $I_1, I_2, \ldots, I_m$ have input $\mathbf{x_i} \in \mathbb{F}$ respectively. Input Clients $I_{inp}$ also knows the function $f$ which they want to compute their data upon. 
% \begin{enumerate}
%     \item A trusted third party performs Keygen for a keyed trapdoor commitment family and distributes the keys $ck_i$ to each data client.
%     \item Each data-client computes commitment to the inputs ($C_{ck_i}(x_i)$) and posts it on the bulletin board
%     \item The input-client then choose a function $f$ to compute upon and submits it to the MPC servers for desired computation.
%     \item Each data-client provides the input to the MPC servers. The MPC servers obtain $\share{x_i}$ and for each client check whether the inputs are consistent with the commitment $C_{ck_i}(x_i)$ posted on the bulletin board. 
%     \item If the commitments do not match or a party fails to open the commitment, then abort. Otherwise proceed with MPC computation. 
% \end{enumerate}
% \end{boxedminipage}
% \caption{MPC input independent processing}
% \label{fig:input-independence}
% \end{figure*}

\subsection{Extractable Commitments}
Our construction for auditable MPC relies on a stronger variant of commitment schemes known as extractable trapdoor commitments. For space, we define these in the appendix.

\subsection{Polynomial Commitments}
\label{sec:prelim_polycommit}
Polynomial commitments~\cite{kate2010constant} allow a prover to commit to a polynomial, and later reveal evaluations of the polynomial and prove they are correct without revealing any other information about the polynomial. 
%We use the definition of polynomial commitment  definition from KZG~\cite{kate2010constant} is not immediately sufficient for a SNARK~\cite{maller2019sonic} and
%Marlin~\cite{chiesa2019marlin} gave a formal definition sufficient for standalone use which we state next.
Following Marlin~\cite{chiesa2019marlin}, we define Polynomial Commitments(\textsf{PC}) over $\mathbb{F}$ by a set of algorithms \textsf{PC} = $(\mathsf{Setup, Trim, Commit, Open, Check})$. 
%The definitions in marlin also consider separate degree bounds for a committing to a batch of (optionally hiding) polynomials with opening proofs at multiple different query points.
We only state the definition for creating a hiding commitment to a single polynomial for a single evaluation point with only one maximum degree(Omitting the $\mathsf{PC.Trim}$) bound that is necessary for our application.

\subsubsection{Polynomial Commitment Definitions}
\begin{itemize}[leftmargin=10pt]
    \item $\mathsf{Setup}(1^\lambda, D)$ $\rightarrow \mathsf{ck, rk}$ On input a security parameter $\lambda$ (in unary), and a maximum degree bound $D \in \mathbb{N}$, $\mathsf{Setup}$ samples some trapdoor $\mathsf{td}$ and outputs some public parameters $\mathsf{ck, rk}$ for supporting maximum degree bound $\mathsf{D}$.
    \item $\mathsf{Commit(ck}, \phi; \omega) \rightarrow c$ Given input committer key $\mathsf{ck}$, univariate polynomial $\phi$ over a field $\mathbb{F}$, $\mathsf{Commit}$ outputs commitments $c$ to the polynomial $\phi$ using randomness $\omega$.
    \item $\mathsf{Open(ck}, \phi, q; \omega) \rightarrow v, \pi$ On inputs $\mathsf{ck}$, uni-variate polynomial $\phi$ over a field $\mathbb{F}$, a query point $q \in F$, $\mathsf{Open}$ outputs an evaluation $v$ and evaluation proof $\pi$.  The $\omega$ used must be consistent with the one used in $\mathsf{Commit}$.
    \item $\mathsf{Check(rk}, c, q, v, \pi) \rightarrow \{0, 1\}$: On input reciever key $\mathsf{rk}$, commitment $c$, a query $q \in  F$, claimed evaluation $v \in \mathbb{F}$ at $q$ and evaluation proof $\pi$, $\mathsf{Check}$ outputs 1 if $\pi$ attests that all the claimed evaluation corresponding the committed polynomial.
\end{itemize}

Note that we will later refer to this definition as "plain" polynomial commitments, in comparison to the polynomial evaluation commitments (Section ~\ref{sec:pec_def}).

% \todo{Refer to formal definition in appendix}
% Informally, the $\mathsf{PC}$ scheme must satisfy the following properties:
% \begin{itemize}
%     \item \textbf{Completeness}: Honest proofs for $\mathsf{PC.Commit}$ produced by $\mathsf{PC.Open}$ verify successfully for $\mathsf{PC.Check}$.
%     \item \textbf{Extractability}: For all adversaries $\Adv$ that can produce a commitment $c$ and a evaluation proof $\pi$ for evaluation $v$ at query point $q$ such that $\mathsf{PC.Check}$ accepts, there exists an extractor which is able to output a polynomial $\phi$ such that $deg(\phi) < D$ $\phi(q) = v$. 
%     \item \textbf{Hiding}: The commitment $c$ and evaluation proofs from $\mathsf{PC.Commit}$ and $\mathsf{PC.Open}$  reveals no information about the committed polynomial apart the opening evaluations. 
% \end{itemize}

\subsubsection{Construction in AGM}
\label{cons:pc_agm}
We next describe the Polynomial Commitment construction from Marlin, which is a variation of KZG~\cite{kate2010constant} adapted for the Algebraic Group Model~(AGM)~\cite{fuchsbauer2018algebraic}. In particular it relies on a pairing-based group with the Strong Diffie-Hellman assumption~(\textsf{SDH}), for which a formal definition is given in the Appendix~\ref{app:assumptions}.
%~\ref{app:algebraic} describes the algebraic group model and %~\ref{app:sdh} shows the \textsf{SDH} assumption. 

\begin{itemize}[leftmargin=10pt]
    \item $\mathsf{Setup:}$ Upon input $\lambda$ and $D$, Setup samples random elements in $\mathbbm{F}_q$ and outputs $\mathsf{ck}:= (\langle \mathsf{group} \rangle, \mathbf{\Sigma})$ and $\mathsf{rk}:= (D, \langle \mathsf{group} \rangle, g^\gamma, h^\alpha)$ where $\mathbf{\Sigma}$ is sampled as follows:
    \begin{equation}
    \label{eq:srs}
    \Sigma := \left(
                \begin{array}{llllll}
                    g & g^\alpha & g^{\alpha^2} & \ldots & g^{\alpha^D}\\
                    g^{\gamma} & g^{\alpha\gamma} & g^{\gamma\alpha^2} & \ldots & g^{\gamma\alpha^D} \\
                \end{array}
              \right)
    \end{equation}
    \item $\mathsf{Commit}$: On input $\mathsf{ck}$, univariate polynomial $\phi$ and randomness $\omega$, $\mathsf{Commit}$ operates as follows: If $deg(\phi) > D$, abort. Else, sample a random polynomial $\overbar{\phi}$ of deg($\phi$) according to randomness $\omega$. Output $c := g^{\phi(\alpha)}g^{\gamma\overbar{\phi}(\alpha)}$
    \item $\mathsf{Open}$: On inputs $\mathsf{ck}$, uni-variate polynomial $\phi$ over a field $\mathbb{F}$, a query point $q \in \mathbbm{F}$, $\mathsf{Open}$ outputs an evaluation $v := \phi(q)$ and evaluation proof $\pi:= (\mathsf{w}, \overbar{v})$ as follows: Compute $w(x) = \frac{\phi(X) - \phi(q)}{X - q}$ and $\overbar{w}(x) = \frac{\overbar{\phi}(X) - \overbar{\phi}(q)}{X - q}$ and set $\mathsf{w} := g^{w(X)}g^{\gamma\overbar{w}(X)}$, $\overbar{v} := \overbar{\phi}(q)$.
    \item $\mathsf{Check}$: On input receiver key $\mathsf{rk}$, commitment $c$, a query $q \in  \mathbb{F}$, claimed evaluation $v \in \mathbb{F}$ at $q$ and evaluation proof $(\mathsf{w}, \overbar{v}) = \pi$, $\mathsf{Check}$ outputs as 
    $e(c/(g^vg^{\gamma\overbar{v}}), h ) \stackrel{?}{=} e(\mathsf{w}, h^{\alpha}/h^q ) $
\end{itemize}

\ignore{
\subsection{zkSNARKs}

We informally state the three properties for brevity:
\todo{refer to formal definition n appendix}

\begin{itemize}
\item Correctness/Completeness: If the statement is true, the honest verifier will be convinced of this fact by an honest prover after a successful protocol execution.
\item Soundness: If the statement is false, no cheating prover can convince the honest verifier that it is true, except with some negligible probability. 
\item Zero-knowledge: If the statement is true, no verifier learns anything other than the fact that the statement is true. 
\end{itemize}
}

\subsection{zkSNARKs for R1CS Indexed Relations:}
\label{sec:index_relations}
A zkSNARK is an efficient proof system where a prover demonstrates knowledge of a satisfying witness for some statement in an NP language.
We focus on zkSNARKs for a generic family of computations, based on R1CS relations, a well known generalization of arithmetic circuits.  
For performance, we are interested in succinct schemes where the proof size and verification time are sublinear (or indeed constant, as with our construction) in the number of gates or constraints.

Following Marlin, we define indexed relations $\mathcal{R}$ as a set of triples ($\mathbbm{i}$, $\mathbbm{x}$, $\mathbbm{w}$) where $\mathbbm{i}$ is the index, $\mathbbm{x}$ is the statement instance and w is the corresponding witness. The corresponding language $\mathcal{L}(\mathcal{R})$ is then defined by the set of pairs $(\mathbbm{i}, \mathbbm{x})$ for which there exists a witness $\mathbbm{w}$ such that (($\mathbbm{i}$, $\mathbbm{x}$), $\mathbbm{w}$) $\in \mathcal{R}$. In standard circuit satisfaction case, the $\mathbbm{i}$ corresponds to the description of the circuit, $\mathbbm{x}$  corresponds to the partial assignment of wires (also known as public input) and $\mathbbm{w}$ corresponds to the witness. 

%Rank 1 Constraint System (R1CS) indexed relations are a set of triples
%$(\mathbbm{i}, \mathbbm{x}, \mathbbm{w}) = \left((\mathbbm{F}, n, m, A, B, C), \mathbbm{x, w}\right)$
%where $\mathbbm{F}$ is a finite field, $A, B, C$ are $n \times n$ matrices over $\mathbbm{F}$ with at most $m$ non zero elements, and $\mathbf{z} := (\mathbbm{x} || \mathbbm{w})$ is a vector in $\mathbbm{F}^n$ such that $A\mathbf{z} \circ B\mathbf{z} = C\mathbf{z}$. 

\subsection{Universal Structured Reference Strings}
The vast majority of zkSNARK schemes rely on a common reference string \textsf{crs}, which must be sampled from a given distribution at the outset of the protocol. 
 In a perfect world, we would only need to sample reference strings from the uniform distribution over a field (a \textsf{urs}), in which case it can be sampled using public randomness~\cite{syta2017scalable}.
 However, most practical SNARKs require sampling the reference string from a structured distribution (an \textsf{srs}), which requires a (possibly distributed) trusted setup process~\cite{bowe2017scalable,ben2015secure,bowe2018multi}.

As a practical compromise, we aim to use a universal structured reference string (\textsf{u-srs}), which allows a single setup to support all circuits of some bounded size.
A deterministic or public coin procedure can specialize the trusted setup to a given circuit. This avoids the need to perform the trusted setup each time a new circuit is desired. 
Some \textsf{u-srs} constructions (like the one we use) are also updatable\cite{groth2018updatable}, meaning an open and dynamic set of participants can contribute secret randomness to it indefinitely. Throughout this paper, we refer to \textsf{u-srs} as \textsf{srs} as the universality is clear from the context. 

A zkSNARK with an \textsf{srs} is a tuple of algorithms  $\mathsf{ARG = (G, I, P, V)}$. 
The setup $\mathsf{G}$ samples the \textsf{srs}, supporting arbitrary circuits up to a fixed size. 
The indexer $\mathsf{I}$ is a deterministic polynomial-time algorithm that uses \textsf{srs} and circuit index $\mathbbm{i}$ satisfying the \textsf{srs} constraint bound, outputs an index proving key $\mathsf{ipk}$ and a verification key $\mathsf{ivk}$. 
The Prover $\mathsf{P}$ uses $\mathsf{ipk}$ to provide a proof $\pi$ for indexed relation $\mathcal{R}$.
The Verifier $\mathsf{V}$ then checks $\pi$ using $\mathsf{ivk}$.

% The structured reference string, also is known as \textsf{srs} model, captures the setup in which all involved parties get access to the same string, which has some structure and is generated by trapdoors. This string is often called the \textsf{srs}. No parties must have the knowledge of trapdoors used in constructing these structured reference strings. If the reference string structure is specific to a computation(circuit), it is referred as "circuit specific \textsf{srs}". On the other hand, 

% \subsubsection{Metrics for Comparison of Proof Systems}
% We describe some metrics to compare different proof systems. 
% \begin{itemize}
% \item \textbf{Trusted setup, and it is type:} There can be three broad categories for trusted setup. 1) No trusted setup, 2) circuit-specific trusted setup, and 3) Universal Trusted setup. 
% \item \textbf{Prover time:} The amount of time required by the prover to generate a proof. The common proving time for circuits are $O(n), O(n \log{n})$
% \item \textbf{Verification time:} The amount of time required by the verifier to validate a proof. The common proving time for circuits are $O(1), O(\log{n}), O(n)$. In standard literature, we say that verification is fast if the at-most $O(\log{n})$
% \item \textbf{Proof size:} The size of non-interactive proof generated by the prover. The common proof sizes for circuits are $O(n), O(\sqrt{n}), O(1), O(\log{n})$
% \end{itemize}

\subsubsection{Review of Marlin's construction}
As our construction closely builds on Marlin, we reuse most of its notation, and review its construction here.
The Marlin construction is centered around an interactive ``holographic proof'' technique~\cite{babai1991checking}, combined with polynomial commitments and Fiat-Shamir to make it non-interactive. In a holographic proof, the verifier does not receive the circuit
description as an input but, rather, makes a small number of queries to an encoding of it.
%\todo{check this exlpanation}. 
%These queries are in addition to the the verifier makes to the proofs provided by the prover.  
This deterministic algorithm responsible for this encoding is referred to as the indexer $\mathsf{I}$. 
Marlin focuses on the setting where the encoding of the circuit description and the proofs consist of low-degree polynomials. 
Another way to look at this is that this imposes a requirement that honest and malicious provers are ``algebraic'' (See Appendix~\ref{app:algebraic}).

% The core contribution of Marlin was an AHP for R1CS indexed relations which we describe next. 
In brief, the Marlin protocol proceeds in four rounds, where in each round the verifier $\mathsf{V}$ sends a challenge and prover $\mathsf{P}$ responds back with one or more polynomials; after the interaction, $\mathsf{V}$ probabilistically queries the polynomials output by the indexer $\mathsf{I}$ as well as those polynomials output by the prover $\mathsf{P}$, and then accepts
or rejects. The verifier does not receive circuit index $\mathbbm{i}$ as input, but instead queries the polynomials output by $\mathsf{I}$ that
encode $\mathbbm{i}$. For our construction, we require a MPC version of Marlin protocol  shown in Appendix~\ref{app:mpc_prover}.

%-------------------------------------------------------------------------------
% Overview
%-------------------------------------------------------------------------------
\section{Overview of Our Construction}
\label{sec:auditable_overview}
%In this section we give a description of the auditable MPC-as-a-Service system model. 
%We explain from first principles, since while auditable MPC is not new, we think it is underappreciated in distributed system design. In part this is because the model is inherently nuanced, since the auditability guarantees only kick in when the ordinary assumptions for the underlying MPC fail to hold.
%We therefore aim to give an accurate but intuitive explanation in terms of graceful degradation of service, and relate the system-level explanation to an ideal functionality model that serves as our formal definition.

\subsection{Motivating application: Auction}
\label{sec:motivate_auction}
We start by explaining an auction application that we use as a running example throughout.
We envision a distributed service that accepts private bids from users, and keeps a running tally of the current best price, but both the bids and the price are only stored in secret shared form.
Finally after all users have submitted bids, the servers publish the winning price.
This application can be summarized with the following two procedures, where secret sharing notation $\share{x}$ indicates that $x$ is confidential:
\begin{figure}[h!]
\begin{itemize}[leftmargin=4pt]
\small
\item Initialize state: $\share{\textsf{bestprice}} := \share{0}$
\item \textsf{ProcessBids}(inputs: $\share{x_i}$ from $P_i$,  state:$\share{\textsf{bestprice}}$):
  \begin{itemize}
    \item[] for each $\share{x_i}$ that has not been processed:
       \begin{itemize}
         \item[] $\share{\textsf{newprice}} := \max(\share{x_i}, \share{\textsf{bestprice}})$
         \item[] return $\texttt{OK}, \share{\textsf{newprice}}$
       \end{itemize}
   \end{itemize}
\item \textsf{Finalize}(inputs: $\emptyset$, state=$\share{\textsf{bestprice}})$:    
  \begin{itemize}
    \item[] return $\share{\textsf{bestprice}}.\textsf{open}(),\bot$
  \end{itemize}
\end{itemize}
\end{figure}

Note that we write our example to process arbitrary-size batches of user-submitted bids at a time. This is to illustrate the flexibility of our construction, since it supports reactive computations (each computation can provide public output as well as secret shared output carried over to the next operation) as well as support for large circuits.
In our example, the \textsf{Finalize} procedure also discloses the current state. %This model allows us to capture the scenarios where an input to another completely different style of MPC might be an output of the previous computation. %Let $S_{old}$ and $S_{new}$ denote the state of MPC computation after computing a circuit $C$.
 %$x_i$ denotes the client inputs, $y_j$ denotes the output of the previous computations.
%
In general, each procedure can be characterized by the following quantities: $X$, the total size of secret inputs; $K$, the number of distinct clients providing input in each invocation; and $M$, the total number of gates needed to express the procedure as an arithmetic circuit. 
In our example, each $\textsf{ProcessBid}$ invocation receives $K$ constant-size bids from different parties, so $X=\Theta(K)$, and the circuit comprises a comparison for each bid, so $M=\Theta(Kc)$ where $c$ is the number of gates for each comparison subcircuit.
%To abstract a bit from our example: more generally we consider that the MPC servers maintain some current secret shared values as application state, which can be update by each operation. Each operation can operate on 1) private client inputs, 2) public inputs provided at the time (e.g., from a smart contract), and 3) the current state. 
Since we are building auditable MPC from SNARKs, we are primarily interested in \emph{witness succinctness}, meaning the verification cost is independent of the circuit size $M$, although it will in general depend on $K$ and $X$ (as we explain more in Section~\ref{sec:construction}). When the verification cost is also independent of $X$, we call it \textit{statement succinct}.

\ignore{
Each server would have a share of state $\share{S}$, the share of previous computations output $y_j$, 
a share of a secret client of input $x_i$, and public inputs $x_{pub}$. The MPC state update rule would then be shown by:
    \begin{equation*}
      \share{S_{new}} = \mathsf{MPC}(\textsf{C}, \share{S_{old}}, \share{x_i}, \share{y_j}, x_{pub})  
    \end{equation*}
    The auditor can, at any time, check whether the outputs and the latest state of the MPC are consistent with the operations carried out MPC. 
    }

\subsection{System overview of Auditable MPCaaS}
Auditable MPC is a distributed system architecture for performing secure computations over inputs provided by clients.
 The computation is organized into several phases. For simplicity we describe these as occurring one after the other, though in the general (reactive) setting each phase can occur multiple times and may run concurrently with each other. 
 
\textbf{One-time Setup Phase.} 
The offline phase of our auditable MPC consists of two components. First is the one-time setup for the underlying SNARK and client input commitment scheme; this setup needs only be carried out once, regardless of the circuit programs to evaluate.
% The one-time setup is the source of toxic waste in our scheme and must be carried out by a trusted party or a collection of trusted parties~\cite{bowe2018multi}. 
The second is translating a circuit description into an index format. This is deterministic and anyone can publicly recompute and check this computation.
%In practice, one  assume that proving and verification keys for standard circuits are readily available and would not require recomputing the keys again.

\textbf{Commit Inputs.}
We have data client parties $I_i$ which provide inputs $x_i$ to the computation.
The data-input parties $I_i$ provide commitments of their input on the bulletin board for availability.

\textbf{Define Program.}
The input party $I_{inp}$ which provides the computation function $f$.
We model this as a separate party, but in general this would be chosen through a transparent process, such as through a smart contracts.
Marlin Indexed circuit generated indexer prover and verification keys: The indexer should be run every time there is a request for a new computation indexing or an update to an existing computation. 

\textbf{MPC Pre-processing.} In order to facilitate fast online multiplication of MPC servers, it is typically necessary to prepare offline  Beaver triples and random element shares~\cite{beaver1991efficient}.
% It is possible to have an auditable MPC proof for robust offline phase as done in~\cite{baum2014publicly}.

\textbf{Compute phase.}
Next, data clients post secret shares $\share{x_i}$ their input values to the MPC servers. 
The online phase includes interaction between MPC servers $P_j$ to compute the desired user function and generation of proof of correct execution. The auditor can collect the proofs from the bulletin board and verify that the computation was carried out correctly. Figure~\ref{fig:overview} shows the high-level overview describing the online phase of auditable MPC. 
The servers carry out MPC protocols to compute the function $f(x_i, y_{prev}) = y_k, \textsf{Com}(y_j)$, where $y_k$ is the public output and $\mathsf{Com}(y_j)$ is a commitment to a secret output along with a SNARK proof $\pi$. 

\subsubsection{Audit phase}
The auditor receives the output and verifies that it is correct.
Finally, the auditor verifies computation was carried out correctly by collecting all input commitments $\mathsf{Com}(x_i)$, secret outputs $\mathsf{Com}(y_j)$, public outputs $y_k$ and the proof $\pi$. To completely audit the computation, one would need to verify the MPC pre-processing, circuit indexing along with execution proof. We only consider the costs for verifying proofs because the indexing cost be amortized over multiple uses and because we operate robust offline pre-processing model.

  \section{Polynomial Evaluation Commitments}
\label{sec:pec_def}
\label{def:pec}
The main building block for our auditable MPC construction is a new variant of polynomial commitments called  Polynomial Evaluation Commitments ($\mathsf{PEC}$).
In the original polynomial commitment definition~\cite{kate2010constant}, the committer must have chosen a polynomial before calling the $\text{Commit}$ procedure.
To adapt these for use in MPC, our extended PEC definition supports an alternative, distributed way to create the polynomial commitments:
Each party starts with a commitment to just an evaluation point on the polynomial. 
Next, the evaluation commitments are combined and interpolated to form the overall polynomial commitment. 
The procedure for generating evaluation proofs is similarly adapted.
% create commitments in a distributed fashion. In a PEC, each party knows one evaluation point of the polynomial, creates commitments to these independently, and the resulting evaluation commitments are combined to form the final polynomial commitment.
%
%
%\subsubsection{Definitions}
In our definition below, the changes to plain polynomial commitments (in Section~\ref{sec:prelim_polycommit}) are highlighted $\textcolor{blue}{blue}$. 
Briefly, these are 1) $\mathsf{Setup}$ outputs additional commitment evaluation keys $\textcolor{blue}{\mathbf{ck_{e}}}$ and 2) the $\mathsf{Commit}$ operation is split into $\textcolor{blue}{\mathsf{Commit_{eval}}}$ and $\textcolor{blue}{\mathsf{Interpolate}}$. 
More formally, our polynomial evaluation commitment scheme over a field $\mathbb{F}$ is defined by the following set of algorithms $\mathsf{PEC}$ = $(\mathsf{Setup, Commit_{eval}, Interpolate, Open, Check})$. We index parties by $i$ and the evaluation of the polynomial by $j$, $e_{ij}$ denotes the $j$th evaluation by the $i$th party.
\begin{itemize}[leftmargin=10pt]
    \item $\mathsf{Setup}(1^\lambda, D, \textcolor{blue}{K})$ $\rightarrow \mathsf{\textcolor{blue}{\mathbf{ck_{e}}, ck_{p}, rk_{p}}, td}:$ On input a security parameter $\lambda$, and a maximum degree bound $D \in \mathbb{N}$, number of parties $K \in \mathbb{N}$ , $\mathsf{Setup}$ samples some trapdoor $\mathsf{td}$ and outputs public parameters $\mathsf{\textcolor{blue}{\mathbf{ck_{e}}, ck_{p}, rk_{p}}}$.
    \item \textcolor{blue}{$\mathsf{Commit_{eval}(ck}_{\mathsf{e}_i}, \vect{e_i}, \vect{p_i}; \omega) \rightarrow c_{e_i}$} Given input evaluation committer key $\mathsf{ck_{e_i}}$, univariate polynomial evaluations $\vect{e_i} = [e_{ij}]_{j=1}^d$ at evaluation point $\vect{p_i} = [p_{ij}]_{j=1}^d$($d \leq \frac{D}{K}$) over a field $\mathbbm{F}$, $\mathsf{Commit_{eval}}$ outputs commitment $c_{e_i}$ to the evaluations $\vect{e_i}$ using randomness $\omega$.
    \item \textcolor{blue}{$\mathsf{Interpolate(ck_{p}}, \mathbf{c_{e}}, \mathbf{p}) \rightarrow c$}: On input $\mathsf{ck_{p}}$ and commitment to evaluations $\mathbf{c_{e}} = [c_{e_i}]_{i=1}^k$ at $\mathbf{p} = [\mathbf{p_i}]_{i=1}^k$, $\mathsf{Interpolate}$ outputs a commitment $c$ to the interpolated polynomial $\phi(.)$ corresponding the evaluations of the committed in $\mathbf{c_{e}}$.  
    \item $\mathsf{Open(ck_{p}}, \phi, q; \vect{\omega}) \rightarrow v, \pi$ Same as the $\mathsf{PC.Open}$ as discussed in Section~\ref{sec:prelim_polycommit} where $\phi$ is the polynomial interpolated by the evaluations $\mathbf{e} = [[e_{ij}]^d_{j=1}]^k_{i=1}$. Note that the $\boldsymbol{\omega} = [\omega_i]$ where $\omega_i$ must be the same as the one used in $\mathsf{Commit_{eval}}$ for point at index $i$. 
    \item $\mathsf{Check(rk_{p}}, c, q, v, \pi) \rightarrow \{0, 1\}$: Same as the $\mathsf{PC.Check}$ as discussed in Section~\ref{sec:prelim_polycommit}.
\end{itemize}
Additionally, a $\mathsf{PEC}$ must satisfy the following properties.  

\begin{itemize}[leftmargin=10pt]
    \item \textbf{Perfect Completeness}: Consider an adversary which chooses evaluations $\mathbf{e} = \mathbf[{e_i}]_{i=1}^k$ at evaluation points $\mathbf{p}= [\mathbf{p_i}]_{i=1}^k$ randomness $[\boldsymbol{\omega}]_{i=1}^k$ and query point $q$. Let $c_{e_i}$ denote the commitments to the evaluations $\mathbf{e}$, $\phi$ the interpolated polynomial at $(\mathbf{e}, \mathbf{p})$ and $\vect{V}_{\mathbf{p}}$ the Vandermonde matrix at evaluation points $\mathbf{p}$ respectively. We say that $\mathsf{PEC}$ is complete if the evaluation proofs created by $\mathsf{Open}$ for $\phi$ at $q$ are correctly verified by $\mathsf{Check}$ with respect to the interpolated commitment $c$ that is generated by $\mathsf{Interpolate}(c_{e_i})$. More formally, we say that $\textsf{PEC}$ is perfect complete if the following probability is 1($\Downarrow$ denotes logic implication).
\end{itemize}
    % \begin{figure*}[!htbp] 
% \begin{minipage}[t]{1\textwidth}
      \[
  \Pr\!\! \left[
    \begin{gathered}
     deg(\phi) \leq \mathsf{D} \\
    \Downarrow \\
    \mathsf{Check(rk_{p}}, c, q, v, \pi)\\
  \end{gathered}
    \middle|\\
        \begin{gathered}
    \mathsf{ck_{e}, ck_{p}, rk_{p}, td} \leftarrow \mathsf{Setup}(1^{\lambda}, \!\mathsf{D,\!K})\\
    [\mathbf{e_i, p_i}, \boldsymbol{\omega}]_{i=1}^k, q \leftarrow \Adv(\mathsf{\mathbf{ck_{e}}, ck_{p}, rk_{p}})\\
    \left[c_{e_i} \!\leftarrow \! \mathsf{Com_{e}(ck_{e_i}}, \mathbf{e_i, p_i}; \boldsymbol{\omega_i)}\right]_{i=1}^k\\
    c \leftarrow \mathsf{Interpolate(ck_{p}},\mathbf{p, c_{e})}\\
    \phi \leftarrow {\vect{V}_{\mathbf{p}}}\mathbf{e}, v \leftarrow \phi(q)\\
    \pi \leftarrow \mathsf{Open}(ck_{p}, \phi, q; \vect{\omega})
    \end{gathered}
  \right] 
%   = 1
  \]
% \end{minipage}
% \caption{\textsf{PEC} completeness definition \label{fig:pec_compl}}
% \end{figure*}

\begin{itemize}[leftmargin=10pt]
    \item \textbf{Extractable:} First, consider an adversary $\Adv_0$ that chooses $k' , (k' \leq k \leq K)$ points $\mathbf{p'} = [\mathbf{p}]_{i=1}^{k'}$ and their evaluations $\mathbf{e'}=[\mathbf{e}]_{i=1}^{k'}$. Next, consider an $\Adv_1$ which upon input setup material ($\mathbf{ck_e}, \mathsf{ck_p}, \mathsf{rk_p}$) and evaluation commitments $\mathbf{c'_e}$(for $\mathbf{e'}$ and $\mathbf{p'}$) chooses commitments to evaluations $\vect{c_e}$ at evaluation points $\mathbf{p}$. Let $c$ denote the interpolated commitment from $\mathbf{ck_e}$. Finally, consider an adversary $\Adv_2$ which upon input state $\mathsf{st}$ from $\Adv_1$ outputs a claimed evaluation $v$ at query point $q$ with a proof $\pi$. We say that $\mathsf{PEC}$ is extractable if evaluations of the polynomial can be extracted from an adversary ($\Adv = (\Adv_1, \Adv_2, \Adv_3)$) generating a valid proof. More formally, $\mathsf{PEC}$ is extractable if for every size bound $\mathsf{D, K} \in \mathbbm{N}$, every efficient adversary $\Adv_1, \Adv_2$ there exists an efficient extractor $\Ext$ such that for every $\Adv_3$ the following probability is $\mathsf{negl}(\lambda)$:
\end{itemize}
      \[
   \Pr\!\!\left[
    \begin{gathered}
    c_{e_i}\!=\!\mathsf{Com_e}(\phi(p_i) ; \omega_i)\\
    \land \mathsf{Check(rk_{p}},\! c,\! q,\! v,\! \pi) \\ 
    \Downarrow \\
    \left( 
    \begin{gathered}
    deg(\phi) \leq \mathsf{D} \ \land \\
    v = \phi(q)  \ \land \\
    [\phi(\mathbf{p'_i}) = \mathbf{e'_i}]^{k'}
  \end{gathered}
    \right)
  \end{gathered}
    \middle|\\
        \begin{gathered}
    \mathsf{\mathbf{ck_{e}},\! ck_{p},\! rk_{p},\! td} \leftarrow \mathsf{Setup}(1^{\lambda},\! \mathsf{D,\!K})\\
    \left[\mathbf{(p'_i, e'_i)}\right]_{i=1}^{k'} \leftarrow \Adv_1(\mathsf{\mathbf{ck_{e}}, ck_{p}}; z)\\
    \left[c_{e_i} \leftarrow \mathsf{Com_{e}(ck_{e_i}}, \mathbf{e'_i, p'_i} ;\boldsymbol{\omega_i})\right]_{i=1}^{k'}\\
    (\mathbf{p, c_{e}}, \mathsf{st}) \!\leftarrow\! \Adv_2(\mathsf{\mathbf{ck_{e}}, ck_{p}, c'_e}; z)\\
    c \leftarrow \mathsf{Interpolate(ck_{p}},\! \mathbf{p||p',\! c_{e} \! || \! c_{e'}})\\
    \phi, \vect{\omega} \leftarrow \Ext^{A_1}(\mathsf{\mathbf{ck_{e}}, ck_{p}, rk_{p}; z}) \\
    v, \pi, q\!\leftarrow\!\!\Adv_3(\mathsf{\mathbf{ck_{e}}, ck_{p}, rk_{p}}, \mathbf{c_e}, \mathsf{st})\\
    \end{gathered}
  \right] 
%   = \mathsf{negl}(\lambda)
  \]
% \end{minipage}
% \caption{\textsf{PEC} extraction definition \label{fig:pec_ext}}
% \end{figure*}
Note that for brevity in the definition, we represent $\left[c_{e_i} = \mathsf{Commit_{eval}}(\mathsf{ck_i}, \phi(p_i), p_i ;\omega_i)\right]^k_{i=1}$ as $c_{e_i} = \mathsf{Com_e}(\phi(p_i) ;\omega_i)$
\begin{itemize}[leftmargin=10pt]
    \item \textbf{Zero knowledge:} We say that $\mathsf{PEC}$ is zero knowledge if the adversary cannot distinguish whether it is interacting with the honest prover or a simulator with trapdoors. More formally, there exists a polynomial-time simulator $\mathsf{S = (Setup, Commit_{eval}, Open)}$ such that, for every maximum degree bound $\mathsf{D, K} \in \mathbbm{N}$, and efficient adversary $\Adv = (\Adv_1, \Adv_2)$, the following distributions are indistinguishable:
\end{itemize}

Real World:
    \[
        \left[
    \mathsf{st}_2,\pi
      \middle|
       \begin{aligned}
  & \mathsf{ck_{e}, ck_{p}, rk_{p}} \leftarrow \mathsf{PEC.Setup}(1^{\lambda}, \mathsf{D,K})& \\
  & [\mathbf{e_i, p_i}]_{i=1}^k,\mathsf{st}_1 \leftarrow \Adv_1(\mathsf{ck_{e}, ck_{p}, rk_{p}}) &\\
  & \left[c_{e_i} \leftarrow \mathsf{Commit_{eval}(ck_{e}}, \mathbf{e_i, p_i}; \boldsymbol{\omega_i})\right]_{i=1}^k&\\
  &\mathbf{c_e}:= [c_{e_i}]_{i=1}^k, \vect{\omega}:= [\boldsymbol{\omega_i}]_{i=1}^k, \phi := \vect{V}_{\mathbf{p}}(\mathbf{e, p})&\\
  & c \leftarrow \mathsf{PEC.Interpolate(ck_{e}}, \mathbf{p, c_{e})}&\\
  & q, \mathsf{st}_2 \leftarrow \Adv_2(\mathsf{st}_1, c) &\\
  &\pi \leftarrow \mathsf{PEC.Open}(ck_{p}, \phi, q; \vect{\omega})&
  \end{aligned}
  \right]
  \]

Ideal World:
    \[
        \left[
    \mathsf{st}_2,\pi
      \middle|
       \begin{aligned}
  &\mathsf{ck_{e}, ck_{p}, rk_{p}, td} \leftarrow \mathsf{S.Setup}(1^{\lambda}, \mathsf{D,K})&\\
  & [\mathbf{e_i, p_i}]_{i=1}^k,\mathsf{st}_1 \leftarrow \Adv_1(\mathsf{ck_{e}, ck_{p}, rk_{p}}) &\\
      & \left[c_{e_i} \leftarrow \mathsf{S.Commit_{eval}(ck_{e}}, \mathbf{p_i}; \boldsymbol{\omega_i})\right]_{i=1}^k&\\
  &\mathbf{c_e}:= [c_{e_i}]_{i=1}^k, \vect{\omega}:= [\boldsymbol{\omega_i}]_{i=1}^k, \phi := \vect{V}_{\mathbf{p}}(\mathbf{e, p})&\\
  & c \leftarrow \mathsf{PEC.interpolate(ck_{e}},\mathbf{p, c_{e})}&\\
  & q, \mathsf{st}_2 \leftarrow \Adv_2(\mathsf{st}_1, c) &\\
  &\pi \leftarrow \mathsf{S.Open}(ck_{poly}, \mathsf{td}, \phi(q) , q; \vect{\omega})&
  \end{aligned}
  \right]
\]

\ignore{
\begin{figure}[!htbp] 
\begin{minipage}[t]{0.5\textwidth}
\centering
\begin{boxedminipage}{0.3\textwidth}
\centering
  Real World
%   \hline
  \begin{flalign*}
  & \mathsf{ck_{e}, ck_{p}, rk_{p}} \leftarrow \mathsf{PEC.Setup}(1^{\lambda}, \mathsf{D})& \\
  & \mathbf{e, p} \leftarrow \Adv_1(\mathsf{ck_{e}, ck_{p}, rk_{p}}) &\\
  & \left[c_{e_i} \leftarrow \mathsf{Commit_{eval}(ck_{e}}, e_i, p_i; \omega_i)\right]_{i=1}^n&\\
  &\mathbf{c_e}:= [c_{e_i}]_{i=1}^n, \vect{\omega}:= [\omega_i]_{i=1}^n, \phi := \mathsf{\vect{V_p}(\mathbf{e, p})}&\\
  & c \leftarrow \mathsf{PEC.Interpolate(ck_{e}}, \mathbf{p, c_{e})}&\\
  & q \leftarrow \Adv_2(\mathsf{st}) &\\
  &\pi \leftarrow \mathsf{PEC.Open}(ck_{p}, \phi, q; \vect{\omega})&
  \end{flalign*}
\end{boxedminipage}
\begin{boxedminipage}{.25\textwidth}
\centering
  Ideal World
  \begin{flalign*}
  &\mathsf{ck_{e}, ck_{p}, rk_{p}, td} \leftarrow \mathsf{S.Setup}(1^{\lambda}, \mathsf{D})&\\
  & \mathbf{e, p} \leftarrow \Adv_1(\mathsf{ck_{e}, ck_{p}, rk_{p}}) &\\
      & \left[c_{e_i} \leftarrow \mathsf{S.Commit_{eval}(ck_{e}}, p_i; \omega_i)\right]_{i=1}^n&\\
  &\mathbf{c_e}:= [c_{e_i}]_{i=1}^n, \vect{\omega}:= [\omega_i]_{i=1}^n, \phi := \mathsf{\vect{V_p}(\mathbf{e, p})}&\\
  & c \leftarrow \mathsf{PEC.interpolate(ck_{e}},\mathbf{p, c_{e})}&\\
  & q \leftarrow \Adv_2(\mathsf{st}) &\\
  &\pi \leftarrow \mathsf{S.Open}(ck_{poly}, \phi(q) , q; \vect{\omega})&
  \end{flalign*}
\end{boxedminipage}
\end{minipage}
\caption{\textsf{PEC} Zero knowledge definition
\todo{Looking for ideas to save space here.}
\label{fig:pec_zk}}
\end{figure}
}

% For simplicity, our definitions stated below support only a single query with a single polynomial of fixed degree bound and fixed evaluation points, although our construction can be generalized using similar from Chiesa et al.~\cite{chiesa2019marlin} c.f. Section 6).

\subsection{PEC Constructions}
\label{sec:cons_pec}
We discuss three constructions for $\mathsf{PEC}$ schemes.
The first, based on Pedersen commitments, is a straightforward approach that involves a commitment to each coefficient of the polynomial.
Naturally, this results in commitments and evaluation proofs that are linear in the degree of the polynomials.
Even still, when used to instantiate auditable MPC in Section~\ref{sec:construction}, this results in a proof and verification time that is \emph{circuit-succinct} (i.e., independent of the circuit size $M$). We defer the details of this scheme to the Appendix.

Towards constructing an auditable MPC that is additionally \emph{statement-succinct}, 
our second approach is to adapt an efficient polynomial commitment scheme such as KZG~\cite{kate2010constant}. However, this turns out to be non-trivial.
Our first attempt was to simply transport the KZG \textsf{Commit} routine ``into the exponent.''
Briefly, (and ignoring zero-knowledge to illustrate the problem even in the simple case) this involves committing to the evaluation $\phi(i)$ with a group element  $g^{\phi(i)}$.
However, we then have now way to obtain $g^{\phi(\alpha)}$, the desired KZG polynomial commitment form.
We can use the CRS to compute interpolation factors $g^{\ell_i(\alpha)}$ for Lagrange polynomials $\ell_i$, but still we cannot combine these with $g^{\phi(i)}$ to get $g^{\phi(\alpha)}$ without breaking the Computational Diffie Hellman assumption in our group (which KZG relies on). 
In particular, the CRS does not allow us to compute $\ell_i(\alpha)$ outside the exponent.
%cannot combine the commitments $g^{\ell_i(X)}$ and $g^{\phi(i)}$ to obtain $g^{\phi(i)\ell_i(X)}$ that would be required for interpolation. Doing so, we would be breaking the cdh assumption. 
To solve this problem, our idea is to have each evaluation commitment take the form $g^{\ell_i(\alpha) \phi(i)}$, i.e., to have each party precompute their Lagrange polynomials when committing to their evaluation points. 
We also need to ensure that corrupt parties cannot perturb the evaluation points committed by honest parties;
we address this by creating separate CRS elements to be used by each party, and proving that each evaluation lies in the span its their assigned CRS elements.
Finally our construction incorporates hiding polynomials to maintain zero-knowledge.
%
%Let $K$ be the maximum of clients that could participate in an auditable MPC protocol. Further let $n_K$
%be the maximum number of each party to. Define the maximum commimtment degree $\mathsf{D} = K*n_K$ and let
%$\mathsf{ck^k}$ denote the "knowledge component"(terms in CRS that contain $\gamma$) of the keys. 
Our succinct PEC construction, $\mathsf{PEC.Succ}$, is defined as follows:

\begin{itemize}[leftmargin=10pt]
    \item $\mathsf{Setup}(1^\lambda, D, K):$ Sample $\mathbf{\Sigma} = [\Sigma_i]^K_{i=1}$  as follows
  
%   \begin{equation}
%   \begin{gathered}
  \begin{equation}
    \Sigma_i := \left(
                \begin{array}{llllll}
                    g^{K*i} & g^{\alpha + K*i} & \ldots & g^{\alpha^{n_K-1} + K*i}\\
                    g^{\gamma_iK*i} & g^{\gamma_i(\alpha + K*i)}  & \ldots & g^{\gamma_i(\alpha^{n_K-1} + K*i)} \\
                    g^{\tau K*i} & g^{\tau\alpha + K*i} & \ldots & g^{\tau\alpha^{n_K-1} + K*i}\\
                    g^{\tau\gamma_iK*i} & g^{\tau\gamma_i(\alpha + K*i)}  & \ldots & g^{\tau\gamma_i(\alpha^{n_K-1} + K*i)} \\
                \end{array}
              \right).
  \end{equation}
    % \end{gathered}
    % \end{equation}
  \begin{equation}
  \begin{gathered}
\mathsf{ck_{p}}:= (\mathsf{\Sigma_K}, n_K), \mathsf{ck_{e}} := (\vect{\Sigma} = [\Sigma_i]_{i=1}^K),\\
\mathsf{rk_{p}}:= (D, g^{\gamma_K}, g^{\tau \gamma_K}, h^{\alpha}), \mathsf{td} := (\tau, \alpha, [\gamma_i]_0^K)\\
  \end{gathered}
  \end{equation}
    \item $\mathsf{Commit_{eval}(ck}_{\mathsf{e}_i}, \vect{e_i}, \vect{p_i}; \omega) \rightarrow c_{e_i}$: With input ${\mathsf{ck}_{\mathsf{e}_i}, \mathsf{ck^k}_{\mathsf{e}_i}} = \Sigma_i$, points $\mathbf{p_i} = [p_{ij}]_{j=1}^d$ with evaluations $\mathbf{e_i} = [e_{ij}]_{j=1}^d$, 
    compute computing $\phi$ at atmost degree $d$ by interpolating $\mathbf{e_i}$ at points $\mathbf{p_i}$. Compute $c_e = \mathsf{KZG.Commit(ck_{e_i}}, \phi; \omega)$. This computes a shifted polynomial commitment using CRS $g^{i*K} \ldots g^{i*K + n_K -1}$. Similarly, compute $c^k_e = \mathsf{Commit(ck^k_{e_i}}, \phi; \omega)$. Return $(c_e, c^k_e)$
    \item $\mathsf{Interpolate(ck_{p}}, \mathbf{c_{e}}, \mathbf{p}) \rightarrow c$ : Parse $\mathbf{c_e} = [c_{e_i}].$ Check the knowledge component: $\forall i \  (e(g, c^k_e)), g$ $\stackrel{?}{=}$ $ e(g^\gamma_i, c_e)$. If check fails, abort, otherwise return $c = \Pi_{i=1}^kc_{e_i}.$ 
    \item $\mathsf{Open(ck_{p}}, \phi, q; \omega) \rightarrow v, \pi$: Same as KZG polycommit open operation as described in Section~\ref{sec:prelim_polycommit}
    \item $\mathsf{Check(rk_{p}}, c, q, v, \pi) \rightarrow \{0, 1\}$: Same as KZG polycommit operation as described in Section~\ref{sec:prelim_polycommit}. 
\end{itemize}

\begin{thm}
\label{thm:pec_succ}
If $\langle group \rangle$ satisfies the SDH assumption(Appendix~\ref{app:sdh}), then the above construction for $\mathsf{PEC.Succ}$ is a $\mathsf{PEC}$ scheme (definition~\ref{def:pec})
\end{thm}

In Appendix~\ref{app:proof_succ}, we provide a proof for the theorem. In Appendix~\ref{app:pec_constructions} we show a third $\mathsf{PEC}$ scheme that offers concrete performance improvements based on Lipmaa's commitments~\cite{lipmaa2017prover}.

%\sanket{Write about the PEC.Succ explanation}
%$g^{l_i(\alpha)*\phi(i)}, h^{\ \phi(i)*\beta_0}$ where $h = g^{}$
%but reduces the concrete comptutation cost (reduces the number of group operations needed).
%uses polynomial commitments that require a trusted setup. 
%On the other hand, the verification in $\mathsf{PEC.Ped}$ requires $n$ exponentiation and $n$ group multiplications whereas $\mathsf{PEC.Poly}$ requires only $n$ group operations. 

%For our construction, it is beneficial to use $\mathsf{PEC.Poly}$ scheme instead of the previous one based on Pedersen ~\ref{sec:pec_cons_ped} commitments because 1) we are already using polynomial commitments in an underlying protocol, so 

%It is also possible to adapt Marlin~\cite{chiesa2019marlin} to use a polynomial commitment scheme that does have a trusted setup at the cost of performance, in which case it would make sense to \textsf{PEC.Ped} that does not have a trusted setup. 
%
%The setup step is already performed and 2) the commitments from $\mathsf{PEC.Poly}$ are compatible with pairing batching techniques that we can use for more efficiency. 

%-------------------------------------------------------------------------------
% Constructions
%-------------------------------------------------------------------------------
\section{Our Auditable MPC Construction}
\label{sec:construction}

\subsection{Adaptive Preprocessing arguments with universal SRS}
\label{sec:adapt_def}

We first give a formal security definition for Adaptive Marlin, our main construction.
Although our final goal is a non-interactive protocol, we follow Chiesa et al. and give an interactive definition and remove interaction with Fiat-Shamir at the end~\cite{chiesa2019marlin}.
We use angle brackets $\langle P(...)  V(...) \rangle$ to denote the output of $V(...)$ when interacting with $P(...)$.

%\paragraph{Indexed relations with Commitments:}
%\label{sec:index_rel}
We extend the indexed relations defined in Section~\ref{sec:index_relations} to the following indexed commitment relations. Let $\mathsf{Setup, Comm}$ be a extractable trapdoor commitment scheme as shown in section~\ref{sec:prelim_commitments}. 
Given indexed relation $\mathcal{R}$ and a commitment key $\mathbf{ck} \leftarrow \mathsf{Setup}(1^\lambda)$
$$\mathcal{R}_{\mathbf{ck}} := \{(\mathbf{C_x, \mathbbm{i}, \mathbf{x}, r, \mathbf{w})}) :$$
$$\forall i. \ C_{x_i} = \mathsf{Comm}(\mathsf{ck}_i, x_i, r_i) \land (\mathbbm{i},\mathbf{x}, \mathbf{w}) \in \mathcal{R}\}$$
Informally, an adaptive preprocessing argument (also refered as adaptive SNARK) for indexed relation is a preprocessing argument for the relation $\mathcal{R}_{\mathbf{ck}}$. We next give the formal definitions for adaptive preprocessing arguments with universal SRS.

Let $\mathsf{C = (Setup, Com)}$ be an extractable commitment scheme. Further, let $\mathsf{C.Setup}(1^\lambda)$ $\rightarrow \mathsf{\mathbf{ck}, rk, td}$.
We define an adaptive SNARKs (referred as preprocessing arguments with universal SRS in Marlin) for extractable trapdoor commitment scheme $\mathsf{C}$ and relation $\mathsf{R_{ck}}$ as a tuple of four algorithms $\mathsf{(G, I, P, V)}$:
\begin{itemize}[leftmargin=9pt]
    \item $\mathsf{G}(1^\lambda, \mathsf{N}) \rightarrow \mathsf{srs, td}$: generator $\mathsf{G}$ is a ppt which when given a size bound $\mathsf{N} \in \mathbbm{N}$, outputs an $\mathsf{srs}$ that supports indices of size up to $\mathsf{N}$ and a trapdoor $\mathsf{td}$.
    \item $\mathsf{I}^{\mathsf{srs}}(\mathbbm{i}) \rightarrow \mathsf{ipk, ivk}$: The indexer $\mathsf{I}$ is a deterministic algorithm that with oracle access to $\mathsf{srs}$ takes in a circuit index $\mathbbm{i} < \mathsf{N}$ outputs proving key $\mathsf{ipk}$ and verification key $\mathsf{ivk}$ specific to the index $\mathbbm{i}$.
    \item $\mathsf{P}(\mathsf{ipk}, \mathbf{ck}, \mathbf{C_{x}}, \mathbf{x}, \mathbf{r}, \mathbf{w}) \rightarrow \pi$: The prover $\mathsf{P}$ is a ppt which on input prover key $\mathsf{ipk}$, committer keys $\mathbf{ck}$, statement $\mathbf{x}$, commitment randomness $\mathbf{r}$ and witness $\mathbf{w}$ outputs a proof $\pi$. 
    \item $\mathsf{V}(\mathsf{ivk}, \mathsf{rk}, \mathbf{C_x}, \pi) \rightarrow \{0, 1\}$ : Verifier $\mathsf{V}$ is a ppt which upon input index verification key $\mathsf{ivk}$, receiver key $\mathsf{rk}$,
    polynomial evaluation commitments $\mathbf{C_x}$ and a proof $\pi$ outputs either $0$ or $1$. 
\end{itemize}

Furthermore, we want adaptive preprocessing agruments to satisfy the following properties:
\begin{itemize}[leftmargin=9pt]
    \item Perfect Completeness: We say that our adaptive preprocessing argument is complete if all adversaries choosing the a tuple $(\mathbf{C_x, \mathbbm{i}, \mathbf{x, r, w})} \in \mathcal{R}_{\mathbf{ck}}$, the interaction between honest prover is always able to convince the honest verifier.
\end{itemize}
      \[
  \Pr \left[
    \begin{gathered}
    % \begin{gather}
    (\mathbf{C_x, \mathbbm{i}, \mathbf{x, r, w})} \notin \mathcal{R}_{\mathbf{ck}} \\
    \lor \\
    \left\langle
    \begin{gathered}
    \mathsf{P}(\mathsf{ipk}, \mathbf{ck}, \mathbf{C_{x}}, \mathbf{x}, \mathbf{r}, \mathbf{w}) \\
    \mathsf{V}(\mathsf{ivk}, \mathsf{rk}, \mathbf{C_x})
    \end{gathered}
    \right\rangle
    % = 1
    % \end{gather}
  \end{gathered}
    \middle|\\
        \begin{gathered}
    \mathsf{srs} \leftarrow \mathsf{G}(\mathsf{1}^{\lambda}, \mathsf{N})\\
    \mathbf{ck} \leftarrow \mathsf{C.Setup}(1^\lambda)\\
    (\mathbf{C_x, \mathbbm{i}, \mathbf{x, r, w)}} \leftarrow \Adv(\mathsf{srs})\\
    \mathsf{ipk, ivk} \leftarrow \mathsf{I}^{\mathsf{srs}}(\mathbbm{i}) \\
    \end{gathered}
  \right] 
%   = 1
  \]
\label{def:adapt_completeness}

\begin{itemize}[leftmargin=9pt]
    \item Extractable: We say that our adaptive preprocessing argument is extractable if for every size bound $\mathsf{N} \in \mathbbm{N}$ and efficient adversary $\Adv$ = $(\Adv_1, \Adv_2)$ there exists an efficient extractor $\Ext$ such that the following probability is $\mathsf{negl}(\lambda)$.
\end{itemize}
      \[
  \Pr \left[
    \begin{gathered}
    % \begin{gather}
    (\mathbf{C_x, \mathbbm{i}, \mathbf{x, r, w})} \notin \mathcal{R}_{\mathbf{ck}} \\
    \land \\
    \langle \Adv_2(\mathsf{st}), \mathsf{V}(\mathsf{ivk}, \mathsf{rk}, \mathbf{C_x})\rangle
    % \end{gather}
  \end{gathered}
    \middle|\\
        \begin{gathered}
    \mathsf{srs} \leftarrow \mathsf{G}(1^{\lambda}, \mathsf{N})\\
    \mathbf{ck} \leftarrow \mathsf{Setup}(1^\lambda)\\
    (\mathbf{C_x, \mathbbm{i}, \mathsf{st})} \leftarrow \Adv_1(\mathsf{srs, \mathbf{ck}; z})\\
    (\mathbf{x, r, w}) \leftarrow \Ext^{A_1, A_2}(\mathsf{srs; z}) \\
    \mathsf{ipk, ivk} \leftarrow \mathsf{I}^{\mathsf{srs}}(\mathbbm{i}) \\
    \end{gathered}
  \right]
%   = \mathsf{negl}(\lambda)
  \]
\label{def:adapt_extraction}

\begin{itemize}[leftmargin=9pt]
    \item Zero Knowledge: We say that our adaptive preprocessing argument is zero knowledge if the adversary is not able to distinguish whether it is interacting with a honest prover or a simulator. More formally, \textsf{ARG} is zero knowledge if for every size bound $\mathsf{N} \in \mathbbm{N}$ there exists a simulator $\mathsf{S = (Setup_1, Setup_2, Prove)}$ such that for every efficient adversary $\Adv$ = $(\Adv_1, \Adv_2)$ the probabilities shown below are equal
\end{itemize}
      \[
  \Pr \left[
    \begin{gathered}
    % \begin{gather}
    (\mathbf{C_x, \mathbbm{i}, \mathbf{x, r, w})} \in \mathcal{R}_{\mathbf{ck}} \\
    \land \\
    \left\langle 
    \begin{gathered}
    \mathsf{P}(\mathsf{ipk}, \mathbf{ck}, \mathbf{C_{x}}, \mathbf{x}, \mathbf{r}, \mathbf{w}) \\ \mathsf{\Adv_2}(\mathsf{st})    
    \end{gathered}
    \right\rangle
    % = 1
    % \end{gather}
  \end{gathered}
    \middle|\\
        \begin{gathered}
    \mathsf{srs} \leftarrow \mathsf{G}(1^{\lambda}, \mathsf{N})\\
    \mathbf{ck} \leftarrow \mathsf{Setup}(1^\lambda)\\
    \mathbf{\mathbbm{i}, \mathbf{x, r, w}, \mathsf{st}} \leftarrow \Adv_1(\mathsf{srs, ck})\\
    \mathsf{ipk, ivk} \leftarrow \mathsf{I}^{\mathsf{srs}}(\mathbbm{i}) \\
    \mathbf{C_x} \leftarrow [\mathsf{Com(ck_i, x_i, r_i)})]_{i=1}^{|\mathbf{x}|} \\
    \end{gathered}
  \right] 
%   = 
  \]
    \[
  \Pr \left[
    \begin{gathered}
    % \begin{gather}
    (\mathbf{\mathbbm{i}, \mathbf{x, r, w})} \in \mathcal{R}_{\mathsf{ck}} \\
    \land \\
    \left\langle 
    \begin{gathered}
    \mathsf{S.Prove}(\mathsf{td_{s}}, \mathsf{td_{c}}, \mathbf{C_{x}}, \mathbbm{i}) \\ \mathsf{\Adv_2}(\mathsf{st})
    \end{gathered}
    \right\rangle
    % = 1
    % \end{gather}
  \end{gathered}
    \middle|\\
        \begin{gathered}
    \mathsf{srs, td_{s}} \leftarrow \mathsf{S.Setup_1}(1^\lambda, \mathsf{N})\\
    \mathsf{\mathbf{ck}, td_{c}} \leftarrow \mathsf{S.Setup_2}(1^\lambda)\\
    \mathbbm{i}, \mathbf{x, r, w}, \mathsf{st} \leftarrow \Adv_1(\mathsf{srs, ck})\\
    \mathsf{ipk, ivk} \leftarrow \mathsf{I}^{\mathsf{srs}}(\mathbbm{i}) \\
    \mathbf{C_x} \leftarrow [\mathsf{Com(ck_i, x_i, r_i)})]_{i=1}^{|\mathbf{x}|} \\
    \end{gathered}
  \right]
  \]
\label{def:adapt_zk}

\begin{figure}[!htbp] 
\begin{boxedminipage}[t]{0.5\textwidth}
  {\centering \textbf{Adaptive Zk-Snark based on Marlin}\\}
% \quad\emph{// Compute the function}
Let $\mathsf{PEC}$ = $\mathsf{Setup, Commit_{eval}, Interpolate, Open, Check}$ and $\mathsf{PEC.Setup}(1^\lambda, D)$ $\rightarrow \mathsf{ck_{e}, ck_{p}, rk_{p}}$. Let ($\mathsf{G_m, I_m, P'_m, V'_m}$) be a augmented pre-processing argument constructed from Marlin for the relation $\mathcal{R'_{\mathbf{ck}}}$. Let $\mathsf{P_m}$ be same as $\mathsf{P'_m}$ with the difference that it knows public commitment $\mathbf{C_x}$ along with the secret $\mathbf{x}$. Similarly, let $\mathsf{V_m}$ be the same as $\mathsf{V'_m}$ with the difference that $\mathsf{V'_m}$ only knows $\mathbf{C'_x}$ instead of $\mathbf{x}$. 
% \begin{enumerate}
\begin{itemize}
    \item $\mathsf{G}(\mathsf{N})$: If $\mathsf{N}> \mathsf{D}$, abort. Return $\mathsf{srs}=\Sigma$ from $\mathsf{ck_e}$
    \item $\mathsf{I}^{\mathsf{srs}}( \mathbbm{i}) \rightarrow \mathsf{(ipk, ivk)}$: Let $\mathbbm{i}' = (\mathbbm{F}, n+1, m, A', B', C')$ where $(A', B', C')=\mathsf{Pad}(A, B, C)$. Return $\mathsf{I_m^{\mathsf{srs}}(\mathbbm{i}')}$. 
    \item $\mathsf{P}(\mathsf{ipk}, \mathsf{ck_{e}}, \mathbf{C_{x}}, \mathbf{x}, \mathbf{r}, \mathbf{w}) \rightarrow \pi$: 
    % \begin{enumerate}
        % \item 
        Sample $x^{\mathsf{b}} \samples \mathbbm{F};$ $\mathbf{C^{\mathsf{b}}_{x}} $ $\leftarrow \mathsf{Commit_{eval}(ck_{e}}, x^{\mathsf{b}},.; \omega )$. $\mathbf{C'_x} = (\mathbf{C_x}, \mathbf{C^{\mathsf{b}}_{x}})$; $\mathbf{x'}$ = $\mathbf{x}||x^{\mathsf{b}}$. call $\mathsf{P_m(ipk,} x||x^{\mathbbm{b}}, w) \rightarrow \pi_m$. 
        % \item 
        Let $\hat{x}'(X) = V_\mathbf{p}\mathbf{x}'$, $\beta_1$ be second verifier challenge in Marlin execution. $\mathsf{PEC.open(ck_{p}}, \hat{x}'(X), \beta_1;\omega)$ $\rightarrow$ $\hat{x'}(\beta_1), \pi_c$.
        return $\pi = (\pi_m, \pi_c, \hat{x}'(\beta_1), \mathbf{C^{\mathsf{b}}_x})$.
    % \end{enumerate} 
    \item $\mathsf{V}(\mathsf{ivk}, \mathsf{rk_{p}}, \mathbf{C_x}, \pi)$: 
    % \begin{enumerate}
        % \item 
        Parse $\pi = (\pi_m, \pi_c, \hat{x'}(\beta_1), \mathbf{C^{\mathsf{b}}_x})$; $\mathbf{C'_x} = (\mathbf{C_x}, \mathbf{C^{\mathsf{b}}_{x}})$
        % \item 
        Invoke the Marlin verifier routine by replacing $\hat{x}$ with constant function $\hat{x}{(\beta_1)}$ as $b_m \leftarrow \mathsf{V_m(ivk}, \hat{x'}(\beta_1), \pi_m)$. $C_{\hat{x}}$ $\leftarrow$ $\mathsf{PEC.Interpolate(ck_p}\mathbf{C'_x}, .)$
        % \item 
        Invoke $b_c$ $\leftarrow$ $\mathsf{PEC.check(vk_{poly}}, C_{\hat{x'}},\beta_1, \hat{x'}(\beta_1), \pi_c)$.
        % \item 
        
        return $b_m \stackrel{?}{=} 1$  and $b_c \stackrel{?}{=} 1$.
    % \end{enumerate}
\end{itemize}
% \end{enumerate}
\end{boxedminipage}
\caption{Adaptive preprocessing arguments using Marlin%\sanket{Improve this based on Space constraints, I think it's best probably written as text only instead of figure?}
}
\label{fig:adapt-snark}
\end{figure}

\subsection{Construction of Adaptive Preprocessing arguments with Universal SRS}
Our construction closely follows Marlin's except for two main modifications to the underlying Algebraic Holographic Proof (AHP). The full dsecription of Marlin is in Appendix ~\ref{app:mpc_prover}, so here we only highlight the differences.
In the Marlin prover algorithm, the verifier is assumed to have the entire statement $\mathbf{x}$ and hence it can construct for itself $\hat{x}$, the polynomial encoding of the statement (querying this polynomial at random challenge points is roughly what makes the scheme "holographic).  In our setting, the verifier does not have $\mathbf{x}$, only a commitment to it $\mathbf{C_x}$, so the prover must additionally supply $\hat{x}$.
We must check that the prover supplied $\hat{x}$ matches the commitment $\mathbf{C_x}$, which can be addressed using $\mathsf{PEC}$. Additionally, statement $\mathbf{x}$ must be kept zero knowledge. We can achieve this the same way as Marlin keeps the witness zero knowledge, namely by padding the degree of $\hat{x}$ by a margin of $\mathsf{b}$ so that learning $\mathsf{b}$ challenge points of $\hat{x}$ reveals nothing about $\mathbf{x}$.
As with Marlin, it suffices to set $\mathsf{b} = 1$, but we stick to $\mathsf{b}$ for consistency of notation. 
%To accommodate this free variable, we also change the $\mathbbm{i}$ by adding a dummy constraint ($0 \times 0 = 0$).

In more detail, we consider an augmented relation $\mathcal{R'_{\mathbf{ck}}} = \{(\mathbf{C'_x},$
$\mathbbm{i}'$, $\mathbf{x'}$, $\mathbf{r'},\mathbf{w})\}$:
$(\mathbf{C_x},$ $\mathbbm{i}$, $\mathbf{x}$, $\mathbf{r},\mathbf{w}) \in \mathcal{R_{\mathbf{ck}}}$, 
$\mathbf{C'_x} = \mathbf{C_x||C_x^{\mathsf{b}}}$, $\mathbbm{x'}$ = $\mathbf{x}||x^\mathsf{b}$, $\mathbf{r'} = \mathbf{r}||r^\mathsf{b}$, $\mathbf{C'_x} = \mathsf{Commit_{eval}(ck_{|x|+1}}, x^\mathsf{b}, r^{\mathsf{b}})$ and $\mathbbm{i'} = \mathsf{Pad(}\mathbbm{i})$ by padding $\mathbbm{i}$ with dummy constraint. In more detail, let
$\mathbbm{i} = (\mathbbm{F}, n, m, A, B, C)$ such that $\mathbf{z} := (\mathbf{x}, \mathbf{w}) \in$ $\mathbbm{F}^n$ such that $A\mathbf{z} \circ B\mathbf{z} = C\mathbf{z}$, compute
$\mathbbm{i}' = (\mathbbm{F}, n+1, m, A', B', C')$  
$\mathbf{z'} := (\mathbf{x'}, \mathbf{w})$ is a vector in $\mathbbm{F}^{n+1}$ such that $A'\mathbf{z'} \circ B'\mathbf{z'} = C'\mathbf{z'}$. This is done by padding matrices $A, B, C$ with dummy constraint ($0 \times 0 = 0$) on free variable $x^\mathsf{b}$ to obtain $A', B', C'$. In simpler words, we add a free statement variable to the indexed constraint system.

Finally, we use the compiler from Marlin to compile the above modified AHP and polynomial commitment scheme from Marlin(different from \textsf{PEC}) to result in pre-processing arguments that are adaptive. 

Our construction for an adaptive Preprocessing arguments \textsf{ARG} = $\mathsf{(G, I, P, V)}$ with universal SRS for extractable trapdoor commitment scheme $\mathsf{PEC}$ and relation $\mathsf{R_{\mathbf{ck}}}$ is shown in ~\ref{fig:adapt-snark}.

We state our construction with a generic $\mathsf{PEC}$, it is possible to instantiate with any of $\mathsf{PEC.Ped}$, $\mathsf{PEC.Lipmaa}$ or $\mathsf{PEC.Succ}$. 
Our routine Generator $\mathsf{G}$ uses the same $\mathsf{srs}$ from the $\mathsf{PEC}$ scheme whereas our Indexer $\mathsf{I}$ operates on the $\mathbbm{i'}$. 
Our prover algorithm first samples additional element $x^{\mathsf{b}} \in \mathbb{F}_q$ element to compute the augmented statement $\mathbf{x}'$. First, the prover computes $\mathbf{C^{\mathsf{b}}_{x}} \leftarrow \mathsf{PEC.Commit_{eval}(PEC.ck_{eval}, x^{\mathsf{b}}}, p_{|x|+1}; \omega )$ to compute the evaluation commitment at point with index $|x|+1$ keeping the randomness $\omega$. It then runs the modified Marlin prover $\mathsf{P_m(\mathsf{ipk}, \mathbf{x}', \mathbf{w})}$ to obtain a proof $\pi_m$. 
Let $\hat{x}'$ be the low degree extension(LDE) of $\mathbf{x}'$ and $\beta_1$ be the second round challenge in the Marlin protocol. Then, our prover routine computes $\mathsf{PEC.Open(ck_p}, \hat{x}', \beta_1, \omega)$ obtain a evaluation $\hat{x}'(\beta_1)$ and opening proof $\pi_c$. Finally, the proof is returned as $\pi = (\pi_m, \pi_c, \hat{x}'(\beta_1), \mathbf{C^{\mathsf{b}}_x})$

The auditor reconstructs the augmented statement from $\mathbf{C'_x} = (\mathbf{C_x}, \mathbf{C^{{b}}_{x}})$ and verifies that $\pi_m$ and $\pi_c$ are correct.

\begin{thm}
\label{thm:snark}
If $\mathsf{PEC}$ is a extractable PEC scheme then the above construction $\mathsf{ARG = (G, I, P, V)}$ is an adaptive zkSNARK according to definition~\ref{def:adapt_zk}.
\end{thm}
We prove this theorem in Appendix~\ref{app:proof_snark}

\subsection{Prover algorithm using MPC}
\label{sec:mpc_prover}
Next, we describe how to implement the above prover $\mathsf{P}$ algorithm by using MPC where the servers compute the proof from shares of the witness and statement.
Our prover, just like Marlin, proceeds in multiple rounds. In each round, the verifier sends some challenges and the prover responds back commitments to polynomials. After the conclusion of rounds, the prover provides the evaluations proofs of the committed polynomials. 
In the MPC setting, this translates to servers knowing shares of the polynomials and computing the evaluation proofs through MPC. We provide our constructions for MPC versions of $\mathsf{PC.Open_{MPC}}$, $\mathsf{PC.Commit_{MPC}}$ and $\mathsf{Eval_{MPC}}$ (Evaluate a polynomial) in Appendix~\ref{app:mpc_polycommit}.

We note that only the first two rounds in Marlin rely on secret shared values; the last two rounds operate on public values and so do not require MPC. To initialize the prover, the servers first compute the commitment to the statement polynomial from shares of that polynomial using $\mathsf{PC.Commit_{MPC}}$. We again use protocol $\mathsf{PC.Commit_{MPC}}$ for the first two Marlin rounds to obtain commitments to polynomials from two rounds from their respective secret shares. After the rounds are concluded, the servers provide evaluations the secret shared polynomials using $\mathsf{Eval_{MPC}}$ and provide an evaluation proof using $\mathsf{PC.Open_{MPC}}$.  Full details are in Appendix~\ref{app:mpc_prover}.

%Our MPC prover algorithm works in five rounds. In the first round, the MPC creates commitments to the statement polynomial from shares of statements, while the last four rounds proceed exactly as Marlin except the polynomial commitment scheme is replaced with the MPC version (Figure~\ref{app:mpc_polycommit}).

\subsection{Construction of auditable MPC}

\begin{figure}
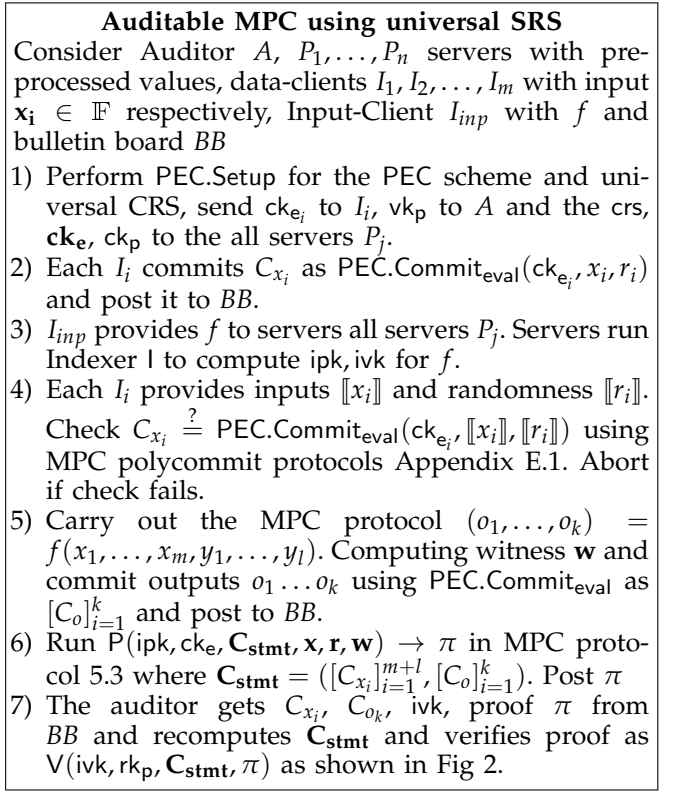

\begin{boxedminipage}[t]{0.5\textwidth}
  {\centering \textbf{Auditable MPC using universal SRS }\\}
% \quad\emph{// Compute the function}
Consider Auditor $A$, $P_1, \ldots, P_n$ servers with pre-processed values, data-clients $I_1, I_2, \ldots, I_m$ with input $\mathbf{x_i} \in \mathbbm{F}$ respectively, Input-Client $I_{inp}$ with $f$ and bulletin board \textit{BB}
\begin{enumerate}[leftmargin=12pt]
    \item Perform $\mathsf{PEC.Setup}$ for the $\mathsf{PEC}$ scheme and universal CRS, send $\mathsf{ck}_{\mathsf{e}_i}$ to $I_i$, $\mathsf{vk_{p}}$ to $A$ and the $\mathsf{crs}$, $\mathbf{ck_{e}}$, $\mathsf{ck_{p}}$ to the all servers $P_j$.
    \item Each $I_i$ commits $C_{x_i}$ as $\mathsf{PEC.Commit_{eval}(ck}_{\mathsf{e}_i}, x_i, r_i)$ and post it to \textit{BB}.
    \item $I_{inp}$ provides $f$ to servers all servers $P_j$. Servers run Indexer $\mathsf{I}$ to compute $\mathsf{ipk}, \mathsf{ivk}$ for $f$.
    \item Each $I_i$ provides inputs $\share{x_i}$ and randomness $\share{r_i}$. Check $C_{x_i} \stackrel{?}{=} \mathsf{PEC.Commit_{eval}(ck}_{\mathsf{e}_i}, \share{x_i}, \share{r_i})$ using MPC polycommit protocols Appendix~\ref{app:sec_mpc_polycommit}. Abort if check fails.
    \item Carry out the MPC protocol $(o_1, \ldots, o_k) = f(x_1, \ldots, x_m, y_1, \ldots, y_l)$. Computing witness $\mathbf{w}$ and commit outputs $o_1 \ldots o_k$ using $\mathsf{PEC.Commit_{eval}}$ as $[C_o]^k_{i=1}$ and post to \textit{BB}.
    \item Run $\mathsf{P}(\mathsf{ipk}, \mathsf{ck_{e}}, \mathbf{C_{stmt}}, \mathbf{x}, \mathbf{r}, \mathbf{w}) \rightarrow \pi$ in MPC protocol~\ref{sec:mpc_prover} where $\mathbf{C_{stmt}}=([C_{x_i}]^{m+l}_{i=1}, [C_o]^k_{i=1})$. Post $\pi$
    \item The auditor gets $C_{x_i}$, $C_{o_k}$, $\mathsf{ivk}$, proof $\pi$ from \textit{BB} and recomputes $\mathbf{C_{stmt}}$ and verifies proof as $\mathsf{V}(\mathsf{ivk}, \mathsf{rk_{p}}, \mathbf{C_{stmt}}, \pi)$ as shown in Fig~\ref{fig:adapt-snark}.
\end{enumerate}
\end{boxedminipage}
\caption{Auditable MPC with universal SRS}
\label{fig:final_protocol}
\end{figure}

Figure ~\ref{fig:final_protocol} shows the auditable MPC protocol using the constructions for adaptive zk-SNARKS, Marlin based MPC prover and Polynomial Evaluation commitment schemes as the underlying commitment scheme. In Appendix~\ref{app:uc} we show a UC~\cite{canetti2001universally} proof that the construction follows the functionality shown previously in Figure~\ref{fig:FAuditable}.

In Step 1, a trusted party performs the setup for \textsf{PEC}(same as the setup for Marlin) the evaluation commitment keys $\mathsf{ck_e}$ to input parties, verification key $\mathsf{vk_p}$ for the polynomial commitment scheme to auditor and $\mathsf{ck_p}$ to the MPC servers. Each input client $I_i$ samples an input $x_i$ with randomness $r_i$ and creates a commitment $C_{x_i} = \mathsf{Commit_{eval}(ck}_{\mathsf{e}_i}, x_i, r_i)$. Because each client has it's own committer key, we can gurantee input independence (Refer ~\ref{sec:inp_ind}) as clients can only use thier own commitment keys. Next, the input client $I_{inp}$ provides the computation function $f$ to MPC system. For simplicity, we assume that computation $f$ itself specifies the desired R1CS computation. If the circuit is not already pre-processed, the MPC servers compute the indexer proving and verification keys, $\mathsf{index_{pk}}$ and $\mathsf{index_{vk}}$ and post them on the bulletin board.

In step 4, the clients again submit the inputs to the MPC system $\share{x_i}$ and randomness $\share{r_i}$. The servers check that the $C_{x_i}$ is consistent with the shares $\share{x_i}$ and $\share{r_i}$ using MPC polycommit check as listed in Appendix~\ref{app:mpc_polycommit}. If the commitments do not match, the servers abort the computation. Looking in the UC proof (Appendix ~\ref{app:uc}) for $f<t$ case, this aborting helps us guarantee Zero knowledge property as the auditor might learn the instance is under Indexed relations with Commitments~\ref{sec:index_relations}.

In Step 5, the servers compute the MPC operation to get the output of the computation $o_k$ and post evaluation commitments to the bulletin board. These can be treated as inputs for the next round of MPC computations, allowing us the reactive functionality. Finally, the MPC servers compute a adaptive Snark proof $\pi$ by using our prover algorithm construction detailed in Figure~\ref{fig:adapt-snark} and post it on the bulletin board. The auditor collects the evaluation commitments to the inputs and outputs from the bulletin board to construct the commitment to $\hat{x}$. The auditor then uses the verify routine in from our adaptive Snark construction as listed in Figure~\ref{fig:adapt-snark}.

%\footnote{Veeningen reports zero communication for single-shot MPC, but extending it to reactive setting would require one final round of communication.} Despite gaining the advantage of Universal setup, our prover complexity is comparable to Veeningen. Our proof size for constant compared to linear in $K$(number of clients) in Veeningen's construction. Finally, $\mathsf{PEC.Succ}$ auditor time is strict improvement over Veeningen but still has $K$ pairings, while $\mathsf{Pec.Ped}$ offers constant pairing cost but suffers in linear group operations.
 
\section{Evaluation}
\label{sec:evaluation}
%In this section, we describe our implementation and evaluation for auditable MPC protocol.
Our rust implementation for auditable MPC based on Marlin~\cite{code:marlin} can found at https://github.com/randomcyrptobuddy/auditablempc. 

% Our code is available at \textsf{https://github.com/sanket1729/auditable\_mpc} as a rust application. We strive to make our benchmarks completely reproducible by dockerizing our setup and one line command to generate all the graphs in this section. The reader can refer to the README for details.

We simulate the MPC behaviour as described in section~\ref{sec:mpc_prelim} by using artificial delays(Appendix \ref{app:applications}) in communication latency of 200ms and a uplink speed of 200Mbits/per second~\cite{lu2019honeybadgermpc}. We report our performance numbers on a single thread machine with a modern CPU processor with 1200 MHz and use the bls12-381 curve\cite{bowefaster} for pairing friendly and fft optimizations. For all the experiments, we use the faster $\mathsf{PEC.Lipmaa}$ (Appendix~\ref{app:pec_constructions}) version of \textsf{PEC} scheme. For sampling random R1CS circuits with $m$ gates, we sample matrices $A$, $B$ and solve for $C$ to obtain a solvable QAP of $m$ constraints.

We first report the performance of our construction over random circuits in terms of prover cost, auditor cost, communication cost, proof size. 
We vary our statement size from [$2^2$, $2^6$], our constraint size from [$2^{10}$, $2^{20}$] and number of MPC servers from [$2^2$, $2^5$].
In the Appendix we also implement and evaluate two applications: 1) Publicly auditable auction and 2) Logrank Test from Vee'17~\cite{veeningen2017pinocchio} for direct comparison with previous work. 
%\begin{figure}[h!]
%    \begin{minipage}{1\textwidth}
%    \centering
%    \begin{minipage}{0.45\textwidth}
    \begin{figure}
        \centering
        \includegraphics[width=0.9\columnwidth,height=1.2in]{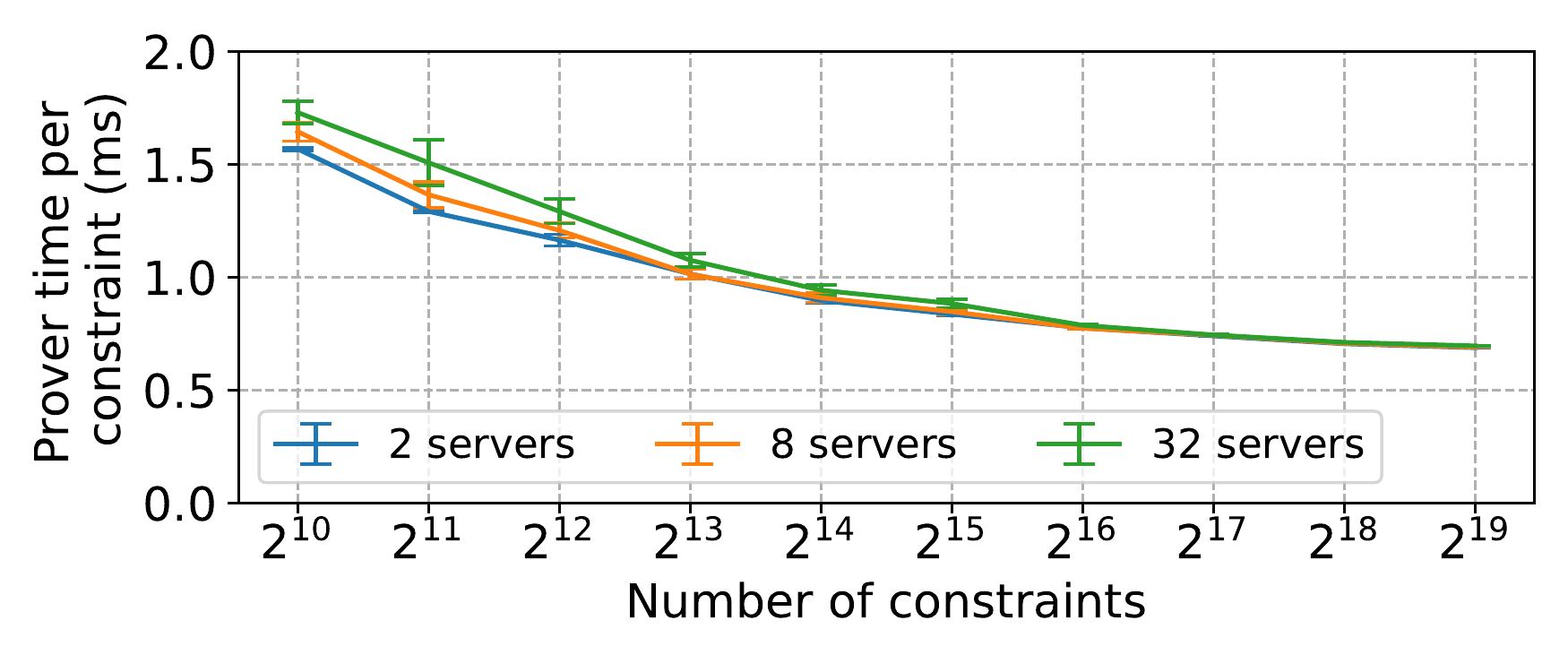} % first figure itself
        \caption{Prover cost per constraint as a function of number of constraints. The different lines indicate number of MPC servers. Error bars report 95\% confidence interval over 10 trials for constraints less than $2^{16}$, 3 trails for $2^{17}$, $2^{18}$ and single trail stretch run for $2^{19}$ and $2^{20}$.}
        \label{fig:prover_cost}
    \end{figure}
%    \end{minipage}\hfill
%    \begin{minipage}{0.45\textwidth}
   \begin{figure}
        \centering
        \includegraphics[width=0.9\columnwidth,height=1.2in]{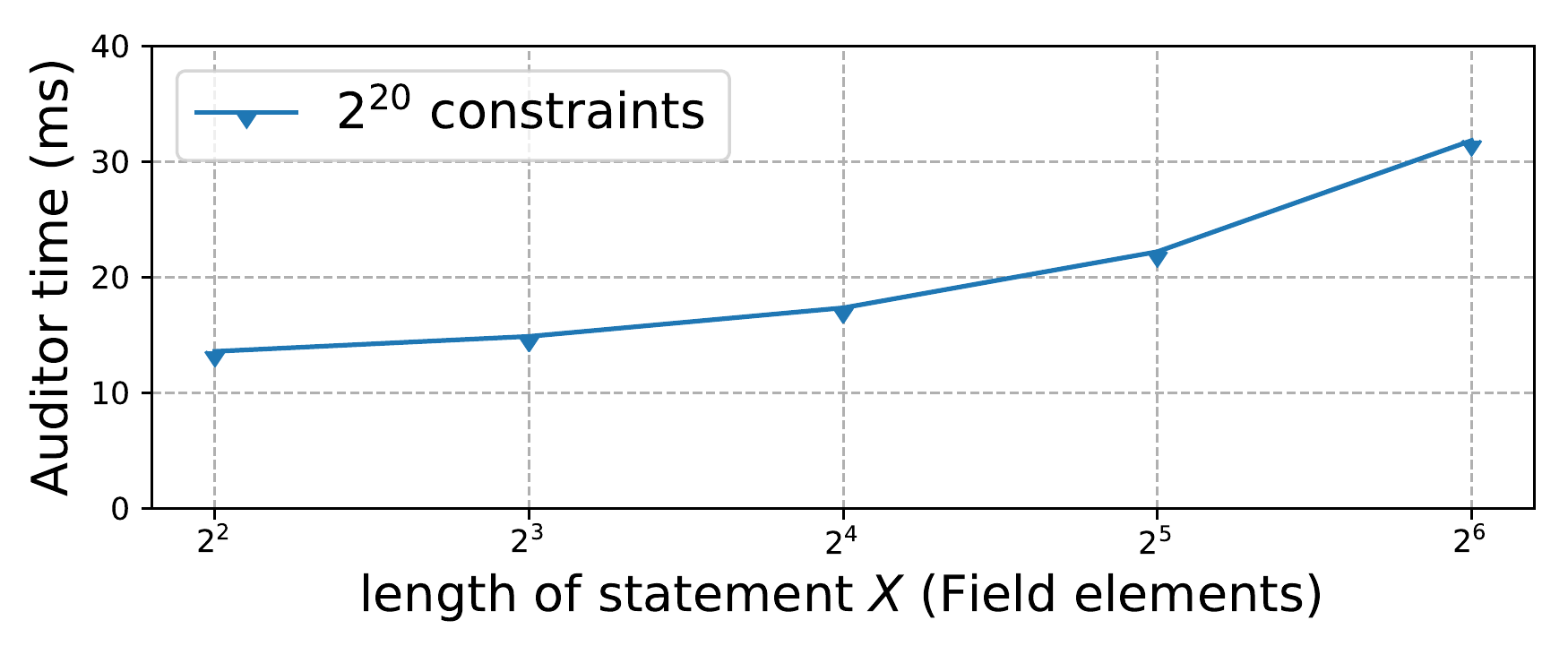} % second figure itself
        \caption{Auditor cost as a function of statement length. Values are averaged over 10 iterations}
        \label{fig:auditor_cost}
   % \end{minipage}
%    \end{minipage}
\end{figure}

\begin{figure}
    \centering
    \includegraphics[width=0.9\columnwidth]{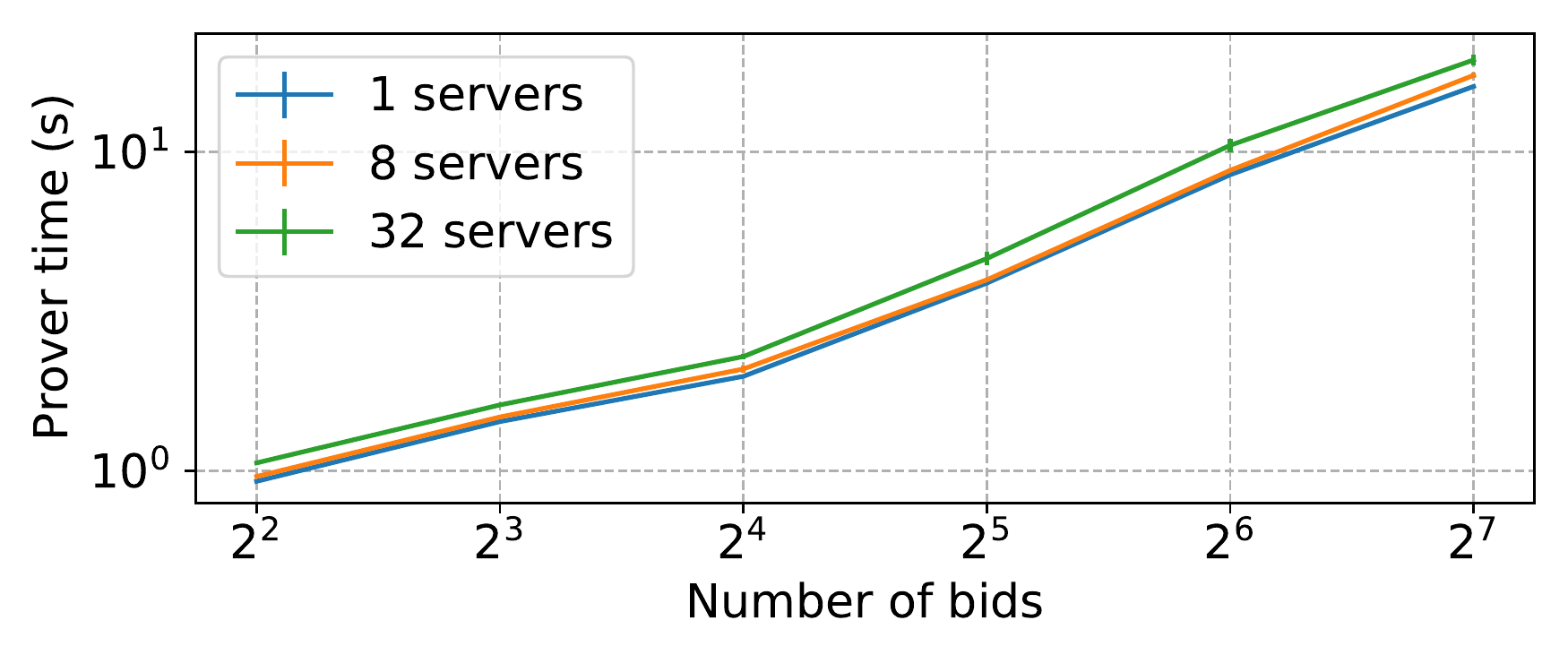} % first figure itself
    \vspace{-5mm}
    \caption{Auction Application Results: Prover cost as a function of number of bids for the auction application. Multiple lines denote the number of MPC servers involved in proof creation. Error bars show 95\% Confidence intervals over 10 iterations. 
    \vspace{-1mm}}
    \label{fig:auction_prover_cost}
\end{figure}

\label{sec:rand_circuits}
%\sanket{Need a better title for this subsection}
%We show how prover time, auditor time, proof size and communication cost are affected by statement size, number of constraints, and number of MPC servers involved. 

\subsubsection{Prover cost}
Server computation consists of two main components: 1) Computing the witness and output wire values at each server and 2) computing the Marlin proof by doing another MPC amongst the servers. When we report prover time benchmarks, we only consider the time for the second component. As our prover works in round-wise synchronous fashion, we consider a round time for MPC as the worst time amongst all provers in that round. As shown in Figure~\ref{fig:prover_cost} prover cost decreases logarithmically ($\frac{1}{\log{m}}$). 
The different lines show the prover time with different number of servers. Because of the linear communication cost in statement size, the prover time overhead due to the extra communication cost is marginal compared to the Marlin prover cost. This is the reason why the different lines start separate and eventually converge with additional constraints.
\subsubsection{Auditor cost}
The auditor computation mainly consists of two parts: 1) Verifying the marlin proof and 2) Carrying out the input consistency check. The first involves a constant number of pairings to verify that the claimed evaluations are consistent with the commitments, while the second involves interpolating a polynomial in the exponent which is linear in statement size.
Neither of these computations depend on the number of MPC servers used, or on the circuit size. Figure~\ref{fig:auditor_cost} shows the auditor cost as a function of statement length.%
\footnote{The x-label on graph shows statement size of $2^k$, it is actually $2^k$ - 2 where one statement is value 1, and one auxiliary value for hiding the commitment}

% \textsf{PEC.Lipmaa} the cost is linear is group exponentiation because it relies of the interpolate operation.  

\subsubsection{Proof size}
Our auditable MPC proof is just a Marlin proof combined with statement commitments. Normally in SNARKS, the statement is not considered a part of proof because the verifier is assumed to have access to it. Similarly when reporting proof size, we assume that auditor already has client input and server output commitments. Concretely speaking our proof size is 17 $\mathcal{G}_1$ elements and 23 $\mathbbm{F}_q$ elements which for bls12-381~\cite{bowefaster} corresponds to 1.5 KB (1552 bytes). (See Table~\ref{table:complexity} for details) 

\subsubsection{Communication Cost}
Our construction involves sequential(round-wise) communication; one  round for checking client inputs are correct against commitments and four additional rounds for the marlin prover algorithm. Recall that before executing the prover, the servers execute an MPC to check whether the inputs are consistent with the commitments provided by the clients. This step involves opening $|X|$ partial commitments while Marlin prover incurs an additional 9 openings in $\mathbbm{G}_1$ and 9 openings in $\mathbbm{F}_q$ across $4$ sequential rounds where a round communication between $n$ parties. For bls381 curve, with $n=32$ and a naive broadcast reconstruction algorithm, the total communication cost across all servers amounts to $\approx 700$ KB.

\section{Conclusion and future work}
Our work shows that public auditability can be practically added to existing MPC protocols, by adapting Marlin, a SNARK construction with universal trusted setup, to the MPC setting. Future work is to explore other SNARKs with different tradeoffs, including transparent SNARKs that avoid trusted setup altogether. Also in our prototype implementation we had to write our application program twice: once for MPC and again for the SNARK that checks the MPC's work. Publicly auditable MPC would benefit from a unified programming framework that targets both MPC and SNARKs in one program.

\bibliography{main.bbl}

\begin{thebibliography}{10}
\providecommand{\url}[1]{#1}
\csname url@samestyle\endcsname
\providecommand{\newblock}{\relax}
\providecommand{\bibinfo}[2]{#2}
\providecommand{\BIBentrySTDinterwordspacing}{\spaceskip=0pt\relax}
\providecommand{\BIBentryALTinterwordstretchfactor}{4}
\providecommand{\BIBentryALTinterwordspacing}{\spaceskip=\fontdimen2\font plus
\BIBentryALTinterwordstretchfactor\fontdimen3\font minus
  \fontdimen4\font\relax}
\providecommand{\BIBforeignlanguage}[2]{{%
\expandafter\ifx\csname l@#1\endcsname\relax
\typeout{** WARNING: IEEEtran.bst: No hyphenation pattern has been}%
\typeout{** loaded for the language `#1'. Using the pattern for}%
\typeout{** the default language instead.}%
\else
\language=\csname l@#1\endcsname
\fi
#2}}
\providecommand{\BIBdecl}{\relax}
\BIBdecl

\bibitem{massacci2018futuresmex}
F.~Massacci, C.~N. Ngo, J.~Nie, D.~Venturi, and J.~Williams, ``Futuresmex:
  secure, distributed futures market exchange,'' in \emph{2018 IEEE Symposium
  on Security and Privacy (SP)}.\hskip 1em plus 0.5em minus 0.4em\relax IEEE,
  2018, pp. 335--353.

\bibitem{cartlidge2019mpc}
J.~Cartlidge, N.~P. Smart, and Y.~Talibi~Alaoui, ``Mpc joins the dark side,''
  in \emph{Proceedings of the 2019 ACM Asia Conference on Computer and
  Communications Security}, 2019, pp. 148--159.

\bibitem{alexopoulos2017mcmix}
N.~Alexopoulos, A.~Kiayias, R.~Talviste, and T.~Zacharias, ``Mcmix: Anonymous
  messaging via secure multiparty computation,'' in \emph{26th $\{$USENIX$\}$
  Security Symposium ($\{$USENIX$\}$ Security 17)}, 2017, pp. 1217--1234.

\bibitem{lu2019honeybadgermpc}
D.~Lu, T.~Yurek, S.~Kulshreshtha, R.~Govind, A.~Kate, and A.~Miller,
  ``Honeybadgermpc and asynchromix: Practical asynchronous mpc and its
  application to anonymous communication,'' in \emph{Proceedings of the 2019
  ACM SIGSAC Conference on Computer and Communications Security}, 2019, pp.
  887--903.

\bibitem{lapets2016secure}
A.~Lapets, N.~Volgushev, A.~Bestavros, F.~Jansen, and M.~Varia, ``Secure mpc
  for analytics as a web application,'' in \emph{2016 IEEE Cybersecurity
  Development (SecDev)}.\hskip 1em plus 0.5em minus 0.4em\relax IEEE, 2016, pp.
  73--74.

\bibitem{rajan2018callisto}
A.~Rajan, L.~Qin, D.~W. Archer, D.~Boneh, T.~Lepoint, and M.~Varia, ``Callisto:
  A cryptographic approach to detecting serial perpetrators of sexual
  misconduct,'' in \emph{Proceedings of the 1st ACM SIGCAS Conference on
  Computing and Sustainable Societies}, 2018, pp. 1--4.

\bibitem{williamson2018aztec}
Z.~J. Williamson, ``The aztec protocol,'' \emph{URL: https://github.
  com/AztecProtocol/AZTEC}, 2018.

\bibitem{keller2020mp}
M.~Keller, ``Mp-spdz: A versatile framework for multi-party computation.''
  \emph{IACR Cryptol. ePrint Arch.}, vol. 2020, p. 521, 2020.

\bibitem{chida2018fast}
K.~Chida, D.~Genkin, K.~Hamada, D.~Ikarashi, R.~Kikuchi, Y.~Lindell, and
  A.~Nof, ``Fast large-scale honest-majority mpc for malicious adversaries,''
  in \emph{Annual International Cryptology Conference}.\hskip 1em plus 0.5em
  minus 0.4em\relax Springer, 2018, pp. 34--64.

\bibitem{wang2017global}
X.~Wang, S.~Ranellucci, and J.~Katz, ``Global-scale secure multiparty
  computation,'' in \emph{Proceedings of the 2017 ACM SIGSAC Conference on
  Computer and Communications Security}, 2017, pp. 39--56.

\bibitem{barak2018end}
A.~Barak, M.~Hirt, L.~Koskas, and Y.~Lindell, ``An end-to-end system for large
  scale p2p mpc-as-a-service and low-bandwidth mpc for weak participants,'' in
  \emph{Proceedings of the 2018 ACM SIGSAC Conference on Computer and
  Communications Security}, 2018, pp. 695--712.

\bibitem{abraham2020blinder}
I.~Abraham, B.~Pinkas, and A.~Yanai. (2020) Blinder--mpc based scalable and
  robust anonymous committed broadcast.

\bibitem{damgaard2009asynchronous}
I.~Damg{\aa}rd, M.~Geisler, M.~Kr{\o}igaard, and J.~B. Nielsen, ``Asynchronous
  multiparty computation: Theory and implementation,'' in \emph{International
  workshop on public key cryptography}.\hskip 1em plus 0.5em minus 0.4em\relax
  Springer, 2009, pp. 160--179.

\bibitem{damgaard2013practical}
I.~Damg{\aa}rd, M.~Keller, E.~Larraia, V.~Pastro, P.~Scholl, and N.~P. Smart,
  ``Practical covertly secure mpc for dishonest majority--or: breaking the spdz
  limits,'' in \emph{European Symposium on Research in Computer
  Security}.\hskip 1em plus 0.5em minus 0.4em\relax Springer, 2013, pp. 1--18.

\bibitem{stadler1996publicly}
M.~Stadler, ``Publicly verifiable secret sharing,'' in \emph{International
  Conference on the Theory and Applications of Cryptographic Techniques}.\hskip
  1em plus 0.5em minus 0.4em\relax Springer, 1996, pp. 190--199.

\bibitem{baum2014publicly}
C.~Baum, I.~Damg{\aa}rd, and C.~Orlandi, ``Publicly auditable secure
  multi-party computation,'' in \emph{International Conference on Security and
  Cryptography for Networks}.\hskip 1em plus 0.5em minus 0.4em\relax Springer,
  2014, pp. 175--196.

\bibitem{veeningen2017pinocchio}
M.~Veeningen, ``Pinocchio-based adaptive zk-snarks and secure/correct adaptive
  function evaluation,'' in \emph{International Conference on Cryptology in
  Africa}.\hskip 1em plus 0.5em minus 0.4em\relax Springer, 2017, pp. 21--39.

\bibitem{parno2013pinocchio}
B.~Parno, J.~Howell, C.~Gentry, and M.~Raykova, ``Pinocchio: Nearly practical
  verifiable computation,'' in \emph{2013 IEEE Symposium on Security and
  Privacy}.\hskip 1em plus 0.5em minus 0.4em\relax IEEE, 2013, pp. 238--252.

\bibitem{bowe2017scalable}
S.~Bowe, A.~Gabizon, and I.~Miers, ``Scalable multi-party computation for
  zk-snark parameters in the random beacon model.'' \emph{IACR Cryptol. ePrint
  Arch.}, vol. 2017, p. 1050, 2017.

\bibitem{bowe2018multi}
S.~Bowe, A.~Gabizon, and M.~D. Green, ``A multi-party protocol for constructing
  the public parameters of the pinocchio zk-snark,'' in \emph{International
  Conference on Financial Cryptography and Data Security}.\hskip 1em plus 0.5em
  minus 0.4em\relax Springer, 2018, pp. 64--77.

\bibitem{ben2015secure}
E.~Ben-Sasson, A.~Chiesa, M.~Green, E.~Tromer, and M.~Virza, ``Secure sampling
  of public parameters for succinct zero knowledge proofs,'' in \emph{2015 IEEE
  Symposium on Security and Privacy}.\hskip 1em plus 0.5em minus 0.4em\relax
  IEEE, 2015, pp. 287--304.

\bibitem{chiesa2019marlin}
A.~Chiesa, Y.~Hu, M.~Maller, P.~Mishra, N.~Vesely, and N.~Ward, ``Marlin:
  Preprocessing zksnarks with universal and updatable srs,'' Cryptology ePrint
  Archive, Report 2019/1047, 2019, https://eprint. iacr. org~…, Tech. Rep.,
  2019.

\bibitem{campanelli2019legosnark}
M.~Campanelli, D.~Fiore, and A.~Querol, ``Legosnark: modular design and
  composition of succinct zero-knowledge proofs,'' in \emph{Proceedings of the
  2019 ACM SIGSAC Conference on Computer and Communications Security}, 2019,
  pp. 2075--2092.

\bibitem{lund1992algebraic}
C.~Lund, L.~Fortnow, H.~Karloff, and N.~Nisan, ``Algebraic methods for
  interactive proof systems,'' \emph{Journal of the ACM (JACM)}, vol.~39,
  no.~4, pp. 859--868, 1992.

\bibitem{shamir1979share}
A.~Shamir, ``How to share a secret,'' \emph{Communications of the ACM},
  vol.~22, no.~11, pp. 612--613, 1979.

\bibitem{beaver1991efficient}
D.~Beaver, ``Efficient multiparty protocols using circuit randomization,'' in
  \emph{Annual International Cryptology Conference}.\hskip 1em plus 0.5em minus
  0.4em\relax Springer, 1991, pp. 420--432.

\bibitem{beerliova2008perfectly}
Z.~Beerliov{\'a}-Trub{\'\i}niov{\'a} and M.~Hirt, ``Perfectly-secure mpc with
  linear communication complexity,'' in \emph{Theory of Cryptography
  Conference}.\hskip 1em plus 0.5em minus 0.4em\relax Springer, 2008, pp.
  213--230.

\bibitem{damgaard2007scalable}
I.~Damg{\aa}rd and J.~B. Nielsen, ``Scalable and unconditionally secure
  multiparty computation,'' in \emph{Annual International Cryptology
  Conference}.\hskip 1em plus 0.5em minus 0.4em\relax Springer, 2007, pp.
  572--590.

\bibitem{kate2010constant}
A.~Kate, G.~M. Zaverucha, and I.~Goldberg, ``Constant-size commitments to
  polynomials and their applications,'' in \emph{International Conference on
  the Theory and Application of Cryptology and Information Security}.\hskip 1em
  plus 0.5em minus 0.4em\relax Springer, 2010, pp. 177--194.

\bibitem{fuchsbauer2018algebraic}
G.~Fuchsbauer, E.~Kiltz, and J.~Loss, ``The algebraic group model and its
  applications,'' in \emph{Annual International Cryptology Conference}.\hskip
  1em plus 0.5em minus 0.4em\relax Springer, 2018, pp. 33--62.

\bibitem{syta2017scalable}
E.~Syta, P.~Jovanovic, E.~K. Kogias, N.~Gailly, L.~Gasser, I.~Khoffi, M.~J.
  Fischer, and B.~Ford, ``Scalable bias-resistant distributed randomness,'' in
  \emph{2017 IEEE Symposium on Security and Privacy (SP)}.\hskip 1em plus 0.5em
  minus 0.4em\relax Ieee, 2017, pp. 444--460.

\bibitem{groth2018updatable}
J.~Groth, M.~Kohlweiss, M.~Maller, S.~Meiklejohn, and I.~Miers, ``Updatable and
  universal common reference strings with applications to zk-snarks,'' in
  \emph{Annual International Cryptology Conference}.\hskip 1em plus 0.5em minus
  0.4em\relax Springer, 2018, pp. 698--728.

\bibitem{babai1991checking}
L.~Babai, L.~Fortnow, L.~A. Levin, and M.~Szegedy, ``Checking computations in
  polylogarithmic time,'' in \emph{Proceedings of the twenty-third annual ACM
  symposium on Theory of computing}, 1991, pp. 21--32.

\bibitem{lipmaa2017prover}
H.~Lipmaa, ``Prover-efficient commit-and-prove zero-knowledge snarks,''
  \emph{International Journal of Applied Cryptography}, vol.~3, no.~4, pp.
  344--362, 2017.

\bibitem{canetti2001universally}
R.~Canetti, ``Universally composable security: A new paradigm for cryptographic
  protocols,'' in \emph{Proceedings 42nd IEEE Symposium on Foundations of
  Computer Science}.\hskip 1em plus 0.5em minus 0.4em\relax IEEE, 2001, pp.
  136--145.

\bibitem{code:marlin}
\BIBentryALTinterwordspacing
``Marlin: rust library for preprocessing zksnarks,'' 2019. [Online]. Available:
  \url{https://github.com/scipr-lab/marlin}
\BIBentrySTDinterwordspacing

\bibitem{bowefaster}
S.~Bowe. Faster subgroup checks for bls12-381.

\bibitem{aliasgari2013secure}
M.~Aliasgari, M.~Blanton, Y.~Zhang, and A.~Steele, ``Secure computation on
  floating point numbers.'' in \emph{NDSS}, 2013.

\bibitem{de2012design}
S.~J.~A. de~Hoogh. (2012) Design of large scale applications of secure
  multiparty computation: secure linear programming.

\bibitem{mantel1966evaluation}
N.~Mantel, ``Evaluation of survival data and two new rank order statistics
  arising in its consideration,'' \emph{Cancer Chemother. Rep.}, vol.~50, pp.
  163--170, 1966.

\bibitem{boneh2004short}
D.~Boneh and X.~Boyen, ``Short signatures without random oracles,'' in
  \emph{International conference on the theory and applications of
  cryptographic techniques}.\hskip 1em plus 0.5em minus 0.4em\relax Springer,
  2004, pp. 56--73.

\bibitem{bernstein2002pippenger3s}
D.~J. Bernstein. (2002) Pippenger$^3$s exponentiation algorithm.

\bibitem{pedersen1991non}
T.~P. Pedersen, ``Non-interactive and information-theoretic secure verifiable
  secret sharing,'' in \emph{Annual international cryptology conference}.\hskip
  1em plus 0.5em minus 0.4em\relax Springer, 1991, pp. 129--140.

\bibitem{rotemalgebraic}
L.~Rotem and G.~Segev. Algebraic distinguishers: From discrete logarithms to
  decisional uber assumptions.

\bibitem{ben2019aurora}
E.~Ben-Sasson, A.~Chiesa, M.~Riabzev, N.~Spooner, M.~Virza, and N.~P. Ward,
  ``Aurora: Transparent succinct arguments for r1cs,'' in \emph{Annual
  international conference on the theory and applications of cryptographic
  techniques}.\hskip 1em plus 0.5em minus 0.4em\relax Springer, 2019, pp.
  103--128.

\end{thebibliography}

\appendices

\section{Applications:}
\label{app:applications}
For 32 servers with circuits over 1 million multiplication gates, the pre-processing takes less than a minute. The benchmarks do not consider the offline cost of pre-processing as it can be carried out previously without knowledge of any particular computation description. A full auditable MPC should also carry out the audit the offline phase, but we believe our ideas can be directly combined with Blinder\cite{abraham2020blinder} to obtain a full robust specification. 
% \begin{figure*}[ht]
% \begin{minipage}{1\textwidth}
%     \centering
%     \begin{minipage}{0.45\textwidth}
%         \centering
%         \includegraphics[width=0.9\textwidth]{evaluation/auction_prover.pdf} % first figure itself
%         \caption{Prover cost as a function of number of bids. Multiple lines denote the number of MPC servers involved in proof creation. Error bars show 95\% Confidence intervals over 10 iterations. }
%         \label{fig:auction_prover_cost}
%     \end{minipage}\hfill
%     \begin{minipage}{0.45\textwidth}
%         \centering
%         \includegraphics[width=0.9\textwidth]{evaluation/auction_auditor.pdf} % second figure itself
%         \caption{Auditor cost as a function of number of bids. Multiple lines denote the number of MPC servers involved in proof creation. Error bars show 95\% Confidence intervals over 100 iterations. }
%         \label{fig:auction_auditor_cost}
%     \end{minipage}
% \end{minipage}
% \end{figure*}

\subsubsection{Auction Application}

We also evaluate our implementation on the motivation auction application functionality listed in section~\ref{sec:motivate_auction}. Our auction application critically relies on comparison operation for which we implement a comparison circuit. We use techniques from \cite{aliasgari2013secure}\cite{de2012design} to compute a share of $\share{a>b}$ and prove that it was computed correctly. 
% Appendix~\ref{app:comparison} shows the details of the R1CS comparison circuit. 

Our application proceeds in a reactive manner processing a fixed of number of bids $k$ each round. At the end of each round, the servers maintain a secret shared state of the highest bid which can be then used as an input for subsequent rounds. Figure~\ref{fig:auction_prover_cost} show the prover cost for the auction application when varied across $k$, the number of bids processed in one round. As expected, the prover cost increases almost linearly because an increase $k$ results in a linear increase in the number of constraints. Finally, the auditor cost exhibits a similar graph as figure ~\ref{fig:auditor_cost}.

\subsubsection{Logrank Test}

% Logrank test algorithm is detailed in Appendix~\ref{app:logrank_alg}.
% Logrank Algorithm uses fixed point operations  Appendix~\ref{app:logrank_fixed} shows the construction of efficient R1CS constraints for fixed point operations from\cite{aliasgari2013secure}\cite{de2012design}.
-
Table~\ref{table:complexity} shows a theoretical comparison of our work with Veeningen, while Table~\ref{table:logrank} shows the performance comparison for the Logrank Test (See Appendix~\ref{app:logrank}) involving fixed point operations. $\mathbf{C_x}$ denotes the total number of statement commitments. The first shows performance numbers reported by Veeningen~\cite{veeningen2017pinocchio}. %They had a block-size parameter that controlled a a tradeoff between prover cost and auditor cost. Their total coste total cost for the application across all blocks, we report the performance numbers in per-block basis for a more direct comparison of programs of similar complexity. 

Our prover cost is empirically close to Veeningen's which is surprising because the Marlin prover is more expensive Pinocchio prover. We think this is likely because of our use of FFT friendly curves like BLS12-381\cite{bowefaster}. We expect our prover to be about 3-4 times slower than adaptive SNARKs with circuit specific setup. Our auditor 
performs significantly better than Veeningen~\cite{veeningen2017pinocchio} because of constant pairing cost compared to linear pairing cost from Veeningen. Our auditor(\textsf{PEC.Lipmaa}) also incurs a linear cost in group exponentiation operations, but in practice the cost of those operations is small compared to pairing operations. 

%---------------------------------------------------------------------
\begin{table}[ht!]
 \caption{Prover and auditor cost comparison of this work with Veeningen.}
\resizebox{\columnwidth}{!}{%
\begin{tabular}{ c c|c |c| c  }
\hline
  & & $|\mathbf{C_x}|$ = 1 & $|\mathbf{C_x}|$ = 7 &  $|\mathbf{C_x}|$ = 175\\
 \hline
 \multirow{2}{1cm}{Vee'17} & Prove  & 0.4s & 3.2s & 73.5s \\
                                     & Audit & 0.0s & 0.3s & 4.9s \\
\hline
 \multirow{2}{1cm}{This work} & Prove  & 1.46 $\pm$ 0.02s & 2.23 $\pm$ 0.02 & 85.74 $\pm$ 0.98s \\
                                     & Audit  & 12.4 $\pm$ 0.04  ms & 18.8 $\pm$ 0.42 ms & 40.5 $\pm$ 0.37 ms \\
 \hline
%  Afghanistan	& AF	&AFG&	004\\
%  Aland Islands&	AX	& ALA	&248\\
%  Albania	&AL	& ALB&	008\\
%  Algeria	&DZ	& DZA&	012\\
%  American Samoa&	AS	& ASM&016\\
%  Andorra&	AD	& AND	&020\\
%  Angola&	AO	& AGO&024\\
%   \hline
 \end{tabular}
 }
 \begin{center}
 \end{center}
 \label{table:logrank}
 \end{table}
%--------------------------------------------------------------------

\section{Logrank}
\label{app:logrank}
\subsection{Logrank Test}

Mantel-Haenzel Logrank test\cite{mantel1966evaluation} is a statistical test to decide whether there is a significant difference in survival rate between the two populations. It is widely used in clinical trials to establish the efficacy of a new treatment in comparison with a control treatment. The survival data about a population is represented by a set of tuples $(n_j , d_j )$, where $n_j$ is the number of patients still in the study just before time $j$ and $d_j$ is the number of deaths at time $j$. The populations are distributed across multiple hospitals and each hospital commits to it's value of $(n_j, d_j)$ tuple. 

The null hypothesis for the logrank test, i.e., the distributions represent the same “survival function”, corresponds to $X \sim \chi^2_1$. This null hypothesis is
rejected if $1 - \mathsf{cdf}(X) > \alpha$, where
$\mathsf{cdf}$ is the cumulative density function of the $\chi^2_1$ distribution. Logrank test involves fixed point operations of division, multiplication. 
Similar to Geppetri~\cite{veeningen2017pinocchio}, we use MPC to compute $X$, and then apply the $\mathsf{cdf}$ in the clear. Figure~\ref{app:logrank_alg} show the algorithms implemented. Veeningen also had a block-size parameter that controlled a a tradeoff between prover cost and auditor cost. Their total cost total cost for the application across all blocks, in table~\ref{table:logrank} we report the performance numbers in per-block basis for a more direct comparison of programs of similar complexity. 

\subsubsection{Logrank Algorithm}
\label{app:logrank_alg}
\begin{algorithm*}
    \caption{Logrank computation for each time step}
    \begin{algorithmic}[1]
        \Require \share{\mathbf{d}_{i,1}},\share{\mathbf{d}_{i, 1}},\share{\mathbf{d}_{i, 2}},\share{\mathbf{n}_{i, 1}},\share{\mathbf{n}_{i, 2}} \text{survival data at time point} $i$
        \Ensure $(\share{e_{i}}^{f},\share{v_{i}}^{f},\share{d_{i}})$ contributions to $\sum_{j} E_{j, 1}, \sum_{j} V_{j}, \sum_{j} d_{j, 1}$ for test statistic
        
        \Function{BLOCK}{\share{\mathbf{d}_{i,1}},\share{\mathbf{d}_{i, 2}},\share{\mathbf{n}_{i, 1}},\share{\mathbf{n}_{i, 2}}}
        
        \State $\share{a c} \leftarrow \share{ \mathbf{d}_{i, 1}}+\share{\mathbf{d}_{i, 2}}$
        
        \State $\share{b d} \leftarrow\share{\mathbf{n}_{i, 1}}+\share{\mathbf{n}_{i, 2}}$
        
        \State $\share{f r c}^{f} \leftarrow \share{a c}/\share{b d}$
        
        \State $\share{e_{i}}^{f} \leftarrow \share{frc}^{f} \cdot\share{\mathbf{n}_{i, 1}}$
        
        \State $\share{v n} \leftarrow\share{\mathbf{n}_{i, 1}} \cdot\share{\mathbf{n}_{i, 2}} \cdot \share{a c} \cdot (\share{b d}-\share{a c})$
        
        \State $\share{v d} \leftarrow \share{b d} \cdot \share{b d} \cdot(\share{b d} -1)$
        \State $\share{v_{i}}^{f} \leftarrow \share{vn} /\share{v d}$
        
        \State \textbf{return} $(\share{e_{i}}^{f},\share{v_{i}}^{f},\share{d_{i}})$
        
        \EndFunction
% \State $\mathrm{B} \mathrm{LOCK}\left(\share{\mathrm{d}_{i, 1}},\share{\mathbf{d}_{i, 2}},\share{\mathbf{n}_{i, 1}},\share{\mathbf{n}_{i, 2}}\right)$

    \end{algorithmic}
\end{algorithm*}

\begin{algorithm*}
    \caption{Logrank final computation}
    \begin{algorithmic}[1]
        \Require $\share{e s},\share{v s},\share{d s}:$ summed-up values required to compute $X$

        \Ensure $\share{{chi}}^{f}$ test statistic comparing two curves; supposedly ${chi} \sim \chi_{1}^{2}$
        
        \Function{FIN}{\share{e s},\share{v s},\share{d s}}
        
        \State $\share{d s}^{f} \leftarrow \share{d s} \ll$ PRECISION
        
        \State $\share{d m i}^{f} \leftarrow \share{d s}^{f}-\share{v s}^{f}$
        
        \State $\share{{chi}}^{f} \leftarrow \share{{dmi}}^{f} /\share{{vs}}^{f}$
        
        \State $\share{{chi}}^{f} \leftarrow \share{{chi} }^{f} \cdot \share{{dmi}}^{f}$
        
        \State \textbf{return} $\share{c h i}^{f}$
        
        \EndFunction
% \State $\mathrm{B} \mathrm{LOCK}\left(\share{\mathrm{d}_{i, 1}},\share{\mathbf{d}_{i, 2}},\share{\mathbf{n}_{i, 1}},\share{\mathbf{n}_{i, 2}}\right)$

    \end{algorithmic}
\end{algorithm*}

$$E_{j, 1}=\frac{\left(d_{j, 1}+d_{j, 2}\right) \cdot n_{j, 1}}{n_{j, 1}+n_{j, 2}}$$

$$V_{j}=\frac{n_{j, 1} n_{j, 2}\left(d_{j, 1}+d_{j, 2}\right)\left(n_{j, 1}+n_{j, 2}-d_{j, 1}-d_{j, 2}\right)}{\left(n_{j, 1}+n_{j, 2}\right)^{2} \cdot\left(n_{j, 1}+n_{j, 2}-1\right)}$$

$$X=\frac{\sum_{j} E_{j, 1}-\sum_{j} d_{j, 1}}{\sum_{j} V_{j}}$$

% \subsubsection{Fixed point operations}
% \label{app:logrank_fixed}
% \subsection{Auction}
% \label{app:auction}
% Our auction application critically relies on comparison circuit. We logic fom circom.js library. for completeness, we state the 
\section{Cryptographic Assumptions:}
\label{app:assumptions}
\subsection{Strong Diffie-Hellman:}
\label{app:sdh}

Let 
$\langle \mathsf{group} \rangle = 
(\mathbb{G}_1, \mathbb{G}_2, \mathbb{G}_T, q, g, h, e)$ where $\mathbb{G}_1, \mathbb{G}_2, \mathbb{G}_T$
are groups of a prime order $q$, $g$ generates $\mathbb{G}_1$,
$h$ generates $\mathbb{G}_2$, and $e: \mathbb{G}_1 \times \mathbb{G}_2 \rightarrow \mathbb{G}_T$ is a (non-degenerate) bilinear map with security parameter $\lambda$. The $\textbf{Strong Diffie-Hellman}$\cite{boneh2004short}\cite{chiesa2019marlin} assumption states that for every efficient adversary $\Adv$ and a degree $d \in \mathbbm{N}$ the following holds:

      \[
  \Pr \left[
    \begin{gathered}
     C = g^{\frac{1}{\alpha + c}} \\
  \end{gathered}
    \middle|\\
        \begin{gathered}
    \alpha \leftarrow \mathbbm{F}_q\\
    \mathbf{\Sigma} \leftarrow (\{g^{\alpha^i}\}^d_{i=0}, h^\alpha)\\
    c, g^c \leftarrow \Adv(\langle \mathsf{group} \rangle, \mathbf{\Sigma})\\
    \end{gathered}
  \right] 
  = \mathsf{negl}(\lambda)
  \]
  
We note that this assumption is a stronger assumption that the $q$ dlog assumption as it additionally allows the adversary to pick $c$. If c is pre-specified $\mathsf{SDH}$ reduces to q-dlog.

\subsection{Algebraic group Model:}
\label{app:algebraic}

In order to achieve additional efficiency, marlin papers shows how to construct polynomial commitment schemes in the Algebraic Group Model(AGM)~\cite{fuchsbauer2018algebraic}, which replaces specific knowledge assumptions (such as Power Knowledge of Exponent assumptions) with simpler assumptions like $\mathsf{SDH}$. 
Let $\mathbbm{G}$ be a cyclic group of prime order $p$. An algorithm $A_{alg}$ algebraic if whenever $A_{alg}$ outputs a group element $Z \in \mathbbm{G}$, it also outputs
a ``representation'' $\vect{z}$ = $(z_1, \ldots, z_t) \in \mathbbm{Z}^t_p$ such that $\mathbf{Z} = \Pi_iL_i^{z_i}$ where $\mathbf{L} = (L_1, \ldots, L_t)$ is the list of all group elements that were given to $A_{alg}$ during it's execution so far.

In AGM, we model adversaries as algebraic, which means that whenever an adversary $\Adv$ outputs a group element $G$, $\Adv$ must also output an ``explanation'' or ``representation" of $G$ in terms of the group elements that it has seen beforehand.

\subsection{Extractable Commitments}
\label{sec:prelim_commitments}
%\todo{These definitions are informal anyway, this may as well go to appendix  or inline when needed}
An extractable commitment scheme consists of a pair of probabilistic polynomial time algorithms $\mathsf{Setup, Comm}$. The setup algorithm \textsf{Setup}$(1^\lambda, \mathsf{D})\rightarrow \mathsf{crs, [ck_i]_{i=1}^D, rk, td}$ generates committer keys $\mathsf{ck_i}$ and a receiver key $\mathsf{rk}$ for the scheme and some trapdoor $\mathsf{td}$ for a given security parameter $\lambda$ and a bound $\mathsf{D}$. The commitment algorithm $\mathsf{Comm}$ defines a function $\mathsf{Comm(ck_i}, m; r)$ outputs a commitment $c$ to the message $m$ with randomness $r$ using committer key $\mathsf{ck_i}$.  
Additionally, the trapdoor commitment scheme must satisfy the following properties:

\textbf{Computational Binding:} For all ppt $\Adv$ $\mathsf{Pr}[\mathsf{(crs, \mathbf{ck}, .)}$ $\leftarrow
\mathsf{Setup}(1^\lambda, \mathsf{D}), (m, r, m', r, \mathsf{i})$ $\leftarrow \Adv(\mathsf{crs}, \mathbf{ck})$, $\mathsf{Comm({ck_i}}, m; r)$ $= \mathsf{Comm({ck_i}}, m'; r')] \approx \mathsf{negl}(\lambda)$. 

\textbf{Trapdoor property:} There exist ppts $T_1,T_2$ such that: $\mathsf{(crs, \mathbf{ck}, rk, td)}$ $\leftarrow
\mathsf{Setup}(1^\lambda, \mathsf{D})$, $(c_1, \tau) \leftarrow T_1(crs, td)$, $r \leftarrow T_2(c_1, \tau, m)$, then $c_1$ is identically distributed to real commitments and ${\mathsf{Comm(ck_i, m; r)} = c_1}$.

\textbf{Perfect hiding:} $\mathsf{(crs, \mathbf{ck}, .)}$ $\leftarrow
\mathsf{Setup}(1^\lambda, \mathsf{D})$, ($r, r') \samples \mathbbm{F}$, $\forall (m, m'):$ $\mathsf{Comm(ck_i}, m; r)$ is identically distributed to $\mathsf{Comm(ck_i}, m'; r')$.

\textbf{Extractability:} $\forall$ ppt $\Adv$, there exists ppt $\Ext^\Adv$ such that $\mathsf{Pr}[\mathsf{(crs, \mathbf{ck}, .)}$ $\leftarrow$ $\mathsf{Setup}(1^\lambda, \mathsf{D}), (c||m;r)\leftarrow (\Adv||\Ext^\Adv)(\mathsf{crs}$, $\mathbf{ck})$, $c \in \mathsf{Range(Comm(ck_i, .)) \land c \neq \mathsf{Comm}(\mathbf{ck}_i, m;r)}] \approx \mathsf{negl}(\lambda)$. $\mathsf{Range}$ denotes the range of commitments.

\section{Cost analysis}
Table \ref{table:complexity} shows the theoretical comparison between our constructions($\mathsf{PEC.Ped}$ and $\mathsf{PEC.Succ}$) and the construction from Veeningen~\cite{veeningen2017pinocchio}. Our construction for $\mathsf{PEC.Ped}$ and $\mathsf{PEC.Succ}$ incurs a five round communication overhead compared to one round in Veeningen's construction.
As shown in Table~\ref{table:complexity} the prover cost is proportional to (Variable base Multi-scalar exponentiation) $v-MSM(m)$ that can be calculated in $\frac{m}{\log{m}}$ by using Peppinger's algorithm~\cite{bernstein2002pippenger3s}. 
%---------------------------------------------------------------------
\begin{table*}[ht!]
\begin{minipage}{1\textwidth}
 \caption{\textmd{$N:$ number of MPC parties; $K:$ number of clients; $X:$total statement size(all inputs + output); $m:$ total number of gates, $n:$ number of multiplication gates; Comm: Communication cost, $\pi$ proof size. v-MSM($m$) denotes variable-base multi-scalar multiplications (MSM) each of size $m$. }}
\resizebox{\columnwidth}{!}{%
\tiny
\begin{tabular}{ c c|c c|c c |c  }
\hline
& & \multicolumn{2}{|c|}{size/cost(bytes)} & \multicolumn{2}{c|}{Time Complexity}\\
\hline
  & & Comm & $|\pi|$ &  Prover & Auditor & Setup\\
 \hline
 \multirow{3}{1cm}{Vee'17} & $\mathbb{G}_1$ & $K\  \mathsf{open}$ & 3$K$ + 8 & $\bigO{(n)}$ & 6$K$ + 12 pair. & \textsf{circ}  \\
                                     & $\mathbb{G}_2$ & - & - & - & - \\
                                     & $\mathbb{F}_q$ & - & - & $\bigO(m + n\log n)$ &- &  \\
\hline
 \multirow{3}{1cm}{Ours\\(Pec.Ped)} & $\mathbb{G}_1$ & $X + 9 \ \mathsf{open}$ & 17 & 21 v-MSM($3m$) + 9 v-MSM($N$) &  $X + 2$ grp operations & \textsf{univ}  \\
                                     & $\mathbb{G}_2$ & - & - & - & 2 pairings\\
                                     & $\mathbb{F}_q$ & $9 \ \mathsf{open}$ & $23$ & $\bigO(N + m\log m)$ & $\bigO(\log m)$  \\
 \hline
  \multirow{3}{1cm}{Ours\\(Pec.Suc)} & $\mathbb{G}_1$ & $K + 9 \ \mathsf{open}$ & 17 & 21 v-MSM($3m$) + 9 v-MSM($N$) &  $K + 3$  pair& \textsf{univ}  \\
                                     & $\mathbb{G}_2$ & - & - & - & -\\
                                     & $\mathbb{F}_q$ & $9 \ \mathsf{open}$ & $23$ & $\bigO(N + m\log m)$ & $\bigO(\log m)$  \\
%  \hline
%  Afghanistan	& AF	&AFG&	004\\
%  Aland Islands&	AX	& ALA	&248\\
%  Albania	&AL	& ALB&	008\\
%  Algeria	&DZ	& DZA&	012\\
%  American Samoa&	AS	& ASM&016\\
%  Andorra&	AD	& AND	&020\\
%  Angola&	AO	& AGO&024\\
  \hline
 \end{tabular}
 }
 \begin{center}
 \end{center}
 \label{table:complexity}
   \end{minipage}
 \end{table*}
%--------------------------------------------------------------------
% \input{appendix/comm_cost_details}
\section{Polycommits using MPC}
\label{app:mpc_polycommit}
We observe that all polycommit operations are MPC friendly, that is one can create commitments, provide evaluation proofs for local secret shared polynomials and then later interpolate them. $\mathsf{Eval_{MPC}}$ thus requires regular Lagrange interpolation while providing evaluation proofs, polynomial commitments requires interpolating in the exponent. For simplicity, we state the protocols for single commitment and single evaluation, although batching can be supported. Infact, our implementation makes use of such batching.
% \subsection{Robust interpolate in the exponent:}
% \label{sec:robust_exp}
% Note that all the Lagrange polynomials are fixed for the given set of $n$ parties. 

% \begin{figure}
% \begin{boxedminipage}[t]{0.5\textwidth}
%   {\centering \textbf{PolyCommits from shares of evaluations}\\}
\subsection{\texorpdfstring{$\mathsf{PC.Commit_{MPC}}$}{}: PolyCommits from shares of evaluations}
\label{app:sec_mpc_polycommit}
\textbf{Input:} shares of evaluation $\share{x_{i_1}}, \ldots, 
\share{x_{i_k}}$ of $\hat{x}$ 

\textbf{Output:} Hiding polynomial commitment $C_{\hat{x}} = g^{\hat{x}}h^{\hat{r}}$.
\textbf{Procedure:}(For each server $P_i$)
% \quad\emph{// Compute the function}
\begin{enumerate}
    \item $\share{\hat{x}}$ = $\mathbf{V}(\share{x_{i_1}}, \ldots, \share{x_{i_k}})$ where $\mathbf{V}$ is the Vandermonde matrix over the evaluation domain corresponding to $k$. 
    \item Similarly, Compute $\share{\hat{r}}=$ $\mathbf{V}(\share{r_{i_1}}, \ldots, \share{r_{i_k}})$ where $\share{r_{i_j}}$ are pre-processed random shares.
    \item Compute $C_{\share{\hat{x}}} = \mathsf{PC.commit(ck,} \share{\hat{x}}, \share{\hat{r}})$ and send the commitment $C_{\share{\hat{x}}}$ to all other parties.
    \item After receiving all the shares of commitments $C_{\share{\hat{x}}}$, we check easily the evaluate the commitment to candidate interpolation result polynomial $\Pi_{i=1}^{t+1} (C_{\share{\hat{x}}})^{\ell_i(v)}$ at any point $v$ we desire. We can use that to check whether $2t + 1$ shares agree on some polynomial in the exponent. After finding such polynomial, we evaluate it at $0$ in the exponent to obtain the $C_{\hat{x}}$.
\end{enumerate}
% \end{boxedminipage}
% \caption{MPC Polynomial Commitments from shares of evaluations}
% \label{fig:mpc_polycommit}
% \end{figure}

\subsection{\texorpdfstring{$\mathsf{PC.Open_{MPC}}$}{}: MPC Polynomial evaluation proofs}
\label{sec:mpc_polycommit_createwitness}
The protocol for $\mathsf{PC.Open_{MPC}}$ proceeds similar to protocol for creating polynomial commitments as in Appendix~\ref{app:sec_mpc_polycommit} with two minor changes. Instead of $\mathsf{PC.Commit}$ in step $3$, we use $\mathsf{PC.Open}$ and instead of sampling randomness in step $2$, we use the same randomness as the ones used in $\mathsf{PC.Commit}$. 

\section{Constructions of Polynomial Commitments}
\label{app:pec_constructions}
The versions of $\mathsf{PEC.Ped}$ and $\mathsf{PEC.Lipmaa}$ are useful only when there is a single input per party(as with the auction application). Hence, while stating the below constructions, we state them with $K = D$.
\subsection{Construction using Pedersen Commitments}
\label{sec:pec_cons_ped}
Our polynomial evaluation commitment scheme $\mathsf{PEC.Ped}$ over a cyclic group $\mathbbm{G}$ is constructed as follows: %$(\mathsf{Setup, Commit_{eval}, Interpolate, Open, Check})$. 

\begin{itemize}
    \item $\mathsf{Setup}(1^\lambda, D)$ $\rightarrow \mathsf{ck_{e}, ck_{p}, rk_{p}, td}$: Sample random generators $g$ and $h = g^{\mathsf{td}}$ and return $\mathsf{ck_{eval}} = (g, h) , \mathsf{ck_{poly}} = (g, h) ,\mathsf{rk_{poly}} = (g, h) , \mathsf{td}$.
    \item $\mathsf{Commit_{eval}(ck_{e}}, e, p; \omega) \rightarrow c_{e}$: Parse $\mathsf{ck_{e}} = (g, h)$. Then the commitment to a evaluation $e$ is  $c_{e}$ = $g^{e_i}h^{r_i}$ where $r_i$ is sampled randomly according to randomness $\omega$.
    \item $\mathsf{Interpolate(ck_{p}}, \mathbf{c_{e})} \rightarrow \mathbf{c}$: $\mathbf{c}$ = $\mathbf{c_{e}}$. The interpolate operation is same as collection of commitments and thus is not a succinct commitment.
    \item $\mathsf{Open(ck_{p}}, \phi, q; \vect{\omega}) \rightarrow v, \pi$: Obtain the interpolated randomness $r_i$ from $\boldsymbol{\omega} = [\omega_i]$ (must the be same as the one used for $\mathsf{Commit_{eval}}$ at evaluation point $p_i$) and must be interpolation of committed values $e_i$. Interpolate $\mathbf{r}$, polynomial with evaluations $r_i$ and return $(v, \pi) = (\mathbf{p}(q), \mathbf{r}(q))$.
    \item $\mathsf{Check(rk_{poly}}, \mathbf{c}, q, v, \pi) \rightarrow \{0, 1\}$: Parse $\mathsf{rk_{poly}} = (g, h)$ and check
    \begin{equation}
      \Pi_{i=1}^n\left(c_i\right)^{\ell_i(q)} \stackrel{?}{=} g^vh^\pi  
    \end{equation}
    where $\ell_i(X)$ denotes the Lagrange polynomial at evaluation point $i$ (or $\omega^i$ in case of FFT). 
\end{itemize}

We defer the proof of $Pec.Ped$ to the original paper from Pedersen\cite{pedersen1991non} for secret sharing. 
\subsection{Auditor based on Lipmaa commitments}
\label{sec:pec_cons_poly}

Our key idea is to commit to an evaluation (share) of the polynomial by committing to a Lagrange polynomial multiplied by the share. We can later homomorphically combine the commitments to shares to obtain a commitment to the polynomial, which we can provide polynomial evaluation proofs. 

We first list some preliminaries that we use in this scheme.
Let $\mathbf{a} = (a_1, a_2, \ldots, a_{n})$ be a evaluation vector of length $n$ which we wish to commit. For simplicity, we assume $n$ is a power of two, and let $\omega$ be the n-th primitive root of unity in a field $\mathbb{F}_p$.
%     \item $v_H(X)$ denote the vanishing polynomial over the multiplicative subgroup
% set $H$ defined by $\Pi_{i=1}^{i=n}(X- \omega^i)$ = $X^n -1$. Note that $v_H(\omega^i) = 0 \ \forall i \in \mathbb{N}$. 
Further, let $\ell_i(X)$ be $\Pi_{j \neq i} \frac{X - \omega^{j}}{\omega{i} - \omega^j}$ be $ith$ Lagarange polynomial that is unique and has degree $n-1$ such that $\ell_i(\omega^i) = 1$ and for $\ell_i(\omega^j) = 0$ for $j \neq i$.

Clearly, we can evaluate the interpolated polynomial by viewing the $a_i$ as evaluations of the polynomials. 
$L_{\textbf{a}}(X) = \Sigma_{i =1}^n a_i\ell_i(X)$.
Lipmaa's construction provided an extractable interpolating commitment scheme, which we extend to create an extractable polynomial commitment. We define our scheme based on Lip'16~\cite{lipmaa2017prover}'s construction with two major adaptations. 
First, any combination of linearly independent polynomials can be chosen for creating the CRS for a vector commitment scheme. For interpolating efficiency, Lip'16 schemes use evaluations of $\ell_i(\beta)$ for a trapdoor secret $\beta$. We instead use the more $1, \beta, \beta^2$ and create a polynomial commitment using Kate style polynomial commitments. Since our ZKsnark for Marlin uses the same CRS, it allows us to use common polycommit proof batching techniques for efficiency.
Second, the Lip'16 scheme is based on PKE assumptions and hence requires double the elements in CRS and double the commitment size. Although it is possible to adapt our scheme to use plain model under knowledge assumptions, similar to Marlin, we use AGM to obtain an efficient construction. 

We define $\mathsf{PEC.Lipmaa}$ as follows:
\begin{itemize}
    \item $\mathsf{Setup}(1^\lambda, D):$ Sample $\mathbf{\Sigma}$  as follows
  
%   \begin{equation}
%   \begin{gathered}
  \begin{equation}
    \Sigma := \left(
                \begin{array}{llllll}
                    g & g^\alpha & g^{\alpha^2} & \ldots & g^{\alpha^D}\\
                    g^{\gamma} & g^{\alpha\gamma} & g^{\gamma\alpha^2} & \ldots & g^{\gamma\alpha^D} \\
                \end{array}
              \right).
  \end{equation}
    % \end{gathered}
    % \end{equation}
  \begin{equation}
  \begin{gathered}
\mathsf{ck_{p}}:= (\mathsf{\Sigma}, D), \mathsf{ck_{e}} := ([g^{\ell_i(\alpha)}]_{i=1}^n, \vect{\Sigma}),\\
\mathsf{rk_{p}}:= (D, g^{\gamma}, h^\alpha), \mathsf{td} := (\alpha, \gamma)\\
  \end{gathered}
  \end{equation}
    \item $\mathsf{Commit_{eval}(ck_{e}}, e, p_i; \omega) \rightarrow c_{e}$: With input ${\mathsf{ck}_{\mathsf{e}_i} = g^{\ell_i(\alpha)}}$, point $p_i$ at index $i$, 
    compute the commitment to evaluation $e$ as: 
    $$c_{e} := ((g^{\ell_i(\alpha)})^{e}g^{\gamma\overbar{\phi}(\alpha)}, K, s, \overbar{\phi_s})$$ 
    where $\overbar{\phi}(X)$ is a polynomial sampled according to randomness $\omega$ such that $\textsf{deg}(\ell_i(X)) = \textsf{deg}(\overbar{\phi}(X))$, and
    $$K := (g^{\ell_i(\alpha)})^{r}g^{\gamma\overbar{\phi_r}(
    \alpha)}; ~~
    s := r + \beta e; ~~\overbar{\phi_s} = \overbar{\phi_r} + \beta {\overbar{\phi}}$$ 
    for a random challenge $\beta$ (which can be sampled by Fiat Shamir in a non-interactive version).
    \item $\mathsf{Interpolate(ck_{p}}, \mathbf{c_{e}}) \rightarrow c$: Parse $\mathbf{c_e} = [c_{e_i}, K, s, \overbar{\phi_s}]_{i=1}^n$. Check $\forall i \  (g^{\ell_i(\alpha)})^{s}g^{\gamma\overbar{\phi_s}(\alpha)}$ $\stackrel{?}{=}$ $K(c_{e_i})^{\beta}$ where $\beta$ is a verifier chosen challenge(or Fiat Shamir in Non-interactive setting). If check fails, abort, otherwise return $c = \Pi_{i=1}^nc_{e_i}$
    which is a polynomial commitment to the polynomial whose shares are committed by $\mathbf{c_{e}}$.
    \item $\mathsf{Open(ck_{p}}, \phi, q; \omega) \rightarrow v, \pi$: Same as regular polycommit open operation as described in Section~\ref{sec:prelim_polycommit}
    \item $\mathsf{Check(rk_{p}}, c, q, v, \pi) \rightarrow \{0, 1\}$: Same as regular polynomial commitment operation as described in Section~\ref{sec:prelim_polycommit}. 
\end{itemize}
\begin{thm}
\label{thm:pec}
If $\langle group \rangle$ satisfies the SDH assumption(Appendix~\ref{app:sdh}), then the above construction for $\mathsf{PEC.Lipmaa}$ is a $\mathsf{PEC}$ scheme (definition~\ref{def:pec})
\end{thm}

We prove this theorem in Appendix~\ref{app:proof_pec}
\section{Prover algorithm in MPC}
In this section, we describe how the Marlin prover algorithm is computed using MPC. We first describe a round the Marlin protocol without the MPC version, then we describe the corresponding round in it's MPC version. All the public values are written in $\textcolor{blue}{blue}$ color.
\label{app:mpc_prover}
\subsection{Prover Initialize}
\begin{enumerate}
    \item MPC servers engage in protocol to create commitments to $\textsf{PEC.Poly}$ to the outputs from the shares of the output. The servers post the output commitments to the bulletin board. The bulletin board now has commitments for all statement elements(input and output).
    \item Each servers computes a share of solution vector $\share{z_i} = (\share{x_i}, \share{w_i})$ to the QAP derived from R1CS. 
    \item For achieving zero knowledge purposes of underlying Holographic proof, we additionally need to have a additional degree $\mathsf{b} = 1$ into the $\hat{x}$ (x interpolated) polynomial. We do this by adding an extra dummy output statement chosen by the server. Note that $\hat{x}(X) \in \mathbb{F}^{|I|+ |O|+ b}[X]$ where $|I|$, $|O|$ denote the input domain size and output domain size of the MPC. Similar to $\hat{x}(X)$, the provers also construct a $\hat{r}(X)$ polynomial(with a hiding bound) consisting of random masks given by the client. 
\end{enumerate}

\subsection{Marlin First Round}
With the prover initialized, we now engage in Marlin protocol. Instead of the statement being the public value, in this case, our public input is a commitment to the statement elements.

In the Marlin protocol, the prover first engages in a rowcheck protocol to attest the relation $\hat{z}_A(X)\hat{z}_B(X) - \hat{z}_C(X) = h_0(X)v_H(X)$. The prover computes $z := (x, w)$, $z_A := Az$, $z_B := Bz$ and $z_C := Cz$. It computes a $\hat{w}(X) \in \mathbb{F}^{|w| + b}[X]$, 
$\hat{z}_{A}(X) \in \mathbb{F}^{|H| + b}[X]$, 
$\hat{z}_{B}(X) \in \mathbb{F}^{|H| + b}[X]$, 
$\hat{z}_{C}(X) \in \mathbb{F}^{|H| + b}[X]$.

The prover then computes $h_0(X)$ such that 
\begin{equation}
  \hat{z}_A(X)\hat{z}_B(X) - \hat{z}_C(X) = h_0(X)v_H(X)  
\end{equation}
Sample $s(X) \in \mathbb{F}^{2|H| + b - 1}[X]$ and compute $\sigma_1 := \sum_{\kappa \in H} s(\kappa)$. This $s(X)$ would help us in achieving zero knowledge.

In an MPC setting however, all servers have access to shares of $\share{z} =
(\share{x}, \share{w})$ which was previously computed in prover Initialize step. Therefore, we directly
operate on shares locally to compute shares of
polynomials. Each server computes $\share{z_A} := A\share{z}$, $\share{z_B} := B\share{z}$ and $\share{z_C} := C\share{z}$. It computes a $\share{\hat{w}(X)} \in \mathbb{F}^{|w| + b}[X]$, 
$\share{\hat{z}_{A}(X)} \in \mathbb{F}^{|H| + b}[X]$, 
$\share{\hat{z}_{B}(X)} \in \mathbb{F}^{|H| + b}[X]$, 
$\share{\hat{z}_{C}(X)} \in \mathbb{F}^{|H| + b}[X]$.
Recall the $\mathsf{b}$ is an additional degree added for zero-knowledge. 

The MPC servers then computes $\share{h_0(X)}$ such that 
\begin{equation}
  \share{h_0(X)} = \frac{\share{\hat{z}_A(X)}\share{\hat{z}_B(X)} - \share{\hat{z}_C(X)}}{\textcolor{blue}{v_H(X)}}  
\end{equation}

Ideally, the multiplication for $\share{\hat{z}_A(X)}\share{\hat{z}_B(X)}$ would require beaver multiplication. But we the same optimization from Marlin to force $\hat{z}_C(X)$ to be equal to
$\hat{z}_A(X)\hat{z}_B(X)$. 
The way to think about this is that all MPC servers combined together act as a single prover with each server having individual share elements. So, all optimizations to Marlin are still applicable to at MPC servers as a whole, but not an individual share level.
Similarly, we use another optimization in Marlin to sample
$\share{s(X)}$ such that $\share{\sigma_1}$ is 0. Both of these optimizations combined allow us to skip
the above row check and $\sigma_1$ round.

The servers use protocol from  App~\ref{app:mpc_polycommit} to create the commitments $C_{w(X)}, C_{z_A(X)},$  $C_{z_B(X)}$,$C_{s(X)}$ from local evaluations of the respective polynomial shares and publish it to the
blockchain. 

To Summarise, in the first round, the servers:
\begin{enumerate}
    \item Create the polynomials $\share{\hat{z}_A(X)},\share{\hat{z}_B(X)}$, $\share{\hat{w}(X)},\share{\hat{s}(X)}$ using the methods described above.
    \item Send the commitments $C_{w(X)}, C_{z_A(X)},$  $C_{z_B(X)}$,$C_{s(X)}$ to the bulletin board.
    \item Verifier sends a challenge $\alpha$, $\eta_A, \eta_B, \eta_C$ $\in \mathbb{F}$. In non-interactive proof, the prover computes himself using random oracle using a transcript that contains the commitments to the statement elements and the above four polynomial commitments.
\end{enumerate}

\subsection{Marlin Second Round}

Prover computes the polynomial 
\begin{multline}
  q_1(X) = s(X) + r(\alpha, X)\left(\sum_{M}\eta_M\hat{z}_M(X)\right) - \\
\left(\sum_{M}\eta_Mr_m(\alpha, X)\right)\hat{z}(X))  
\end{multline}
Prover then divides $q_1(X)$ by $v_H(X)$ to get $h_1(X)$ and $g_1(X)$ such that 
\begin{equation}
  q_1(X) = h_1(X)v_H(X) + g_1(X)X  
\end{equation}

In the marlin variant, each MPC server computes the share of the polynomial $\share{q_1(X)}$ as follows:
\begin{multline}
    \share{q_1(X)} = \share{s(X)} + \textcolor{blue}{r(\alpha, X)}\left(\sum_{M}\textcolor{blue}{\eta_M}\share{\hat{z}_M(X)}\right) - \\
\textcolor{blue}{\left(\sum_{M}\eta_Mr_M(\alpha, X)\right)}\share{\hat{z}(X)}
\end{multline}
The quantities in \textcolor{blue}{blue} represent the public polynomials which don't rely on any secret data. Note that $\textcolor{blue}{r(X, Y) = \frac{X^{|H|} - Y^{|H|}}{X - Y}}$ is the derivative polynomial as defined earlier, \textcolor{blue}{$\alpha, \eta_A, \eta_B, \eta_C$} are challenges which are public and finally \textcolor{blue}{$r_M(X, Y) = \sum_{\kappa \in H}r(X, \kappa)\hat{M}(\kappa, Y)$} which is also a public polynomial since \textcolor{blue}{$r(X,Y)$, $\hat{M}(X, Y)$} are both public 
polynomials. Recall that \textcolor{blue}{$\hat{M}$} is a low degree extension of the R1CS matrix M where $M \in \{A, B, C\}$ and hence public.

Finally, each server then divides $\share{q_1(X)}$ by $\textcolor{blue}{v_H(X)}$ using the divmod algorithm in fft to get $\share{h_1(X)}$ and $\share{g_1(X)}$ such that:
\begin{equation}
  \share{q_1(X)} = \share{h_1(X)}\textcolor{blue}{v_H(X)} + \share{g_1(X)}\textcolor{blue}{X}  
\end{equation}
 and $deg(\share{g_1(X)}\textcolor{blue}{X}) < deg(\textcolor{blue}{v_H(X)}$). Recall that $\textcolor{blue}{\sigma_1}$ was chosen to be zero as an optimization in the previous round. 
Again, as before the MPC servers compute $C_{h_1(X)}, C_{g_1(X)}$ using protocol from App~\ref{app:mpc_polycommit} from the local shares of $\share{h_1(X)}, \share{g_1(X)}$

To summarize, the MPC servers in the second round.
\begin{enumerate}
    \item Prover carries our Marlin locally to compute $h_1(X), g_1(X)$ using the challenges from previous round.
    \item Use Appendix~\ref{app:mpc_polycommit} commit operation to create $C_{h_1(X)}, C_{g_1(X)}$ from local polynomial shares.
    \item Server challenge $\beta_1$ is sampled from $\mathbb{F} \backslash H$ based on the transcript of first transcript plus the commitments $C_{h_1(X)}, C_{g_1(X)}$.
\end{enumerate}

\subsection{Marlin Third Round}

Note that the third and fourth rounds do not use any secret shared input, and thus this protocol can be
thoroughly carried out in the open. All the polynomials in this round \textcolor{blue}{$r(X, Y),
\hat{M(X, Y)}$} are all public, and hence there is no difference between the marlin protocol and the
secret shared version. We state the third and fourth rounds from Marlin for completeness. 
Each MPC computes the polynomial 
\begin{equation}
    \textcolor{blue}{q_2(X) = r(\alpha, X)\left(\sum_{M}\eta_M\hat{M}(X, \beta_1)\right)}
\end{equation}

and the sum-check result 
\begin{equation}
\textcolor{blue}{\sigma_2 = \sum_{\kappa \in H} r(\alpha, \kappa) \left(\sum_{M \in A,B,C}\eta_M\hat{M}(\kappa, \beta_1)\right)}
\end{equation}
Each server then divides \textcolor{blue}{$q_2(X)$} by \textcolor{blue}{$v_H(X)$} to get \textcolor{blue}{$h_2(X)$} and \textcolor{blue}{$g_2(X)$} such that 
\begin{equation}
    \textcolor{blue}{q_2(X) = h_2(X)v_H(X) + g_2(X)X +  \sigma_2/|H|}
\end{equation}
 and $\textcolor{blue}{deg(g_2(X)X) < deg(v_H(X)})$. Such a division is similar to protocol for divmod except that it is carried out in the open instead of secret shared form. Servers can then use standard $\textsf{PC.commit()}$ to create commitments $C_{h_2(X)}, C_{g_2(X)}$ which are extractable and hiding. MPC servers sample $\beta_2$ from $\mathbb{F} \backslash H$ using Fiat Shamir using transcript upto the current round. 

\subsection{Marlin Fourth Round}
Again, as with the previous round, all the operations in this round are public and carried out in the open. So, everything is the same as the Marlin fourth round. 
Sum-check for the term:
\begin{equation}
\textcolor{blue}{\sum_{M \in \{A,B,C\}}\eta_M\frac{v_H{(\beta_2)}v_H{(\beta_1)}\hat{val}_M(X)}{(\beta_2 - \hat{row}_M(X))(\beta_1 - \hat{col}_M(X))}}
\end{equation}
\begin{equation}
\textcolor{blue}{
    \sigma_3 = \sum_{\kappa \in K}\sum_{M \in \{A,B,C\}}\eta_M\frac{v_H{(\beta_2)}v_H{(\beta_1)}\hat{val}_M(\kappa)}{(\beta_2 - \hat{row}_M(\kappa))(\beta_1 - \hat{col}_M(\kappa))}
}
\end{equation}

Compute \textcolor{blue}{$a(X)$} and \textcolor{blue}{$b(X)$} deterministically from indexed
\textcolor{blue}{$\hat{row}(X)$, $\hat{col}(X)$, $\hat{val}(X)$.} 
We ignore the exact details for now, but this is done publicly based on challenges and public indexer values.

Find \textcolor{blue}{$h_3(X)$} and \textcolor{blue}{$g_3(X)$} such that 
\begin{equation}
\textcolor{blue}{h_3(X)v_K(X) = a(X) - b(X)(Xg_3(X) + \sigma_3/|K|)}    
\end{equation}
. Finally, compute $C_{h_3(X)}, C_{g_3(X)}$ using \textsf{PC.commit()} since all polynomials are public. The prover samples sends a challenge $\beta_3$ sampled from $\mathbb{F}$ according to fiat sharmir.

\subsection{Prover Poly Evaluation proofs}
After the four rounds, the prover needs to provide proofs of evaluation of polycommits. 
The prover posts a proof for all the polynomials $p_i(X)$ with commitments $C_{p_i(X)}$ at evaluation points $\beta_j$ using $\mathsf{KZG.Open}$($\beta_j$, $C_{p_i(X)}$, $p_i(X)$). To create a proof for public polynomials \textcolor{blue}{$p(X)$}, we would standard
$\mathsf{KZG.open}$ $(\beta_j, C_{p_i(X)}, \textcolor{blue}{p_i(X)})$. If we want to create an evaluation proof on a secret shared polynomial, we use a create witness protocol described in Appendix~\ref{app:mpc_polycommit}.
    
In standard marlin protocol, the prover and the verifier both had access to the statement, but in our scenario, the auditor does not have that access. Instead, we additionally need to provide the value $\hat{x}(\beta_1)$ proof that the value was correct which is exactly done by $\mathsf{PEC.Open}$. 

\section{Security Proofs}
\label{app:proofs}
In this section, we will show that completeness, extraction, and zero-knowledge properties of PEC scheme construction~\ref{sec:pec_cons_poly} and our adaptive zk-snarks described in section ~\ref{fig:adapt-snark}.

\subsection{Proof of theorem~\ref{thm:pec}}
\label{app:proof_pec}
\textbf{Perfect Completeness} By inspection

\textbf{Extractability:} We first provide a construction for extractor and then claim that if the adversary is able to pass the extractability game(see \ref{def:pec}) with non-negligible probability, the our construction for extractor fails with only negligible probability. 
Recall that in AGM~\ref{app:algebraic}, the adversary would output a representation of $\mathbf{c_e}$ in terms of $\mathsf{crs}$ elements from $\vect{\Sigma}$. 

The idea is to extract thw polynomial from the Sigma proof with AGM. This proof is similar to Uber assumptions in AGM~\cite{rotemalgebraic}, but repeat it here with our notation. 
For each commitment($c_{e_i}, K_i, s_i, \overbar{\phi_s}$), $c_{e_i} = $ $(g^{\phi_i{\alpha}})^{e_i}g^{\gamma\overbar{\phi_i}(\alpha)}$ 
and $K = $ $(g^{\phi_{r_i}{\alpha}})g^{\gamma\overbar{\phi_{r_i}}(\alpha)}$. Obtain $r_i$ from the adversary representation. 
Then compute $(\phi_s(X))/\ell_i(X)$, if the division fails, our extractor aborts. Finally, compute $e_i = \frac{s - r}{\beta}$. Next, the outputs $\phi(X)$ as the interpolated polynomial corresponding to evaluations $e_i$ at $p_i$. The extractor also outputs randomness $\vect{\omega}$ as $\vect{\omega} = \Sigma \overbar{\phi_i(X)}$.

Next, we need show to that if the adversary wins the game then our extractor will only fail with negligible probability and that values output by the extractor are satisfy the evaluation binding property. The first reason the extractor might fail is if the division $\phi_i(X)/\ell_i(X)$ fails. Note that since the adversary succeeds with non-negligible probability, it must satisfy the relation interpolation was carried out correctly. Which implies that the $\forall i \  (g^{\ell_i(\alpha)})^{s}g^{\gamma\overbar{\phi_s}(\alpha)}$ $=$ $K*(c_{e_i})^{\beta_c}$ for a randomly sampled $\beta_c$. Next, consider the algebraic $s*\ell_i(X) = r_i\phi_r(X) + Y*\phi(X)$ which must hold true for all $Y$(with soundness error $1/|\mathbbm{F}|$), which implies both $\phi_r(X)$ and $\phi(X)$ must be a multiple of $\ell_i(X)$. Which means that $\phi(X)$ interpolated by $e_i$ indeed corresponds to the interpolated polycommit $c$.

Now, this polycommit is exactly like any other polycommit generated by KGZ10, but instead of single party creating it, we had different parties create a commitment. Note that Opening and 
The next part of the proof is to show that the extracted polynomial is indeed evaluation binding such that the evaluation $v$ at query point $q$ is the same as \cite{chiesa2019marlin} and KGZ10~\cite{kate2010constant}. At a high level, consider an adversary outputs two proofs $\pi = (\overbar{v}, \mathbf{w})$ and $\pi' = (\overbar{v'}, \mathbf{w})$ for two different evaluations $v$ and $v'$ such that both proofs satisfy the verification equation. Output a pair $$(-q, \frac{1}{v' - v + \gamma(\overbar{v'} - \overbar{v})}(\mathbf{w - w'}))$$ where $q$ is the query point.

We can show the if $q \neq \alpha$(the trapdoor), the pair breaks the SDH assumption~\ref{app:sdh}. The probability that $q = \alpha$, is $\frac{1}{|F|}$ since the trapdoor is not known to the adversary. For detailed version of the proof, we defer the reader to Appendix B of Marlin~\cite{chiesa2019marlin}.

\textbf{Zero Knowledge:} We need show that views of the adversary in Ideal world(when interacting with simulator having trapdoors) and real prover are identically distributed. Following is the construction for our simulator. 
\begin{itemize}
    \item $\mathsf{S.Setup}$ $\rightarrow$ samples $\mathsf{td} = \alpha, \gamma$ and computes $\Sigma$(same as powers of $\alpha$ in equation~\ref{eq:srs}).
    \item $\mathsf{S.Commit_{eval}}$: Sample random evaluation points $e_{r_i}$ and create simulated commitments to it randomness $\omega_i$. Let $\phi_r, c$ denote the hiding polynomial corresponding to interpolated randomness and interpolated commitment $c$. 
    \item $\mathsf{S.Open}$: Using the trapdoor $\alpha$, provide an evaluation proof for the polynomial opening at $\phi(q)$  at $q$ with randomness $\vect{\omega}$ as follows: Compute $\overbar{v} = \phi_r(q) - \phi(q)/\gamma$ and $\mathbf{w}$ as $(c/(g^{\phi(q)-\gamma\overbar{v})}))^{1/(\alpha - z)}$. Return $\pi = (\mathbf{w}, \overbar{v})$
\end{itemize}

The view of the adversary consists of ($\pi, \mathbf{C_e}, \mathsf{ck_e}$) and public values $q, \phi(q), \mathbf{p}$. Since, $\mathsf{S.Setup}$ uses $\mathsf{PEC.Setup}$ the output is distributed identically in both worlds. Similarly, the commitments are identically distributed in both worlds because of blinding polynomial. $c$ is the output of interpolate method which is a deterministic function and thus is identically distributed in both worlds. Finally, substituting the above expression of simulated $\pi$ in $\mathsf{Check}$, it is easy that proofs are also indistinguishable in both worlds.

\subsection{Proof of theorem~\ref{thm:pec_succ}}
\label{app:proof_succ}
\textbf{Perfect Completeness} By inspection

\textbf{Extractability:} We first provide a construction for extractor and then claim that if the adversary is able to pass the extractability game(see \ref{def:pec}) with non-negligible probability, the our construction for extractor fails with only negligible probability. 
Recall that in AGM~\ref{app:algebraic}, the adversary would output a representation of $\mathbf{c_e}$ in terms of $\mathsf{crs}$ elements from $\vect{\Sigma}$.

The extraction proceeds similar to the extraction proof for the $PEC.Lipmaa$~\ref{app:proof_pec}, so we only highlight the differences here. The first part of the proof is to extract the polynomials from the evaluations. Since, we have distributed different trapdoors to different parties, the validity pairing check in interpolate ensures that the parties only use those terms in CRS.

For each commitment ($c_{e_i}, c^k_{e_i}) = $ $(g^{\phi_i{\alpha}})^{e_i}g^{\gamma_i\tau\overbar{\phi_i}(\alpha)}$. Furthermore, we also have the validity condition from the interpolate algorithm $\forall i \  (e(g, c^k_e)), g$ $=$ $ e(g^\gamma_i, c_e)$. Looking at the terms on the right hand side, we see that $c_e$ must only contain CRS terms with the corresponding powers of which have a $\gamma_i$ component. This is because the knowledge components only have terms corresponding to powers of degrees of $\alpha^i$. Since, the interpolate check passes, we can conclude that parties can only commit to the certain degrees of $\alpha_i$ assigned to them by $\mathsf{ck_i}$. Since, all parties have different degrees of $\alpha_i$ allotted by $\mathsf{ck_i}$, the multiplication of those terms produces a commitment to a polynomial of a larger degree $\mathsf{D}$. The first $d$ terms come party with $g^{\alpha^0}$, $g^{\alpha^1}$ $\ldots$ $g^{\alpha^d}$, the next $d$ terms from the second party and so on. Finally, a single combined commitment $c$ of a degree $\mathsf{D} = d*K$ is created by multiplying the commitments. 

Now, this polycommit is exactly like any other polycommit generated by KGZ10, but instead of single party creating it, we had different parties create a commitment. Note that Opening and 
The next part of the proof is to show that the extracted polynomial is indeed evaluation binding such that the evaluation $v$ at query point $q$ is the same as \cite{chiesa2019marlin} and KGZ10~\cite{kate2010constant}. At a high level, consider an adversary outputs two proofs $\pi = (\overbar{v}, \mathbf{w})$ and $\pi' = (\overbar{v'}, \mathbf{w})$ for two different evaluations $v$ and $v'$ such that both proofs satisfy the verification equation. Output a pair $$(-q, \frac{1}{v' - v + \gamma(\overbar{v'} - \overbar{v})}(\mathbf{w - w'}))$$ where $q$ is the query point.

We can show the if $q \neq \alpha$(the trapdoor), the pair breaks the SDH assumption~\ref{app:sdh}. The probability that $q = \alpha$, is $\frac{1}{|F|}$ since the trapdoor is not known to the adversary. For detailed version of the proof, we defer the reader to Appendix B of Marlin~\cite{chiesa2019marlin}.

\textbf{Zero Knowledge:} We need show that views of the adversary in Ideal world(when interacting with simulator having trapdoors) and real prover are identically distributed. Following is the construction for our simulator. 
\begin{itemize}
    \item $\mathsf{S.Setup}$ $\rightarrow$ samples $\mathsf{td} = \alpha, \gamma$ and computes $\Sigma$(same as powers of $\alpha$ in equation~\ref{eq:srs}).
    \item $\mathsf{S.Commit_{eval}}$: Sample random evaluation points $e_{r_i}$ and create simulated commitments to it randomness $\omega_i$. Let $\phi_r, c$ denote the hiding polynomial corresponding to interpolated randomness and interpolated commitment $c$. 
    \item $\mathsf{S.Open}$: Using the trapdoor $\alpha$, provide an evaluation proof for the polynomial opening at $\phi(q)$  at $q$ with randomness $\vect{\omega}$ as follows: Compute $\overbar{v} = \phi_r(q) - \phi(q)/\gamma$ and $\mathbf{w}$ as $(c/(g^{\phi(q)-\gamma\overbar{v})}))^{1/(\alpha - z)}$. Return $\pi = (\mathbf{w}, \overbar{v})$
\end{itemize}

The view of the adversary consists of ($\pi, \mathbf{C_e}, \mathsf{ck_e}$) and public values $q, \phi(q), \mathbf{p}$. Since, $\mathsf{S.Setup}$ uses $\mathsf{PEC.Setup}$ the output is distributed identically in both worlds. Similarly, the commitments are identically distributed in both worlds because of blinding polynomial. $c$ is the output of interpolate method which is a deterministic function and thus is identically distributed in both worlds. Finally, substituting the above expression of simulated $\pi$ in $\mathsf{Check}$, it is easy that proofs are also indistinguishable in both worlds.

\subsection{Proof of theorem~\ref{thm:snark}}
\label{app:proof_snark}
\textbf{Perfect Completeness} By inspection

\textbf{Extractability:}
The output proof from $\Adv_2$ would be of the form ($\mathsf{prf_{mar}, v, \pi_{comm}, \mathbf{C^{\mathbbm{b}}_x}}$).
We construct our extractor $\Ext$ as follows:
\begin{itemize}
    \item Use the Commitment Extractor $\Ext_{comm}$ to extract inputs and randomness $\mathbbm{x,r}$ from augment statement commitments $\mathbf{C_x}, \mathbf{C^{\mathbbm{b}}_x}$.
    \item Use the marlin extractor $\Ext_{Mar}$ with above extracted statement $\mathbbm{x}$ and $\mathsf{prf_{mar}}$ to get witness $\mathbbm{w}$. The constructions of this extractor is described in Section 8.3 of Marlin~\cite{chiesa2019marlin} paper. 
    \item Use the same $\mathbbm{i}$ as the one output by the $\Adv_1$ and output $\mathbbm{(x, r, w, i)}$
\end{itemize}

We need to show that, if $\Adv$ = $(\Adv_1, \Adv_2)$  produces a verifying proof, the extractor fails
only with negligible probability and the returned values are in $\mathcal{R}_{\mathsf{ck}_{eval}}$.

Suppose that $\Ext$ fails with non-negligible probability $\epsilon$, then either two extractors ($\Ext_{comm}$ or $\Ext_{Mar}$) fail or the output $\mathbbm{(x, r, w, i)} \notin \mathcal{R}_{\mathsf{ck}_{eval}}$

\begin{itemize}
    \item If $\Ext_{Mar}$ fails, then we can construct an adversary that makes the extractor fail to either break 1) the soundness of underlying Algebraic Holographic Proof(AHP) or 2) succeed in the extractability game for the PC scheme. We defer the reader to Section 8.3 of Marlin paper~\cite{chiesa2019marlin} for details.
    \item If $\Ext_{comm}$ fails with non-negligible probability, then we can break the extraction game for the $\mathsf{PEC}$ scheme.
\end{itemize}

It remains to show that the values returned by the extractor are in with high probability in $\mathcal{R}_{\mathsf{ck}}$(see section~\ref{sec:index_relations})

By properties of the extractor $\Ext_{comm}$, we know that openings to the commitments $\mathbf{C_x}$ are ($\mathbbm{x, r}$) and now we need to show that $(\mathbbm{x},\mathbbm{i}, \mathbbm{w}) \in \mathcal{R}$. Let the interpolated polynomial for all $\mathbbm{x}$ be $\tilde{x}(X)$.

Note that in regular Marlin execution, the value $\mathsf{v}$ is computed the verifier himself based on the statement $x$. We need show that with high probability that the $\mathsf{v}$ is exactly the same of $\hat{x}(\beta_1)$. Suppose that this were not the same, $\mathsf{v} \neq \hat{x}(\beta_1)$. i.e $\tilde{x}(\beta_1) \neq \hat{x}(\beta_1)$ which means that $\hat{x}(X) \neq \tilde{x}(X)$.

From the  underlying AHP Marlin protocol, we know that 
\begin{equation}
  \hat{z}(X) := \hat{w}(X)v_H(X) + \hat{x}(X)  
\end{equation}
is true for all the values of $X$. Therefore, the equation can only hold true for $\tilde{x}$ with probability $|X|/|\mathbbm{F}|$ for a randomly sampled challenge $\beta_1$ from the verifier. $|X|$ is the statement length and also the degree of the polynomial $\hat{x}$.

\textbf{Zero Knowledge:}
We need to the simulator with access to the trapdoor can generate proofs that can indistinguishable from real proofs. In other words, our simulator $\mathsf{S}$ is given given commitments to the $\mathsf{C_x}$ and needs to construct proofs using the trapdoors from commitment scheme and trapdoor from marlin. Our simulator works as follows:

\begin{figure*}[!htbp] 
\begin{boxedminipage}[t]{1\textwidth}
  {\centering \textbf{Simulator for Adaptive ZK Construction in ~\ref{fig:adapt-snark} }\\}

\begin{itemize}
    \item Recall that our proof consists of four elements ($\mathsf{prf_{mar}, v, \pi_{comm}, \mathbf{C^{\mathbbm{b}}_x}}$). First, the simulator samples a random polynomial $\hat{x}$ of
    $\mathsf{deg(|C_x|)} + \mathsf{deg}(|C^{\mathbbm{b}}_x|)$. 
    Note that all the information simulator needs is the length of the augmented statement.
    \item Using the trapdoors of the $\mathsf{PEC}$, open the commitments $\mathbf{C_x}$ to corresponding evaluations of $\hat{x'}$ over a pre-selected domain. Extend the polynomial by degree $\mathbbm{b}$ to get a resultant polynomial $\hat{x}$. Create a commitment to the additional $\mathbbm{b} $evaluations as $(|C'^{\mathbbm{b}}_x|)$.
    \item Run the marlin simulator with $\hat{x}$ to obtain a fake proof $\mathsf{prf'_{mar}}$ and obtain the evaluation $\hat{x}(\beta_1)$.
    \item Interpolate the commitments $\mathbf{C_x}$ using $\mathsf{PEC.interpolate}$ to obtain a polynomial commitment $c_p$ for the $\hat{x}$ and provide a evaluation proof $\mathsf{\pi'_{comm}}$ for the correct value $\hat{x}(\beta_1)$.
    \item return the proof ($\mathsf{prf'_{mar}},
    \hat{x}(\beta_1)$ 
    $\mathsf{\pi'_{comm}}$, $(|C'^{\mathbbm{b}}_x|)$).
\end{itemize}

\end{boxedminipage}
\caption{Simulator construction for adaptive zk-snark}
\label{fig:sim_adapt_zk}
\end{figure*}
The figure ~\ref{fig:sim_adapt_zk} shows the construction for simulator satisfying the above definition for our construction in ~\ref{fig:adapt-snark}. Next, we provide a proof about the indistinguishably for the ideal and real world for the adversary w.r.t our simulator.

First, we note that in the underlying AHP for $x$ polynomial, we also have introduced a query bound $\mathbbm{b}$ in order to ensure zero knowledge of the evaluations of $x$ upto $\mathbbm{b}$ queries.
Informally, since the prover sampled $\hat{x}$ such that $\mathsf{deg(\hat{x}) = deg(x) + \mathbbm{b}}$ queries less than the query bound $\mathbbm{b}$ information theoretically does not reveal any information about $x$.

Similar to the construction in~\cite{ben2019aurora},  we would construct a simulator $\mathsf{AHP'}$ for the underlying AHP with changes that the first message of the prover also includes an
encoding of statement along with the encoding of witness and encoding of its linear combinations. These encodings are protected against up to $\mathbbm{b}$ queries because the encodings have degree $\mathbbm{b} $more than corresponding encodings. The rest of the simulator for the subsequent rounds proceeds similarly to what is described in Marlin~\cite{chiesa2019marlin}. At the high level, all the subsequent messages are hidden by adding the additional $\mathsf{s}$ polynomial and hence do not reveal any information. 

From the marlin simulator with the trapdoors to $\mathsf{srs}$ which uses $\mathsf{AHP'}$ instead of
$\mathsf{AHP}$ described in marlin, we know that $\mathsf{prf'_{mar}}$ and $\mathsf{prf_{mar}}$ are
indistinguishable. For the last proof element 
$\mathbf{C'^{\mathbbm{b}}_x}$ is indistinguishable from $\mathbf{C^{\mathbbm{b}}_x}$. 
Similarly, using the trapdoors $\mathsf{td_{comm}}$ for $\mathsf{PEC}$, we can simulated proofs for $(\pi'_{comm},
\hat{x}(\beta_1))$ are indistinguishable from $(\mathsf{\pi_{comm}, v})$. 

Although, we have argued about the individual distributions of $\mathsf{prf'_{mar}}$ and $\mathsf{prf_{mar}}$ are same and distributions for $(\mathsf{\pi_{comm}, v, }, $ and $(\pi'_{comm}, \hat{x}(\beta_1))$. Similarly, we also showed that $\mathbf{C'^{\mathbbm{b}}_x}$ is indistinguishable from $\mathbf{C^{\mathbbm{b}}_x}$. The hiding property of the \textsf{PEC} scheme ensures that the simulator by using the trapdoor $\mathsf{trap_{comm}}$ can perfectly simulate the evaluation and the commitments. Since both of these distributions are independent and can individually be simulated we argue that the joint distribution views of prover and simulator are identical. 

\subsection{Auditable MPC ideal functionality}
\label{sec:inp_ind}

We provide an ideal functionality for auditable MPC in the universal composability (UC)  framework \cite{canetti2001universally} in Figure~\ref{fig:FAuditable}.
Our modeling is guided by a few goals: first, the protocol should allow programs to be chosen adaptively, without having to conduct additional trusted setups. 
Second, the ideal functionality to be simple and self-contained.
While we provide a proof that our protocol realizes this ideal functionality in Appendix~\ref{app:uc},
here we focus mainly on explaining what security properties the ideal functionality expresses.
First, notice that while we allow any clients to submit secret inputs, a separate ``Auditor" party is responsible for choosing the arbitrary application circuits and is the only party that receives the output. This is just for simplicity, but in a real system we envision using some public process (like a smart contract) to choose it. All our protocol requires is that the MPC servers and auditors agree on which circuit was chosen. Also in our protocol the audit routine is public coin, so anyone could could re-run the auditing subroutine for themselves.

\begin{figure}
\begin{boxedminipage}[t]{0.5\textwidth}
  {\centering \textbf{$\mathcal{F}_\mathsf{AuditableMPC}$}\\}

(01)  
On init: $\mathsf{inputs} \leftarrow [\ ]$, $\mathsf{new} \leftarrow [\ ]$

(02)  
On input $x_i$ from data-client $C_i$:

(03)
    \quad Append $x_i$ to $\mathsf{new}$
   
(04) 
    \quad Send $C_i$ to $\mathcal{A}$
   
 (05) 
On input circuit $\mathbf{F}$ from Auditor:

(06)
    \quad If $f \le t$:

(07)
        \qquad Append $\mathsf{new}$ to $\mathsf{inputs}$ and set $\mathsf{new} \leftarrow [\ ]$
   
  (08)     
        \qquad Compute $o \leftarrow \mathbf{F}(\mathsf{inputs})$ 
    
(09)
        \qquad $\mathsf{Optionally}$: Send $(\mathbf{F}, o)$ to Auditor
   
   (10)     
        \qquad Send $(\mathbf{F}, o)$ to $\mathcal{A}$
    
    (11)
    \quad Else:
    
    (12)
        \qquad$\mathsf{Optionally}$: 
    
    (13)
            \quad \qquad Append $\mathsf{new}$ to $\mathsf{inputs}$
      
      (14)      
            \quad \qquad Compute $o \leftarrow \mathbf{F}(\mathsf{inputs})$ 
        
(15)
            \quad \qquad $\mathsf{Optionally}$: Send $(\mathbf{F}, o)$ to Auditor
  
  (16)          
            \quad \qquad Send $(\mathbf{F}, \mathsf{new})$ to $\mathcal{A}$ and set $\mathsf{new} \leftarrow [\ ]$
      
     (17)      
        \qquad Send $\mathbf{F}$ to $\mathcal{A}$

\end{boxedminipage}
\caption{UC ideal functionality for Auditable MPC}
\label{fig:FAuditable}
\end{figure}

Integrity of outputs is required in both the $f \leq t$ and $f > t$ settings since the only outputs (lines 8 and 14) are from applying the given circuit $\mathbf{F}$ to the provided inputs.
Confidentiality of inputs is required in the $f \leq t$ setting, since the only information leaked to the adversary is the identity of data-clients (line 4) and the final output of the function (line 10). In an auction application, this corresponds to corrupt servers not learning bids placed by data-clients within the round. If $f > t$ then new inputs are leaked to the adversary in line 16.
Our definition only considers public outputs for simplicity, though the construction can be extended to support private outputs to designated parties as well. 

%%%% This is part of the related work about the ideal functionality
\ignore{
Our functionality definition builds most closely on the one given by Veeningen~\cite{veeningen2017pinocchio}.
The main difference is that we consider a reactive, rather than one-shot case, and explicitly prove security in the UC setting with an arbitrary environment. 
%We also give a more explicit explanation of the real world, including the bulletin board functionality.
The ideal functionality model from Hawk~\cite{kosba2016hawk} is also closely related. 
Besides the above-mentioned support for adaptively chosen programs, our functionality is also simpler than Hawk as it does not mode availability, pseudonymity, and interaction with smart contracts (though it is likely our protocol could support all these).
}
%\todo{Shreyas: Present the ideal functionality here and describe its properties. Defer to the appendix for the UC proof.}

Our ideal functionality implies a subtle security guarantee: input independence.
Notice that even in the $f>t$ case, the inputs of honest parties are leaked only \emph{after} the output is computed. In other words, corrupted parties must commit to and have knowledge of the inputs they provide, even before they learn anything about what honest parties input. 
In the context of our auction application, each time the auction update function is computed (one round of the auction), all of the bids collected during this bound must be independent of each other. 
Put another way, seeing commitments to honest bids does not help corrupt parties create related bids in the same round, and this holds regardless of the number of corruptions. 
Note however that this no longer holds across rounds. In the $f>t$ case, corrupt servers in later rounds of MPC can know the inputs of data-clients from earlier rounds and choose their inputs based on those past inputs. Consider the same auction example as before, where bidding servers maintain a state of the top $k$ bids. In every MPC round, a new party submits new bids and servers update the state to reflect the current top $k$ bids. When all servers are corrupted, data-clients can collude the servers to learn the top bids and bid accordingly.
The use of $\mathsf{optionally}$ in our functionality (lines 9,12,15) express that our functionality does not guarantee availability. The adversary may prevent output from being provided, but if output is delivered it is guaranteed to be correct. % In the context of the auction example, the adversary can decide to not output a winner, but if a winner \textit{is} chosen, it will be per auction specifications.

% In the scenario where all servers are corrupted, the servers will directly know the input and finish the computation for a particular round of MPC. 
% In general, input independence is not possible in reactive MPC across multiple rounds.

% We briefly summarize our modular construction.
% In Marlin, the verifier has access to the statement and can verify the proof with respect to the statement. However, in an auditable MPC scenario, the auditor only has access to commitments of the statements and must make sure that proof is consistent with the commitment. Previous approaches to adaptive zk-SNARKs~\cite{veeningen2017pinocchio} divided the Pinocchio~\cite{parno2013pinocchio} proof into multiple blocks and embedded the statement in each block. 
\section{UC Proof Sketch}
\label{app:uc}
% In interest of space, we suggest the reader to refer to online e-print for detailed sketch of the UC proof.
The universal composability framework \cite{canetti2001universally} is based on the real/ideal-world paradigm. In this setting, the real world consists of interactions between the environment $\mathcal{Z}$, real-world adversary $\mathcal{A}$, and parties $P_1,\dots,P_n$ running a protocol $\pi$. The ideal world consists of interactions between the environment $\mathcal{Z}$, the ideal-world adversary (or simulator) $\mathcal{S}$, \say{dummy} parties $\mathcal{D}_1,\dots,\mathcal{D}_n$, and an ideal functionality $\mathcal{F}$.

Security in this framework is defined by having $\mathcal{Z}$ output a bit after interacting in the real or ideal worlds. If a simulator can be defined so that, for every possible $\mathcal{Z}$, the distribution on this output bit in the real world is indistinguishable from that in the ideal world, it follows that the the real and ideal worlds are indistinguishable. This is usually stating as saying that a protocol $\pi$ UC-realizes a functionality $\mathcal{F}$.

More formally, let $\mathsf{EXEC_{IDEAL}^{\mathcal{F}, \mathcal{S}, \mathcal{Z}}(z, \vec{r})}$ be the output of $\mathcal{Z}$ in the ideal world, after interacting with the simulator $\mathcal{S}$ and the dummy parties $\mathcal{D}_1,\dots,\mathcal{D}_n$ (where dummy parties act as passthrough parties between $\mathcal{Z}$ and $\mathcal{F}$). $\mathsf{z}$ here is the input $\mathcal{Z}$ is initialized with and $\mathsf{\vec{r}}$ is the set consisting of $\mathcal{Z}$, $\mathcal{S}$, and $\mathcal{F}$'s random tapes. We define $\mathsf{EXEC_{IDEAL}^{\mathcal{F}, \mathcal{S}, \mathcal{Z}}(z)}$ to be the random variable after choosing $\mathsf{\vec{r}}$ uniformly at random. 

We define $\mathsf{EXEC}_{\mathsf{REAL}}^{\pi, \mathsf{\mathcal{A}, \mathcal{Z}}}(\mathsf{\vec{r}})$ in a similar manner in the real world, except now considering the execution consisting of the real-world adversary $\mathcal{A}$ and parties $P_1,\dots,P_n$ running a protocol $\pi$.

A protocol $\pi$ is said to UC-realize a functionality $\mathcal{F}$ if, for all adversaries $\mathcal{A}$, there exists a simulator $\mathcal{S}$, such that for all environments $\mathcal{Z}$,
$$\{\mathsf{EXEC}_{\mathsf{REAL}}^{\pi, \mathsf{\mathcal{A}, \mathcal{Z}}}(\mathsf{\vec{r}})\}_{\mathsf{z} \in \{0, 1\}^*} \stackrel{c}{\equiv} \{\mathsf{EXEC_{IDEAL}^{\mathcal{F}, \mathcal{S}, \mathcal{Z}}(z, \vec{r})}\}_{\mathsf{z} \in \{0, 1\}^*}$$

One thing to note is that the every real-world adversary $\mathcal{A}$ can be split into two parts: (1) a logical adversary, which performs the actual computations and so on, and (2) a dummy adversary, which simply receives messages computed by the logical component and routes them to where it is instructed to send messages. As the UC security definition quantifies over all environments, Canetti \cite{canetti2001universally} recognized that it is often simpler to work with the dummy adversary and proved the equivalence with the above definition. Accordingly, we will work with the following, simpler definition:

A protocol $\pi$ is said to UC-realize a functionality $\mathcal{F}$ if there exists a simulator $\mathcal{S}$, such that for all environments $\mathcal{Z}$ and the dummy adversary $\mathcal{D}$,
$$\{\mathsf{EXEC}_{\mathsf{REAL}}^{\pi, \mathsf{\mathcal{D}, \mathcal{Z}}}(\mathsf{\vec{r}})\}_{\mathsf{z} \in \{0, 1\}^*} \stackrel{c}{\equiv} \{\mathsf{EXEC_{IDEAL}^{\mathcal{F}, \mathcal{S}, \mathcal{Z}}(z, \vec{r})}\}_{\mathsf{z} \in \{0, 1\}^*}$$

We now note that our protocol is based in the algebraic group model. To our knowledge, the interaction between the universal composability (UC) framework and the algebraic group model has not previously been explored. We leave in-depth exploration of this relationship to future work. In this work, however, we recognize that we only consider algebraic adversaries and note that this corresponds to the logical component of real-world adversaries. As the UC-security definition in terms of the dummy adversary depends on "moving" the logical component of the adversary to the environment, we note that we therefore only quantify over "algebraic environments." That is, we only consider environments that provide a representation for group elements in their messages to the dummy adversary or simulator. 

We now show that the protocol $\Pi_\mathsf{AuditableMPC}$ (Figure~\ref{fig:PiAuditable}) UC-realizes the functionality $\mathcal{F}_\mathsf{AuditableMPC}$ (Figure~\ref{fig:FAuditable}). We show the indistinguishability between the real and ideal worlds by performing an exhaustive case analysis on the messages sent by $\mathcal{Z}$ in both worlds, and considering the resulting transcript. We consider the $f \leq t$ and $f > t$ cases separately.

\textbf{Case 1: $\mathbf{f \leq t}$:}

We argue that the simulator defined in Figure \ref{fig:UCSimulator_honestmaj} causes indistinguishability between the real and ideal worlds when $f \leq t$. We show that the internal simulation always tracks what occurs in the real world and additionally, the messages that $\mathcal{Z}$ receives in both worlds are indistinguishable.

First, consider the messages that $\mathcal{Z}$ can send to honest parties:

\begin{itemize}
    \item $\mathcal{Z}$ provides an input $x_i$ to an honest data-client $C_i$: 
	
	\textbf{Real World:} By lines 2-6 of Figure~\ref{fig:PiAuditable}, $C_i$ creates a commitment to $x_i$ and sends this to $\mathcal{F}_\mathsf{BulletinBoard}$.
	
% 	By lines 2-7 of Figure~\ref{fig:PiAuditable}, $C_i$ creates a commitment to $x_i$, generates a random $\phi(\cdot)$, and inputs $\phi$ to $\mathcal{F}_\mathsf{ReactiveMPC}$. By line 5 of Figure~\ref{fig:FReactive}, $\mathcal{F}_\mathsf{ReactiveMPC}$ sends the message $C_i$ to $\mathcal{D}$, which forwards it to $\mathcal{Z}$.

	\textbf{Ideal World:} The dummy $C_i$ in the ideal world forwards $x_i$ to $\mathcal{F}_\mathsf{AuditableMPC}$. By line 4 of Figure~\ref{fig:FAuditable}, $\mathcal{F}_\mathsf{AuditableMPC}$ sends the message $C_i$ to $\mathcal{S}$. By lines 6-8 of Figure~\ref{fig:UCSimulator_honestmaj}, $\mathcal{S}$ simulates an internal input of $0$ by $C_i$ and all state transitions.

	In both cases, $\mathcal{Z}$ is activated with no incoming messages and $\mathcal{Z}$ will not be able to distinguish on its activation. Additionally, the internal simulation will be at the same point of the execution as the real world.
	
\end{itemize}

Next, we consider the messages $\mathcal{Z}$ sends to the auditor:

\begin{itemize}
    \item $\mathcal{Z}$ provides a circuit $\mathbf{F}$ to the auditor:
% 	\textbf{Real World:} By line 35 of Figure~\ref{fig:PiAuditable}, the auditor sends $\mathbf{F}$ to $\mathcal{F}_\mathsf{ReactiveMPC}$. $\mathcal{F}_\mathsf{ReactiveMPC}$ then executes lines 19-32 of Figure~\ref{fig:FReactive} and sends $(\mathbf{F}, o)$ and shares of inputs and wires to $\mathcal{D}$, which are forwarded to $\mathcal{Z}$.

    \textbf{Real World:} One of two things can happen here. Either (1) the auditor stores $\mathbf{F}$ or (2) the auditor forwards $\mathbf{F}$ to $\mathcal{F}_\mathsf{ReactiveMPC}$. 
    
    In the case of (1), nothing happens in the real world.
    
    In the case of (2), by line 16 of Figure~\ref{fig:PiAuditable}, $\mathcal{F}_\mathsf{ReactiveMPC}$ schedules an optional codeblock to send $\mathbf{F}$ to each $C_i$.
    
    \textbf{Ideal World:} By line 10 of Figure~\ref{fig:FAuditable}, $\mathcal{F}_\mathsf{AuditableMPC}$ forwards $(\mathbf{F}, o)$ to $\mathcal{S}$. By lines 20-21 of Figure~\ref{fig:UCSimulator_honestmaj}, $\mathcal{S}$ emulates the same actions as in the real world internally. Since the simulated execution is at the same point as the real world execution, this means that the internal emulation does (1) and (2) exactly as above.
    
    In both worlds, $\mathcal{Z}$ is activated with no incoming message. Additionally, the internal simulation of the real world matches the actual real world after this step.
    
    % and sends $(\mathbf{F}, o)$ and shares of inputs and wires to $\mathcal{Z}$. As only $f \leq t$ shares of each of these are given to $\mathcal{Z}$, and $t+1$ shares are required to define a $t$-degree polynomial, from the perspective of $\mathcal{Z}$, real shares and random shares are indistinguishable.
    
    ~\\
    
    In order to explain lines 10-19 of Figure~\ref{fig:UCSimulator_honestmaj}, we note that, in line 34 of Figure \ref{fig:PiAuditable}, $\Pi_\mathsf{Auditable}$ runs a prover algorithm for an auditable zk-SNARK. As proven in Theorem \ref{thm:snark}, this prover algorithm is for an auditable zk-SNARK as defined in Section \ref{sec:adapt_def}. As it satisfies this definition, there is a simulator $\mathcal{S}_\mathsf{zk-SNARK}$ for the zero-knowledge property. The adaptive zk-SNARK zero-knowledge definition makes $\mathcal{S}_\mathsf{zk-SNARK}$ sufficient for $\mathcal{S}$ to run internally: the adversary in the definition chooses strictly more parameters than $\mathcal{Z}$ does in our setting. Additionally, as the definition quantifies over all adversaries, it quantifies over adversaries that run the prover algorithm themselves, for any set of parameters, a polynomial number of times, which is exactly what $\mathcal{Z}$ can do. This means that a simulated transcript and proof generated by $\mathcal{S}_\mathsf{zk-SNARK}$ will be indistinguishable from an actual transcript and proof, from the perspective of $\mathcal{Z}$. \\
    
    On input $\mathbf{F}$, using lines 10-19 of Figure~\ref{fig:UCSimulator_honestmaj}, $\mathcal{S}$ generates a simulated copy of the Marlin transcript and proof using $\mathcal{S}_\mathsf{zk-SNARK}$. $\mathcal{S}$ also generates random honest shares of each polycommit in the transcript and random honest shares of the proof, consistent with the views of corrupt parties. These shares are used in lines 40-43 of Figure~\ref{fig:UCSimulator_honestmaj}, when they are sent to $\mathcal{Z}$. This is okay for the following reason: $\mathcal{Z}$ only knows enough information to reconstruct up to $f$ shares of corrupt parties - however, as $f \leq t$, $\mathcal{Z}$ does not learn enough shares to constrain honest shares of polycommits. We note that $\mathcal{Z}$ (unlike $\mathcal{S}$) knows inputs of honest parties in the protocol - however, as $\mathcal{Z}$ doesn't know the trapdoor or randomness used for polycommits, it does not learn enough information to verify the honest shares of polycommits created by $\mathcal{S}$. This means that the simulated honest shares will be indistinguishable from actual honest shares.

\end{itemize}

Finally, we analyze the messages that $\mathcal{Z}$ can send the adversary.

\begin{itemize}
    \item $\mathcal{Z}$ instructs $\mathcal{D}$ to send a message to $\mathcal{F}_\mathsf{RO}$, $\mathcal{F}_\mathsf{Setup}$, or corrupt $P_i$:

	\textbf{Real World:} $\mathcal{D}$ forwards the message and forwards any incoming message to $\mathcal{Z}$.

	\textbf{Ideal World:} In lines 30-35 of Figure~\ref{fig:UCSimulator_honestmaj}, $\mathcal{S}$ forwards the message to the internal emulation and forwards any generated message to $\mathcal{Z}$. $\mathcal{F}_\mathsf{RO}$ and $\mathcal{F}_\mathsf{Setup}$ will be computationally indistinguishable in both worlds, so any queries will have indistinguishable responses. Additionally, $\mathcal{Z}$ can also instruct a corrupt $P_i$ to send a message to $\mathcal{F}_\mathsf{BulletinBoard}$, but this will happen identically in both worlds. Thus, $\mathcal{Z}$ will not be able to distinguish based on these inputs.
	
	\item $\mathcal{Z}$ instructs $\mathcal{D}$ to send message to a corrupt data-client $C_i$:
	
	\textbf{Real World:} $\mathcal{D}$ forwards the message to $C_i$ and forwards any incoming message to $\mathcal{Z}$

	\textbf{Ideal World:} By lines 24-29 of Figure~\ref{fig:UCSimulator_honestmaj}, if this is related to $C_i$ sending an input to $\mathcal{F}_\mathsf{ReactiveMPC}$, $\mathcal{S}$ instructs the dummy corrupt data-client $C_i$ to forward the same input to $\mathcal{F}_\mathsf{AuditableMPC}$. The reason for this is this is the mechanism by which corrupt inputs are created in the real world; they must be created in the ideal world as well. The thing to note is that $\mathcal{A}$ cannot modify an input after it sends $\phi_{x_i}$ to $\mathcal{F}_\mathsf{ReactiveMPC}$ - this means any computation will happen on this input in the real world, and the computation in the ideal world should also happen on this input. $\mathcal{S}$ then forwards the message to the internal simulation and forwards any generated message to $\mathcal{Z}$ - as the emulation and actual real worlds were indistinguishable before this point, the output message will be indistinguishable in both worlds.
	
	\item $\mathcal{Z}$ instructs $\mathcal{D}$ to send a message to $\mathcal{F}_\mathsf{ReactiveMPC}$:
	
	We consider two separate cases here: (1) message related to triggering the optionally on line 16 of $\mathcal{F}_\mathsf{ReactiveMPC}$ in  Figure~\ref{fig:FReactive} and (2) all other messages.
	
	~\\In the first case:
	
	\textbf{Real World:} $\mathcal{D}$ instructs $\mathcal{F}_\mathsf{ReactiveMPC}$ to deliver $\mathbf{F}$ to a data-client $C_i$. 
	
	If $C_i$ is honest, by lines 7-12 of Figure~\ref{fig:PiAuditable}, $C_i$ sends inputs (that previously committed to in line 5) to $\mathcal{F}_\mathsf{AuditableMPC}$. By line 6, if this results in all the inputs to $\mathbf{F}$ being sent to $\mathcal{F}_\mathsf{ReactiveMPC}$, the $\mathsf{compute}$ subroutine is run and, by line 29 of Figure~\ref{fig:FReactive}, $(\mathbf{F}, o, \mathsf{inputs'}, \mathsf{wires'}, \mathsf{round})$ to $\mathcal{A}$ is sent to $\mathcal{D}$. Else, if it doesn't result in all inputs being sent, by line 7 of Figure~\ref{fig:FReactive}, $\mathcal{F}_\mathsf{ReactiveMPC}$ sends $C_i$ to $\mathcal{D}$. 
	
	Else, if $C_i$ is corrupt, $\mathcal{F}_\mathsf{ReactiveMPC}$ sends $\mathbf{F}$ to $C_i$, which is forwarded to $\mathcal{D}$. In each of these cases, the message to $\mathcal{D}$ is forwarded to $\mathcal{Z}$.
	
	\textbf{Ideal World:} By lines 33-35 of Figure~\ref{fig:UCSimulator_honestmaj}, $\mathcal{S}$ executes its internal simulation of the real world and generates the same types of messages as above. Except for messages of type $(\mathbf{F}, o, \mathsf{inputs'}, \mathsf{wires'}, \mathsf{round})$, these will be the same in both worlds. In the case of $(\mathbf{F}, o, \mathsf{inputs'}, \mathsf{wires'}, \mathsf{round})$ being forwarded to $\mathcal{A}$, we note that only $f \leq t$ shares of each input are given to $\mathcal{Z}$. This means that, although honest inputs were simulated as inputs of $0$, this is okay as $t+1$ shares are required to define a $t$-degree polynomial. From the perspective of $\mathcal{Z}$, shares in $\mathsf{inputs'}$ for actual inputs in the real world and simulated inputs from the emulated real world are therefore indistinguishable.
	
	~\\In the second case:
	
	\textbf{Real World:} $\mathcal{D}$ forwards the message to $\mathcal{F}_\mathsf{ReactiveMPC}$. One of two things might happen: (1) a dummy corrupt party forwards a copy of the output and shares from line 26 of Figure \ref{fig:FReactive} or (2) the auditor produces output. These would, in each case, be forwarded to $\mathcal{Z}$.

	\textbf{Ideal World:} By lines 33-35 of Figure~\ref{fig:UCSimulator_honestmaj}, $\mathcal{S}$ forwards the message to the internal emulation. As a copy of the real world is run internally, one of two things can happen here: the simulated $\mathcal{D}$ forwards a copy of the output and shares from line 26 of Figure \ref{fig:FReactive} to $\mathcal{S}$ or the auditor produces output (line 48-55 of Figure~\ref{fig:PiAuditable}). We maintain indistinguishability in either situation:

	(1) In line 35 of Figure~\ref{fig:UCSimulator_honestmaj}, $\mathcal{S}$ forwards copies of the simulated shares and output to $\mathcal{Z}$. As argued in the first case, these simulated shares are indistinguishable from actual honest shares.

	(2) By lines 45-47 of Figure~\ref{fig:UCSimulator_honestmaj}, $\mathcal{S}$ instructs $\mathcal{F}_\mathsf{AuditableMPC}$ to deliver the output to the auditor, which then forwards it to $\mathcal{Z}$. As $\mathcal{F}_\mathsf{ReactiveMPC}$ and $\mathcal{F}_\mathsf{AuditableMPC}$ generate outputs by evaluating the same function, and the inputs in both worlds are the same, the output will be the same in both worlds. As these outputs by the auditors in both worlds will be the same, we have that $\mathcal{Z}$ cannot distinguish in this case. We note that the outputs in both worlds being identical demonstrates the correctness of this protocol. 
	
	\item $\mathcal{Z}$ instructs $\mathcal{D}$ to send a message to $\mathcal{F}_\mathsf{BulletinBoard}$:
	
	\textbf{Real World:} $\mathcal{D}$ forwards the message to $\mathcal{F}_\mathsf{BulletinBoard}$. One of two things can happen: (1) a dummy corrupt party forwards a copy of the bulletin board to $\mathcal{D}$ (by line 7 of Figure~\ref{fig:FBulletin}) or (2) the auditor produces output (per lines 48-55 of Figure~\ref{fig:PiAuditable}). These would, in each case, be forwarded to $\mathcal{Z}$.

	\textbf{Ideal World:} By lines 36-44 of Figure~\ref{fig:UCSimulator_honestmaj}, $\mathcal{S}$ forwards the message to the internal emulation. As a copy of the real world is run internally, one of two things can happen here: the simulated $\mathcal{D}$ forwards a copy of the bulletin board to $\mathcal{S}$ (by line 7 of the simulated $\mathcal{F}_\mathsf{BulletinBoard}$ in Figure~\ref{fig:FBulletin}) or the auditor produces output (lines 48-55 of the simulated $\Pi_\mathsf{AuditableMPC}$ in Figure~\ref{fig:FAuditable}). We maintain indistinguishability in either case.

	(1) In lines 40-43 of Figure~\ref{fig:UCSimulator_honestmaj}, shares of the Marlin transcript and proof in the bulletin board are replaced with the versions from lines 15-18 of $\mathcal{S}$ in Figure~\ref{fig:UCSimulator_honestmaj}. As argued previously, these shares are indistinguishable in both worlds. We note that the indistinguishability here corresponds to the zero-knowledge property of this protocol.

	(2) By lines 45-47 of Figure~\ref{fig:UCSimulator_honestmaj}, $\mathcal{S}$ instructs $\mathcal{F}_\mathsf{AuditableMPC}$ to deliver the output to the auditor, which then forwards it to $\mathcal{Z}$. As $\mathcal{F}_\mathsf{ReactiveMPC}$ and $\mathcal{F}_\mathsf{AuditableMPC}$ generate outputs by evaluating the same function, and the inputs in both worlds are the same, the output will be the same in both worlds. Similar to before, this demonstrates the correctness of this protocol.

\end{itemize}

As indistinguishability is maintained despite any messages sent by $\mathcal{Z}$, $\mathcal{S}$ as defined in Figure~\ref{fig:UCSimulator_honestmaj} causes $\Pi_\mathsf{AuditableMPC}$ to UC-realize $\mathcal{F}_\mathsf{AuditableMPC}$ when $f \leq t$.

~\\
\textbf{Case 2: $\mathbf{f > t}$:}

We argue that $\mathcal{S}$, as defined in Figure \ref{fig:UCSimulator_dishonestmaj}, creates indistinguishability between the real and ideal worlds. We show that the internal simulation always tracks what occurs in the real world and additionally, the messages that $\mathcal{Z}$ is activated with in both worlds are indistinguishable.

In this setting, zero-knowledge is no longer a concern, as MPC does not guarantee confidentiality when $f > t$. Instead, since when $f > t$, corrupt inputs in line 3 of $\mathcal{F}_\mathsf{ReactiveMPC}$ (Figure~\ref{fig:FReactive}) can be different from those sent to parties in line 34, $\mathcal{S}$ must extract corrupt inputs from corrupt commitments to send to $\mathcal{F}_\mathsf{AuditableMPC}$. Additionally, per the ideal functionality, input independence still holds in this setting and our proof must demonstrate that.

Similar to the $f \leq t$ case, we analyze each possible input by $\mathcal{Z}$.

First, consider the messages that $\mathcal{Z}$ can send to honest parties:

\begin{itemize}
    \item $\mathcal{Z}$ provides an input $x_i$ to an honest data-client $C_i$:

	\textbf{Real World:} By lines 2-6 of Figure-\ref{fig:PiAuditable}, $C_i$ creates a commitment to $x_i$ and sends this to the bulletin board $\mathcal{F}_\mathsf{BulletinBoard}$.

	\textbf{Ideal World:} The dummy $C_i$ in the ideal world forwards $x_i$ to $\mathcal{F}_\mathsf{AuditableMPC}$. By line 4 of Figure~\ref{fig:FAuditable}, $\mathcal{F}_\mathsf{AuditableMPC}$ sends $C_i$ to $\mathcal{S}$. By lines 6-8 of Figure~\ref{fig:UCSimulator_dishonestmaj}, $\mathcal{S}$ simulates an internal input of $0$ by $C_i$ and all state transitions.

	In both cases, $\mathcal{Z}$ is activated with no incoming messages and $\mathcal{Z}$ will not be able to distinguish on its activation. Additionally, the internal simulation will be at the same point of the execution as the real world.
\end{itemize}

Next, consider the messages that $\mathcal{Z}$ can send the auditor.

\begin{itemize}
    \item $\mathcal{Z}$ provides a circuit $\mathbf{F}$ to the auditor:
    
    \textbf{Real World:} One of two things can happen here. Either (1) the auditor stores $\mathbf{F}$ or (2) the auditor forwards $\mathbf{F}$ to $\mathcal{F}_\mathsf{ReactiveMPC}$. 
    
    In the case of (1), nothing happens in the real world.
    
    In the case of (2), by line 16 of Figure~\ref{fig:PiAuditable}, $\mathcal{F}_\mathsf{ReactiveMPC}$ schedules an optional codeblock to send $\mathbf{F}$ to each $C_i$.

% 	\textbf{Real World:} By line 35 of Figure~\ref{fig:PiAuditable}, the auditor sends it to $\mathcal{F}_\mathsf{ReactiveMPC}$. $\mathcal{F}_\mathsf{ReactiveMPC}$ then executes lines 33-40 of the functionality, leaking F and all new inputs for the round of computation to $\mathcal{D}$. $\mathcal{D}$ then forwards these to $\mathcal{Z}$.

	\textbf{Ideal World:} The auditor sends $\mathbf{F}$ to $\mathcal{F}_\mathsf{AuditableMPC}$. By line 17 of Figure~\ref{fig:FAuditable}, $\mathcal{F}_\mathsf{AuditableMPC}$ sends $\mathbf{F}$ to $\mathcal{S}$. By lines 9-10 of Figure~\ref{fig:UCSimulator_dishonestmaj}, $\mathcal{S}$ emulates the same actions as in the real world internally. Since the simulated execution is at the same point as the real world execution, this means that the internal emulation does (1) and (2) exactly as above. In the case of (2), the simulator additionally runs lines 27-41 of Figure~\ref{fig:UCSimulator_dishonestmaj}. What this does is perform an extraction of corrupt inputs. Note that, as the auditor is unable to be corrupted, we have that the auditor only causes an output whenever the zk-SNARK verifies, per lines 52-54 of $\Pi_\mathsf{AuditableMPC}$ (Figure~\ref{fig:PiAuditable}). If we consider the extractability definition of zk-SNARKs in Section~\ref{sec:adapt_def}, we can see that $\mathcal{Z}$ satisfies the conditions for $\mathcal{A}$ in the definition. This means that, even though we consider the UC setting, we can apply Theorem~\ref{thm:snark} in this situation for the result that our zk-SNARK scheme is extractable. But as a result of this scheme being extractable, we know that the verification algorithm will pass only with negligible probability if input commitments are not to the inputs on which computation is performed. That means that, as $\mathcal{Z}$ is algebraic, we can use the representation provided by $\mathcal{Z}$ for corrupt commitments to extract inputs (by the binding property of commitments, this is the only representation that $\mathcal{Z}$, except with negligible probability). \\
	
	This extraction is exactly what lines 28-39 of Figure~\ref{fig:UCSimulator_dishonestmaj} do; lines 40-41 then send these inputs to $\mathcal{F}_\mathsf{AuditableMPC}$, thereby providing corrupt inputs for this round of execution. $\mathcal{S}$ also executes the $\mathsf{optionally}$ statement on line 12 of Figure~\ref{fig:FAuditable}, which results in all honest inputs being leaked to $\mathcal{S}$ in line 16. Once honest inputs are leaked, $\mathcal{S}$ modifies its internal emulation in lines 11-18 of Figure~\ref{fig:UCSimulator_dishonestmaj} to make it consistent with these inputs - as lines 13 and 16 modify polynomials that have not been sent to $\mathcal{Z}$ and line 14-15 use knowledge of the trapdoor to open commitments to new values, $\mathcal{S}$ is able to ensure that its internal execution is the same as the actual real world execution. \\
	
	At this point, the output to $\mathcal{Z}$ will be the same in both worlds, and the internal emulation will be at the same point as (or in case (2), identical to) the execution in the real world, maintaining indistinguishability. 
\end{itemize}

Finally, we analyze the different inputs from $\mathcal{Z}$ to the adversary.

\begin{itemize}
    \item $\mathcal{Z}$ instructs $\mathcal{D}$ to send message to $\mathcal{F}_\mathsf{RO}$, $\mathcal{F}_\mathsf{Setup}$, $\mathcal{F}_\mathsf{ReactiveMPC}$, $\mathcal{F}_\mathsf{BulletinBoard}$, or corrupt $P_i$/$C_i$:

	\textbf{Real World:} $\mathcal{D}$ forwards the message. There are 3 possibilities at this point: either (1) the auditor forwards $\mathbf{F}$ to $\mathcal{F}_\mathsf{ReactiveMPC}$, (2) the auditor generates an output, or (3) $\mathcal{D}$ receives some miscellaneous message.

	\textbf{Ideal World:} In lines 19-26 of Figure~\ref{fig:UCSimulator_dishonestmaj}, $\mathcal{S}$ forwards the message to its internal emulation. We consider each possibility described above in turn:
	
	(1) If the auditor forwards $\mathbf{F}$ to $\mathcal{F}_\mathsf{ReactiveMPC}$ in the real world, this will have been because $\mathcal{Z}$ instructed $\mathcal{D}$ to deliver a copy of the bulletin board to the auditor. By lines 41-42 and 45-47 of Figure~\ref{fig:PiAuditable}, the real world auditor will have received all commitments for inputs. This will also occur in the simulated real world and line 27 of Figure~\ref{fig:UCSimulator_dishonestmaj} will be triggered. Note that, in the ideal world, the auditor will already have sent $\mathcal{F}_\mathsf{AuditableMPC}$ the circuit $\mathbf{F}$. Despite this, we note that at this point, the same situation as described in case (2) of $\mathcal{Z}$ providing the auditor the circuit will happen at this point. $\mathcal{S}$ extracts and provides inputs to $\mathcal{F}_\mathsf{AuditableMPC}$ via lines 27-41 of Figure~\ref{fig:UCSimulator_dishonestmaj} in the same manner, and security still holds for the same reasons.
	
	(2) If the auditor generates an output in the real world, this will be because the verify function on line 52 of Figure~\ref{fig:PiAuditable} passes. As argued previously, the extractability of the zk-SNARK scheme implies that this only passes if the inputs and output are actually as it should be. By lines 42-44 of Figure~\ref{fig:UCSimulator_dishonestmaj}, $\mathcal{S}$ instructs $\mathcal{F}_\mathsf{AuditableMPC}$ to send this same output to the ideal world auditor, which will then forward it to $\mathcal{Z}$. As the outputs in the real and ideal worlds match, indistinguishability follows.
	
	(3) Finally, consider the case that $\mathcal{D}$ outputs any other message in the real world. There will be two types of messages: (a) those sent \textit{before} the auditor sends $\mathbf{F}$ to $\mathcal{F}_\mathsf{ReactiveMPC}$ and (b) those sent after. \\
	
	In the case of (a), note that the only messages that $\mathcal{Z}$ can send $\mathcal{D}$ will be to either query $\mathcal{F}_\mathsf{RO}$ and $\mathcal{F}_\mathsf{Setup}$, instructing appending/delivery of the bulletin board in $\mathcal{F}_\mathsf{BulletinBoard}$, instructing corrupt parties to send a message to $\mathcal{F}_\mathsf{BulletinBoard}$, or instructing corrupt inputs to $\mathcal{F}_\mathsf{ReactiveMPC}$. None of these rely on any secret information, so $\mathcal{S}$ will simulate exactly what occurs in the real world.\\
	
	Now consider case (b). But by this point, $\mathcal{S}$ will have run lines 40-41 in Figure~\ref{fig:UCSimulator_dishonestmaj}. This will have triggered line 16 in Figure~\ref{fig:FAuditable} and by lines 11-18 of Figure~\ref{fig:UCSimulator_dishonestmaj}, $\mathcal{S}$ will already have modified its internal emulation to account for actual honest inputs. This means that the emulation will be exactly like that of the real world, and so any outputs in the simulated real world will be exactly that which happens in the actual real world.
\end{itemize}

Thus, the simulator defined in Figure~\ref{fig:UCSimulator_honestmaj} for $f \leq t$ corruptions and defined in Figure~\ref{fig:UCSimulator_dishonestmaj} for $f>t$ corruptions causes the real and ideal worlds to be indistinguishable to $\mathcal{Z}$. So we have that $\Pi_\mathsf{AuditableMPC}$ UC-realizes $\mathcal{F}_\mathsf{AuditableMPC}$ in the ($\mathcal{F}_\mathsf{Setup}$, $\mathcal{F}_\mathsf{RO}$, $\mathcal{F}_\mathsf{BulletinBoard}$, $\mathcal{F}_\mathsf{ReactiveMPC}$) hybrid world.

\begin{figure}[!htbp]
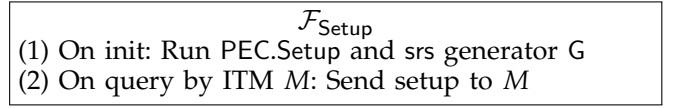
 
\begin{boxedminipage}[t]{0.5\textwidth}
  {\centering \textbf{$\mathcal{F}_\mathsf{Setup}$}\\}
(1)
On init:  Run $\mathsf{PEC.Setup}$ and $\mathsf{srs}$ generator $\mathsf{G}$

(2)
On query by ITM $M$: Send setup  to $M$

\end{boxedminipage}
\label{fig:Fsetup}
\caption{Ideal functionality for trusted setup}
\end{figure}

\begin{figure}[!htbp]
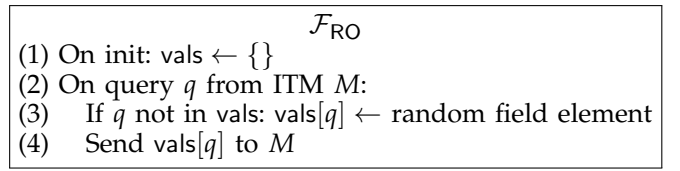
 
\begin{boxedminipage}[t]{0.5\textwidth}
  {\centering \textbf{$\mathcal{F}_\mathsf{RO}$}\\}

(1)
On init: $\mathsf{vals} \leftarrow \{ \}$

(2)
On query $q$ from ITM $M$:

(3)
    \quad If $q$ not in $\mathsf{vals}$: $\mathsf{vals}[q] \leftarrow$ random field element
   
(4)    
    \quad Send $\mathsf{vals}[q]$ to $M$
\end{boxedminipage}
\label{fig:FRO}
\caption{Ideal functionality for random oracle}
\end{figure}

\begin{figure}[!htbp]
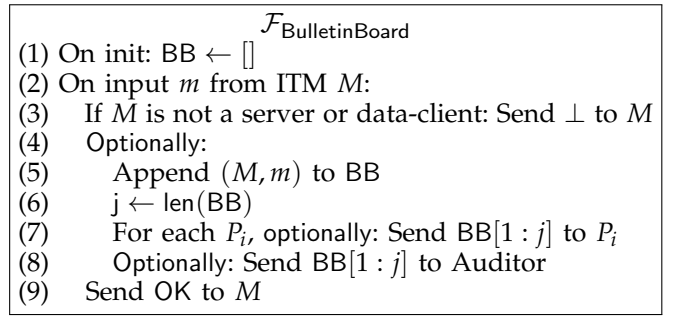
 
\begin{boxedminipage}[t]{0.5\textwidth}
  {\centering \textbf{$\mathcal{F}_\mathsf{BulletinBoard}$}\\}
%\internallinenumbers[1]

(1)
On init: $\mathsf{BB} \leftarrow [ ]$

(2)
On input $m$ from ITM $M$:

(3)
    \quad If $M$ is not a server or data-client: Send $\bot$ to $M$
    
(4)
    \quad $\mathsf{Optionally}$:
    
(5)
        \qquad Append $(M, m)$ to $\mathsf{BB}$
        
(6)
        \qquad $\mathsf{j} \leftarrow \mathsf{len}(\mathsf{BB})$
        
(7)
        \qquad For each $P_i$, $\mathsf{optionally}$: Send $\mathsf{BB}[1:j]$ to $P_i$
        
(8)
        \qquad $\mathsf{Optionally}$: Send $\mathsf{BB}[1:j]$ to Auditor

(9)
    \quad Send $\mathsf{OK}$ to $M$
  
\end{boxedminipage}
\caption{Ideal functionality for bulletin board}
\label{fig:FBulletin}
\end{figure}

\begin{figure}[!htbp] 
\begin{boxedminipage}[t]{0.5\textwidth}
  {\centering \textbf{$\Pi_\mathsf{AuditableMPC}$}\\}
 % \internallinenumbers[1]

\ (1)
\textbf{As data-client $C_i$:}

\ (2)
On input $x_i$ from $\mathcal{Z}$:

    % \quad Send $\mathsf{RAND}$ to $\mathcal{F}_\mathsf{ReactiveMPC}$ and get $r_i$
    
    % \quad $C_{x_i} \leftarrow \mathsf{Commit_{eval}}(\mathsf{ck_{e_i}}, x_i, r_i)$
    
    % \quad Send $C_{x_i}$ to $\mathcal{F}_\mathsf{BulletinBoard}$ and receive $\mathsf{OK}$
    
    % \quad Create random t-degree polynomial $\phi(\cdot)$ s.t. $\phi(0) = x_i$
    
    % \quad Send $\phi(\cdot)$ to $\mathcal{F}_\mathsf{ReactiveMPC}$

\ (3)
    \quad $r_i \xleftarrow{\$} \mathbb{Z}_q$ 
    
\ (4)    
    \quad $C_{x_i} \leftarrow \mathsf{Commit_{eval}}(\mathsf{ck_{e_i}}, x_i, r_i)$

\ (5)    
    \quad Send $C_{x_i}$ to $\mathcal{F}_\mathsf{BulletinBoard}$ and receive $\mathsf{OK}$

\ (6)    
    \quad Store $(x_i, r_i)$
   
\ (7)    
On input $\mathbf{F}$ from $\mathcal{F}_\mathsf{ReactiveMPC}$:

\ (8)
    \quad For each stored $(x_i, r_i)$: 

\ (9)    
        \qquad $\phi_{x_i}(\cdot) \leftarrow$ rand t-deg poly s.t. $\phi(0) = x_i$

(10)        
        \qquad $\phi_{r_i}(\cdot) \leftarrow$ rand t-deg poly s.t. $\phi(0) = r_i$

(11)        
    \quad Unstore all $(x_i, r_i)$

(12)    
    \quad Send $\{(\phi_{x_i}, \phi_{r_i})\}_i$ to $\mathcal{F}_\mathsf{ReactiveMPC}$

(13)    
\textbf{As server $P_i$:}

(14)
On init: $\mathsf{BB} \leftarrow [\ ]$

(15)
On input $\mathsf{BB}'$ from $\mathcal{F}_\mathsf{BulletinBoard}$:

(16)
    \quad If $\mathsf{len(BB)'} > \mathsf{len(BB)}$:

(17)
        \qquad $\mathsf{BB} \leftarrow \mathsf{BB}'$

(18)        
        \qquad Run $\mathsf{process\_inputs}$

(19)        
On $(\mathbf{F}, o, \mathsf{inputs, wires, round})$ from $\mathcal{F}_\mathsf{ReactiveMPC}$:

(20)
    \quad Send $(\mathsf{RANDS}, \mathsf{round})$ to $\mathcal{F}_\mathsf{Reactive}$ and get $\mathsf{rands}$

(21)
    \quad $\share{x_{i_1}}, \ldots, \share{x_{i_k}} \leftarrow$ shares of new inputs

(22)    
    \quad $\share{r_{i_1}}, \ldots, \share{r_{i_k}} \leftarrow$ shares in $\mathsf{rands}$ for new inputs

(23)    
    \quad $\share{\hat{x}} \leftarrow$ interpolation of $\share{x_{i_1}}, \ldots, \share{x_{i_k}}$
    
(24)    
    \quad $\share{\hat{r}} \leftarrow$ interpolation of $\share{r_{i_1}}, \ldots, \share{r_{i_k}}$
    
(25)    
    \quad $C_\share{\hat{x}} \leftarrow \mathsf{PC.commit(ck, \share{\hat{x}}, \share{\hat{r}}})$

(26)    
    \quad Send $C_\share{\hat{x}}$ to $\mathcal{F}_\mathsf{BulletinBoard}$

(27)    
    \quad Store $(\mathbf{F}, o, \mathsf{inputs, wires, round})$

(28)    
    \quad Run $\mathsf{process\_inputs}$

(29)    
Subroutine $\mathsf{process\_inputs}$:

(30)
    \quad For each stored $(\mathbf{F}, o, \mathsf{inputs, wires, round})$:

(31)    
    \qquad $\mathbf{C} \leftarrow$ shares of $C_{\hat{x}}$ for this round of inputs

(32)    
    \qquad $b \leftarrow $ bit for if Step 4 of Appendix~\ref{app:sec_mpc_polycommit} verifies

(33)    
    \qquad If $b$: 

(34)        
        \quad \qquad $\share{\pi} \leftarrow $ prover protocol in Section~\ref{sec:mpc_prover}

(35)        
        \quad \qquad Send $\share{\pi}$ to $\mathcal{F}_\mathsf{BulletinBoard}$ and receive $\mathsf{OK}$

(36)        
        \quad \qquad Unstore $(\mathbf{F}, o, \mathsf{inputs, wires, round})$

(37)    
    \qquad If not $b$ and $\mathsf{len}(\mathbf{C}) == n$: abort

(38)  
\textbf{As Auditor:}

(39)
On init: $\mathsf{BB} \leftarrow [\ ]$

(40)
On input circuit $\mathbf{F}$ from $\mathcal{Z}$: Store $\mathbf{F}$ and run $\mathsf{audit}$

(41)
On input $\mathsf{BB'}$ from $\mathcal{F}_\mathsf{BulletinBoard}$:

(42)
\quad If $\mathsf{len(BB')} > \mathsf{len(BB)}$: $\mathsf{BB} \leftarrow \mathsf{BB'}$ and run $\mathsf{audit}$

(43)
On input $(\mathbf{F}, o, \mathsf{round})$ from $\mathcal{F}_\mathsf{ReactiveMPC}$:

(44)
    \quad Store $(\mathbf{F}, o, \mathsf{round})$ and run $\mathsf{audit}$

(45)
Subroutine $\mathsf{audit}$:

(46)
    \quad For each stored $\mathbf{F}$:
    
(47)
    \qquad Send $\mathbf{F}$ to $\mathcal{F}_\mathsf{ReactiveMPC}$ if $\mathsf{BB}$ has $C_{x_i}$ for all  $x_i$

(48)
    \quad For each stored $(\mathbf{F}, o, \mathsf{round})$:

(49)
    \qquad If $\mathsf{BB}$ contains the entire Marlin proof:

(50)
        \quad \qquad Run verify function described in Fig. \ref{fig:adapt-snark} on the 

(51)        
        \quad \qquad transcript in BB and $(\mathbf{F}, o, \mathsf{round})$

(52)
        \quad \qquad If verify function passes:

(53)
            \qquad \qquad Unstore $(\mathbf{F}, o, \mathsf{round})$ 

(54)
            \qquad \qquad Send $(\mathbf{F}, o)$ to $\mathcal{Z}$

(55)
        \quad \qquad Else: abort

\end{boxedminipage}
\caption{UC protocol for auditable MPC}
\label{fig:PiAuditable}
\end{figure}

\begin{figure}[!htbp] 
\begin{boxedminipage}[t]{0.5\textwidth}
  {\centering \textbf{$\mathcal{F}_\mathsf{ReactiveMPC}$}\\}
%\internallinenumbers[1]

\ (1)
On init: $\mathsf{inputs} \leftarrow [\ ]$, $\mathsf{new\_inputs} \leftarrow [\ ]$, $\mathsf{rands} \leftarrow \{\}$,

\ (2)
$\mathsf{round} \leftarrow 0$

\ (3)
On input $(\phi_{x_i}, \phi_{r_i})$ from data-client $C_i$:

\ (4)
    \quad Append $\phi_{x_i}$ to $\mathsf{new\_inputs}$
   
\ (5)    
    \quad Append $\phi_{r_i}$ to $\mathsf{rands[round]}$

\ (6)    
    \quad If all inputs in $\mathbf{F}$ received: Execute $\mathsf{compute}$

\ (7)    
    \quad Else: send $C_i$ to $\mathcal{A}$

\ (8)    
On input $(\mathsf{RANDS, round'})$ from server $P_i$:

\ (9)
    \quad If $\mathsf{round'} > \mathsf{round}$: Send $[\ ]$ to $P_i$

(10)
    \quad If rand shares for proof for $\mathsf{round'}$ not generated:

(11)    
        \qquad $\phi_{r'_1}, \dots, \phi_{r'_k} \leftarrow $ random t-deg polys for SNARK

(12)        
        \qquad Append $\phi_{r'_1}, \dots, \phi_{r'_k}$ to $\mathsf{rands[round']}$

(13)        
    \quad Send $[\phi(i) \text{ for each $\phi$ in } \mathsf{rands[round']}]$ to $P_i$

(14)
On input circuit $\mathbf{F}$ from Auditor:

(15)
    \quad $\mathsf{round} \leftarrow \mathsf{round} + 1$
    
(16)
    \quad Optionally: For each data-client $C_i$, send $\mathbf{F}$ to $C_i$

(17)    
Subroutine $\mathsf{compute}$:
    
(18)
    \quad If $f \leq t$:

(19)
        \qquad Add $\mathsf{new\_inputs}$ to $\mathsf{inputs}$ and $\mathsf{new\_inputs} \leftarrow [\ ]$
        
(20)    
        \qquad $o \leftarrow \mathbf{F}(\mathsf{inputs})$

(21)
        \qquad $\mathsf{wires} \leftarrow$ rand shares consistent with $\mathbf{F}$

(22)    
        \qquad $\mathsf{Optionally}$: Send $(\mathbf{F}, o, \mathsf{round})$ to Auditor

(23)    
        \qquad For each $P_i$, optionally: 

(24)    
            \qquad \quad $\mathsf{inputs'} \leftarrow $ $P_i$'s shares of each input

(25)        
            \qquad \quad $\mathsf{wires'} \leftarrow $ $P_i$'s shares of each wire

(26)    
            \qquad \quad Send $(\mathbf{F}, o, \mathsf{inputs'}, \mathsf{wires'}, \mathsf{round})$ to $P_i$

(27)    
        \qquad $\mathsf{inputs'} \leftarrow $ corrupt shares of each input

(28)        
        \qquad $\mathsf{wires'} \leftarrow $ corrupt shares of each wire

(29)        
        \qquad Send $(\mathbf{F}, o, \mathsf{inputs'}, \mathsf{wires'}, \mathsf{round})$ to $\mathcal{A}$
    
(30)    
    \quad Else:
    
(31)
        \qquad $\mathsf{Optionally}$ ($\mathcal{A}$ providing $\mathsf{o'}$):
    
(32)    
            \quad \qquad Send $(\mathbf{F}, o', \mathsf{round})$ to Auditor
    
(33)    
         \qquad For each $P_i$:

(34)     
            \quad \qquad Optionally ($\mathcal{A}$ sending $o'/\mathsf{new\_inputs}'/ \mathsf{wires}'$): 

(35)        
                \qquad \qquad Append $\mathsf{new\_inputs}'$ to $\mathsf{inputs[i]}$

(36)    
                \qquad \qquad Send $(\mathbf{F}, o', \mathsf{inputs[i]}, \mathsf{wires'}, \mathsf{round})$ to $P_i$

(37)    
        \qquad Send $(\mathbf{F}, \mathsf{new\_inputs}, \mathsf{round})$ to $\mathcal{A}$

\end{boxedminipage}
\caption{UC ideal functionality for reactive MPC}
\label{fig:FReactive}
\end{figure}

\begin{figure}[!htbp] 
\begin{boxedminipage}[t]{0.5\textwidth}
  {\centering \textbf{$\mathcal{S}_\mathsf{AuditableMPC}$}\\}
%\internallinenumbers[1]

\ (1)
\textbf{Case 1: $\mathbf{f \leq t}$:}

\ (2)
On init: Initialize internal emulation of real world, 

\ (3)
including simulated parties and functionalities. 

\ (4) 
Additionally, $\mathcal{F}_\mathsf{Setup}$ is initialized using $\mathsf{\mathcal{S}_{PEC}.Setup}$ 

\ (5)
and $\mathsf{\mathcal{S}_{zk-SNARK}.Setup}$

\ (6)
On input data-client $C_i$ from $\mathcal{F}_\mathsf{AuditableMPC}$:

\ (7)
    \quad If $C_i$ is honest: 

\ (8)    
        \qquad Simulate $C_i$ being provided an input of $0$

\ (9)        
On input $(\mathbf{F}, o)$ from $\mathcal{F}_\mathsf{AuditableMPC}$:

(10)
    \quad $C_o, C_r \leftarrow $ polycommits to $o$ and extra degree of $\hat{x}$

(11)    
    \quad \ \  from $\Pi_\mathsf{Marlin}$, using internally simulated $r_{1, \dots, k}(\cdot)$

(12)        
    \quad $C_{\hat{x}} \leftarrow [\text{simulated/adv. commits to inputs}, C_o, C_r]$
    
(13)    
    \quad $(\mathsf{transcript}, \pi)$ $\leftarrow$ $\mathsf{\mathcal{S}_{SNARK}.Prove(\mathsf{td}_{srs}, \mathsf{td}_{comm}, C_{\hat{x}}, \mathbb{I})}$

(14)    
    \quad For each polycommit $C$ in $\mathsf{transcript}$:

(15)    
        \qquad Create share of $C$ for each honest $P_i$, consistent 

(16)        
            \quad \qquad with the $f$ shares of corrupt parties and $C$ (1)

(17)            
    \quad Create share of $\pi$ for each honest $P_i$, consistent

(18)    
        \qquad with the $f$ shares of corrupt parties and $\pi$ (2)

(19)        
    \quad Store $(\mathsf{transcript}, \mathsf{shares}, \pi, \mathbf{F}, o)$

(20)    
    \quad Pass $\mathbf{F}$ to the simulated Auditor, simulating 

(21)    
        \qquad $\mathcal{F}_\mathsf{ReactiveMPC}$ substituting $o$ as the output

(22)        
On input $(m, M')$ from $\mathcal{Z}$:

(23)
    \quad If $M'$ is an honest $C_i$ or $P_i$: end activation

(24)    
    \quad If $M'$ is a corrupt data-client $C_i$: 

(25)    
        \qquad If $m$ is sending $(\phi_{x_i}, \phi_{r_i})$ to $\mathcal{F}_\mathsf{ReactiveMPC}$: 
    
(26)    
            \quad \qquad Instruct $C_i$ to send $\phi_{x_i}(0)$ to $\mathcal{F}_\mathsf{AuditableMPC}$

(27)            
            \quad \qquad and wait for message $C_i$ from $\mathcal{F}_\mathsf{AuditableMPC}$

(28)    
        \qquad Send $(m, M')$ to simulated $\mathcal{D}$.

(29)        
        \qquad If $\mathcal{D}$ returns $m$, forward $m$ to $\mathcal{Z}$

(30)        
    \quad If $M'$ is a corrupt server $P_i$: 

(31)    
        \qquad Send $(m, M')$ to simulated $\mathcal{D}$.

(32)        
        \qquad If $\mathcal{D}$ returns $m$, forward $m$ to $\mathcal{Z}$

(33)        
    \quad If $M'$ is $\mathcal{F}_\mathsf{RO}$, $\mathcal{F}_\mathsf{Setup}$, or $\mathcal{F}_\mathsf{ReactiveMPC}$:

(34)        
        \qquad Send $(m, M')$ to simulated $\mathcal{D}$. 

(35)        
        \qquad If $\mathcal{D}$ returns $m$, forward $m$ to $\mathcal{Z}$

(36)        
    \quad If $M'$ is $\mathcal{F}_\mathsf{BulletinBoard}$:

(37)    
        \qquad Send $m$ to simulated $\mathcal{F}_\mathsf{BulletinBoard}$

(38)        
        \qquad If $\mathcal{D}$ returns $(P_j, \mathsf{BB})$ for corrupt $P_j$ and bulletin 

(39)        
            \qquad \quad board $\mathsf{BB}$:

(40)            
                \qquad \qquad Replace honest shares of Marlin commits 

(41)                
                \quad \qquad \qquad in $\mathsf{BB}$ with stored shares from (1)
    
(42)            
                \qquad \qquad Replace honest shares of $\pi$ in $\mathsf{BB}$ with 

(43)    
                \quad \qquad \qquad stored shares from (2)

(44)                
                \qquad \qquad Send $(P_j, \mathsf{BB})$ to $\mathcal{Z}$

(45)                
If the simulated Auditor outputs $(\mathbf{F}, o)$:

(46)
    \quad Instruct $\mathcal{F}_\mathsf{AuditableMPC}$ to execute the $\mathsf{optionally}$

(47)    
    \quad statement to send $(\mathbf{F}, o)$ to the Auditor

\end{boxedminipage}
\caption{Simulator for UC proof when $f \leq t$}
\label{fig:UCSimulator_honestmaj}
\end{figure}

\begin{figure}[!htbp] 
\begin{boxedminipage}[t]{0.5\textwidth}
  {\centering \textbf{$\mathcal{S}_\mathsf{AuditableMPC}$ (continued)}\\}    
%\internallinenumbers[1]

\ (1)
\textbf{Case 2: $\mathbf{f > t}$:}

\ (2)
On init: Initialize internal emulation of real world, 

\ (3)
including simulated parties and functionalities. 

\ (4)
Additionally, $\mathcal{F}_\mathsf{Setup}$ is initialized using $\mathsf{\mathcal{S}_{PEC}.Setup}$ 

\ (5)
and $\mathsf{\mathcal{S}_{zk-SNARK}.Setup}$

\ (6)
On input data-client $C_i$ from $\mathcal{F}_\mathsf{AuditableMPC}$:

\ (7)
    \quad If $C_i$ is honest: 

\ (8)    
        \qquad Simulate $C_i$ being provided an input of $0$

\ (9)        
On input $\mathbf{F}$ from $\mathcal{F}_\mathsf{AuditableMPC}$:

(10)
    \quad Send $\mathbf{F}$ to simulated Auditor

(11)
On input $(\mathbf{F}, \mathsf{new})$ from $\mathcal{F}_\mathsf{AuditableMPC}$:

(12)
    \quad For each honest input $x_i$ in $\mathsf{new}$:

(13)    
        \qquad $\phi_{x_i} \leftarrow $ random $t$-degree consistent with $x_i$

(14)        
        \qquad $C_{x_i} \leftarrow$ internally simulated commitment for $x_i$

(15)
        \qquad $r_i \leftarrow$ randomness to commit $x_i$ to $C_{x_i}$

(16)
        \qquad $\phi_{r_i} \leftarrow $ random $t$-degree consistent with $r_i$

(17)        
        \qquad Modify internal simulation of $\mathcal{F}_\mathsf{Reactive}$ to use

(18)        
        \qquad \quad $\phi_{x_i}$ and $\phi_{r_i}$ for input $x_i$

(19)        
On input $(m, M')$ from $\mathcal{Z}$:

(20)
    \quad If $M'$ is an honest $C_i$ or $P_i$: end activation

(21)    
    \quad If $M'$ is a corrupt $C_i$ or $P_i$:

(22)    
        \qquad Send $(m, M')$ to simulated $\mathcal{D}$.

(23)    
    \quad If $M'$ is $\mathcal{F}_\mathsf{RO}$, $\mathcal{F}_\mathsf{Setup}$, $\mathcal{F}_\mathsf{BulletinBoard}$, or $\mathcal{F}_\mathsf{Reactive}$:

(24)    
        \qquad Send $(m, M')$ to simulated $\mathcal{D}$. 

(25)        
        \qquad When $\mathcal{D}$ returns $m$, forward $m$ to $\mathcal{Z}$, unless the

(26)        
        \qquad if statement below is triggered

(27)        
If the simulated Auditor provides $\mathbf{F}$ to $\mathcal{F}_\mathsf{ReactiveMPC}$:

(28)
    \quad For each input commit $C_{x_i}$ on the bulletin board:
 
(29)   
        \qquad If $x_i$ is not calculated for a corrupt party's $C_{x_i}$:

(30)
            \qquad \quad $a_{1\dots k} \leftarrow$ AGM repr. of $C_{x_i}$ provided by $Z$ in

(31)            
            \qquad \qquad terms of SRS and group elems $\mathcal{S}$ sent $\mathcal{Z}$

(32)            
            \qquad \quad $a'_{1\dots l} \leftarrow$ repr. of $C_{x_i}$ in terms of the SRS and

(33)            
            \qquad \qquad elems created that $\mathcal{S}$ has no repr. for

(34)            
            \qquad \quad For each elem created with no representation:

(35)            
            \qquad \qquad Create a representation w.r.t. the SRS using 

(36)            
            \qquad \qquad the trapdoor

(37)
            \qquad \quad $[\overline{a_1},\dots,\overline{a_m}] \leftarrow$ repr. of $C_{x_i}$ in terms of the SRS

(38)            
            \qquad \quad $x_i, r_i \leftarrow$ $[\overline{a_1},\dots,\overline{a_m}]$

(39)            
            \qquad \quad Store $x_i$ for input to $\mathcal{F}_\mathsf{AuditableMPC}$

(40)            
    \quad Send all $x_i$ to $\mathcal{F}_\mathsf{AuditableMPC}$ via dummy corrupt
    
(41)
    \quad parties and instruct execution of inside optionally 

(42)    
If the simulated Auditor outputs $(\mathbf{F}, o)$:

(43)
    \quad Instruct $\mathcal{F}_\mathsf{AuditableMPC}$ to execute the $\mathsf{optionally}$
    
(44)
    \quad statement to send $(\mathbf{F}, o)$ to the Auditor

\end{boxedminipage}
\caption{Simulator for UC proof when $f > t$}
\label{fig:UCSimulator_dishonestmaj}
\end{figure}

\vspace{12pt}

\end{document}